\documentclass[twocolumn,showpacs,preprintnumbers,amsmath,amssymb]{revtex4}
\usepackage{amsmath}
\usepackage{graphicx}% Include figure files
\usepackage{dcolumn}% Align table columns on decimal point
\usepackage{bm}% bold math

\begin{document}

\newcommand{\etal}{{\it et al.}}
\makeatother

\newcommand{\bra}[1]{\left\langle #1 \right|}
\newcommand{\brared}[1]{\langle #1 ||}
\newcommand{\ee}{\eta^{\ast},\eta}
\newcommand{\product}[2]{\left\langle #1 | #2 \right\rangle}

\newcommand{\kbar}{\bar{k}}
\newcommand{\ket}[1]{\left| #1 \right\rangle}
\newcommand{\ketred}[1]{|| #1 \rangle}
\newcommand{\ahat}{\hat{a}}
\newcommand{\adag}{a^{\dagger}}
\newcommand{\ahatdag}{\hat{a}^{\dagger}}
\newcommand{\Ahat}{\hat{A}}
\newcommand{\Adag}{A^{\dagger}}
\newcommand{\Ahatdag}{\hat{A}^{\dagger}}
\newcommand{\Atdag}{\hat{A}^{(\tau)\dagger}}
\newcommand{\Bdag}{\hat{B}^{\dagger}}
\newcommand{\Bhat}{\hat{B}}
\newcommand{\Bhatdag}{\hat{B}^{\dagger}}
\newcommand{\Btdag}{\hat{B}^{(\tau)\dagger}}
\newcommand{\bhat}{\hat{b}}
\newcommand{\bdag}{b^{\dagger}}
\newcommand{\cdag}{c^{\dagger}}
\newcommand{\chat}{\hat{c}}
\newcommand{\chatdag}{\hat{c}^{\dagger}}
\newcommand{\degree}{^{\circ}}
\newcommand{\sprime}{s^{\prime}}
\newcommand{\Hhat}{\hat{H}}
\newcommand{\Hhatp}{\hat{H}^{\prime}}
\newcommand{\Ihat}{\hat{I}}
\newcommand{\Jhat}{\hat{J}}
\newcommand{\hhat}{\hat{h}}

\newcommand{\fp}{f^{(+)}}
\newcommand{\fpp}{f^{(+)\prime}}
\newcommand{\fm}{f^{(-)}}
\newcommand{\Fhat}{\hat{F}}
\newcommand{\Fhatdag}{\hat{F}^\dagger}
\newcommand{\Fhatp}{\hat{F}^{(+)}}
\newcommand{\Fhatm}{\hat{F}^{(-)}}
\newcommand{\Fhatpm}{\hat{F}^{(\pm)}}
\newcommand{\Fhatdagpm}{\hat{F}^{\dagger(\pm)}}
\newcommand{\Hc}{{\cal H}}
\newcommand{\Hcp}{{\cal H}^{\prime}}
\newcommand{\Ic}{{\cal I}}
\newcommand{\It}{\widetilde{I}}
\newcommand{\ITV}{{\cal I}_{\rm TV}}
\newcommand{\Jc}{{\cal J}}
\newcommand{\jp}{j^{\prime}}
\newcommand{\Qc}{{\cal Q}}
\newcommand{\Pc}{{\cal P}}
\newcommand{\Ec}{{\cal E}}
\newcommand{\Sc}{{\cal S}}
\newcommand{\Rc}{{\cal R}}

\newcommand{\ddg}{d^{\dagger}}

\newcommand{\Nhat}{\hat{N}}
\newcommand{\Nt}{\widetilde{N}}
\newcommand{\Vt}{\widetilde{V}}
\newcommand{\nL}[1]{n_{L_{#1}}}
\newcommand{\nK}[1]{n_{K_{#1}}}
\newcommand{\nKb}{\mbox{\boldmath $n_K$}}
\newcommand{\nLb}{\mbox{\boldmath $n_L$}}

\newcommand{\mubar}{\bar{\mu}}

\newcommand{\Dc}{{\mathscr D}}
\newcommand{\Ddag}{\hat{D}^{\dagger}}
\newcommand{\dhat}{\hat{d}}
\newcommand{\Dhat}{\hat{D}}
\newcommand{\Ghat}{\hat{G}}
\newcommand{\Glambda}{G^{(\lambda)}}
\newcommand{\Gstarlambda}{G^{(\lambda)\ast}}
\newcommand{\Qhat}{\hat{Q}}
\newcommand{\Rhat}{\hat{R}}
\newcommand{\Phat}{\hat{P}}
\newcommand{\Pdag}{\hat{P}^{\dagger}}
\newcommand{\Psihat}{\hat{\Psi}}
\newcommand{\Qdag}{Q^{\dagger}}
\newcommand{\That}{\hat{\Theta}}
\newcommand{\Thatt}{\widetilde{\hat{\Theta}}}
\newcommand{\Tr}{{\rm Tr}}

\newcommand{\ktilde}{\tilde{k}}

\newcommand{\Pcirc}{\stackrel{\circ}{P}}
\newcommand{\Qcirc}{\stackrel{\circ}{Q}}
\newcommand{\Ncirc}{\stackrel{\circ}{N}}
\newcommand{\Tcirc}{\stackrel{\circ}{\Theta}}
\newcommand{\Pcircp}{\stackrel{\circ}{P^{\prime}}}
\newcommand{\Qcircp}{\stackrel{\circ}{Q^{\prime}}}
\newcommand{\Ncircp}{\stackrel{\circ}{N^{\prime}}}
\newcommand{\Tcircp}{\stackrel{\circ}{\Theta^{\prime}}}
\newcommand{\Fp}{F^{(+)}}
\newcommand{\Fm}{F^{(-)}}
\newcommand{\Rp}{R^{(+)}}
\newcommand{\Rm}{R^{(-)}}
\newcommand{\Bt}{\widetilde{B}}
\newcommand{\lambdat}{\widetilde{\lambda}}
\newcommand{\Phatt}{\widetilde{\hat{P}}}
\newcommand{\ab}{\bf a}

\newcommand{\Ab}{\mbox{\boldmath $A$}}
\newcommand{\Abdag}{\mbox{\boldmath $A$}^{\dagger}}
\newcommand{\Bb}{\mbox{\boldmath $B$}}
\newcommand{\cb}{\bf c}
\newcommand{\Db}{\mbox{\boldmath $D$}}
\newcommand{\Nb}{\mbox{\boldmath $N$}}
\newcommand{\Nbhat}{\hat{\mbox{\boldmath $N$}}}
\newcommand{\Qb}{\mbox{\boldmath $Q$}}
\newcommand{\Qhatt}{\widetilde{\hat{Q}}}
\newcommand{\Pb}{\mbox{\boldmath $P$}}
\newcommand{\phit}{\phi(t)}
\newcommand{\pdot}{\dot{p}}
\newcommand{\phix}[1]{\phi(#1)}
\newcommand{\qdot}{\dot{q}}
\newcommand{\phivib}{\phi(\eta^{\ast},\eta)}
\newcommand{\Ts}{{\cal T}}
\newcommand{\del}{\partial}
\newcommand{\eps}{\epsilon}
\newcommand{\beq}{\begin{equation}}
\newcommand{\beqa}{\begin{eqnarray}}
\newcommand{\eeq}{\end{equation}}
\newcommand{\eeqa}{\end{eqnarray}}
\newcommand{\Yb}{${}^{168}$Yb\ }
\newcommand{\Zhat}{\hat{Z}}
\newcommand{\rhodot}{\dot{\rho}}
\newcommand{\Khat}{\hat{K}}
\newcommand{\Kp}{K^{+}}
\newcommand{\Km}{K^{-}}
\newcommand{\Kz}{K^0}

\newcommand{\lb}{\bf l}
\newcommand{\sbold}{\bf s}

\newcommand{\Lp}{L^{+}}
\newcommand{\Lm}{L^{-}}
\newcommand{\Lz}{L^0}

\newcommand{\Mc}{{\cal M}}
\newcommand{\Mchat}{\hat{\cal M}}

\newcommand{\ddeta}{\frac{\partial}{\partial \eta}}
\newcommand{\ddetastar}{\frac{\partial}{\partial \eta^\ast}}
\newcommand{\etastar}{\eta^\ast}
\newcommand{\ketvib}{\ket{\phi (\etastar, \eta)}}
\newcommand{\bravib}{\bra{\phi (\etastar, \eta)}}
\newcommand{\zhateta}{\hat{z}(\eta)}
\newcommand{\zhat}{\hat{z}}
\newcommand{\oo}{\stackrel{\circ}{O}(\etastar,\eta)}
\newcommand{\oodag}{\stackrel{\circ}{O^{\dagger}}(\etastar,\eta)}
\newcommand{\oodagp}{\stackrel{\circ}{O^{\dagger\prime}}(\etastar,\eta)}
\newcommand{\oop}{\stackrel{\circ}{O^{\prime}}(\etastar,\eta)}
\newcommand{\Odag}{\hat{O}^{\dagger}}
\newcommand{\Ohat}{\hat{O}}
\newcommand{\Uinv}{U^{-1}(\etastar, \eta)}
\newcommand{\Uinvp}{U^{-1}(\etastar,\eta,\varphi,n)}
\newcommand{\U}{U(\etastar, \eta)}
\newcommand{\Up}{U(\etastar,\eta,\varphi,n)}
\newcommand{\etader}{\frac{\del}{\del \eta}}
\newcommand{\etastarder}{\frac{\del}{\del \etastar}}

\newcommand{\fb}{\mbox {\bfseries\itshape f}}
\newcommand{\SB}{\mbox {\bfseries\itshape S}}

\newcommand{\vbar}{\bar{v}}

\newcommand{\Udag}{U^{\dagger}}
\newcommand{\Vdag}{V^{\dagger}}

\newcommand{\Wc}{{\cal W}}
\newcommand{\Wcdag}{{\cal W}^{\dagger}}

\newcommand{\Xhat}{\hat{X}}
\newcommand{\Xdag}{\hat{X}^{\dagger}}

\renewcommand{\thanks}{\footnote}
\newcommand\tocite[1]{$^{\hbox{--}}$\cite{#1}}%\cite{xx}\tocite{yy}

\newcommand{\bg}{\beta,\gamma}

%\preprint{APS/123-QED}

\title{Triaxial quadrupole deformation dynamics in $sd$-shell
nuclei around $^{26}$Mg}

\author{Nobuo Hinohara$^1$ and Yoshiko Kanada-En'yo$^2$}
 \affiliation{
 $^1$Theoretical Nuclear Physics Laboratory, RIKEN Nishina Center,
 RIKEN, Wako, Saitama, 351-0198, Japan}

 \affiliation{
 $^2$Department of Physics, Graduate School of Science, Kyoto
 University, Kyoto 606-8502, Japan}

\date{\today}% It is always \today, today,

\begin{abstract}
Large-amplitude dynamics of axial and triaxial quadrupole deformation in $^{24,26}$Mg,
 $^{24}$Ne, and $^{28}$Si
is investigated on the basis of the quadrupole collective Hamiltonian 
constructed with use of the constrained Hartree-Fock-Bogoliubov plus
the local quasiparticle random-phase approximation method.
The calculation reproduces well properties of the ground rotational bands,
and $\beta$ and $\gamma$ vibrations in $^{24}$Mg and $^{28}$Si.
The $\gamma$-softness in the collective states of $^{26}$Mg and $^{24}$Ne are discussed.
Contributions of the neutrons and protons to the transition properties are also
analyzed in connection with the large-amplitude quadrupole dynamics.
\end{abstract}

%Use showkeys class option if keyword

\maketitle

\section{\label{sec:intro}Introduction}

It is known that
collective deformation grows up in the middle of the $sd$-shell region.
The appearance of the prolate ground state of $^{24}$Mg and the oblate ground state of $^{28}$Si \cite{Horikawa19719,Ball1980271,DasGupta1967602,Ragnarsson1970155}
is associated with the shell gaps $N=Z=12$ at the prolate region
and $N=Z=14$ at the oblate region in Nilsson diagram \cite{BMvol2}, respectively.
Because of the shell gaps in the deformed regions,
various shapes are expected to appear in the mass number region around
$^{24}$Mg and $^{28}$Si.

Moreover, triaxial deformation degree of freedom plays very important roles
on the low-lying collective dynamics in this mass region \cite{PhysRevC.5.768}.
In $^{24}$Mg, possibility of the triaxial deformation in the ground states
has been discussed for decades \cite{Koepf1988397,Bonche1987115,0305-4616-14-9-008}.
The low-lying $K=2$ band built on top of the $2_2^+$ state suggests that
the triaxial degree of freedom is activated in the collective dynamics.
In $^{28}$Si, importance of triaxiality has been 
suggested in connection with the 
large-amplitude collective dynamics of the oblate-prolate shape coexistence \cite{PhysRevC.43.2254,Pelet1977277}.

In contrast to the well-developed deformations in $^{24}$Mg and
$^{28}$Si, 
the deformation property of $^{26}$Mg is not yet fully clarified.
Since it is a system with $N=14$ and $Z=12$, neutrons and protons favor different shapes separately.
Indeed, so far many mean-field calculations with use of the 
realistic effective interactions have been performed for $^{26}$Mg
within an axial symmetry restriction, and
they yielded a coexistence of oblate and prolate shapes with an oblate minimum
\cite{RodriguezGuzman2002201,peru:044313,Terasaki1997706}.
On the other hand, the symmetry-unrestricted  mean-field calculations
using a Skyrme density functional (SkM*) \cite{inakura-priv}
or relativistic model \cite{PhysRevC.83.014308} show extremely triaxially soft
potential energy surfaces.

In the study of collective excitations in this mass region,
the quasiparticle random-phase approximation (QRPA) calculations have been systematically performed by employing various effective interactions
\cite{peru:044313,yoshida:064316,PhysRevC.81.064307}.
The QRPA is a standard tool to analyze the collective modes of excitations.
However, in order to discuss the low-lying collective dynamics of nuclei which are very soft against quadrupole deformation,
one should use a microscopic theory of large-amplitude collective motion instead of the small-amplitude theory such as the QRPA.
The generator coordinate method (GCM) with the restriction of axial symmetry \cite{RodriguezGuzman2002201},
and the antisymmetrized molecular dynamics + multi-configuration mixing \cite{PhysRevC.80.044316} have been performed
using the energy density functionals for magnesium isotopes and $^{28}$Si, respectively.
However, $^{26}$Mg is soft against $\beta$ and $\gamma$ directions 
as shown in the potential energy surface \cite{RodriguezGuzman2002201,peru:044313,Terasaki1997706,inakura-priv},
and therefore the triaxial degree of freedom in addition to the axial degree of freedom should be included 
for the description of the low-lying collective dynamics.

Quite recently, the GCM calculations including axial and triaxial generator coordinates
have been performed \cite{bender:024309,PhysRevC.81.064323,PhysRevC.81.044311,PhysRevC.83.014308} for magnesium isotopes.
The first applications are concentrated on the low-lying states of $^{24}$Mg, in which the
small-amplitude description 
in the prolate mean field is rather good.
In Ref.~\cite{PhysRevC.83.014308}, the properties of the yrast states
of the magnesium isotopes are discussed systematically.

The quadrupole collective Hamiltonian provides a powerful 
theoretical tool to investigate the large-amplitude collective motion 
while taking into account the $\beta$ and $\gamma$ degrees of freedom \cite{BMvol2,Kumar1967608,Belyaev196517,0954-3899-36-12-123101}.
Recently, on the basis of the adiabatic self-consistent collective coordinate method \cite{PTP.103.959,PTP.117.451},
a new microscopic method to construct the collective Hamiltonian
has been developed, called the constrained Hartree-Fock-Bogoliubov plus local QRPA (CHFB+LQRPA) \cite{PhysRevC.82.064313}.
In this method, the collective potential is calculated by the CHFB
equation, while the inertial functions for large-amplitude quadrupole
shape vibration and the three-dimensional rotation are
determined from the normal modes on the CHFB state in $(\bg)$ plane.
A new point of this method is that the contributions
from the time-odd mean field are taken into account in evaluating the vibrational
and rotational inertial masses.
So far this CHFB + LQRPA method 
in conjunction with the pairing-plus-quadrupole (P+Q) model \cite{Baranger1968490,Bes-Sorensen}
including the quadrupole-pairing force has been successfully applied to 
the oblate-prolate shape coexistence in proton-rich Se and Kr isotopes
\cite{PhysRevC.82.064313,Sato201153}.

In this paper, we analyze the role of triaxiality in connection with the
large-amplitude collective motion in the low-lying states of $^{24}$Mg, $^{28}$Si, $^{26}$Mg, and $^{24}$Ne
using the quadrupole collective Hamiltonian
calculated by use of the CHFB + LQRPA method with the P+Q model.
We also discuss the roles of neutrons and protons in $N\ne Z$ nuclei on the large-amplitude collective dynamics 
in relation to the electric transition properties.
This article is organized as follows.
In the next section, the formulation of the CHFB+LQRPA
method is briefly recapitulated.
The results of the numerical calculations 
are presented in Sec.~\ref{sec:results}, 
and the role of the triaxial degree of freedom in these nuclei is
discussed in Sec.~\ref{sec:discussion}.
Summary is given in Sec.~\ref{sec:summary}.

\section{Formulation}\label{sec:formulation}

\subsection{CHFB+LQRPA method}

The theoretical approach, the CHFB + LQRPA method is briefly summarized
in this section.
See Ref.~\cite{PhysRevC.82.064313} for detailed description of the method.

The method enables us to derive the five-dimensional quadrupole
collective Hamiltonian of the Bohr-Mottelson type \cite{BMvol2,Kumar1967608,Belyaev196517,0954-3899-36-12-123101}
\begin{align}
\Hc_{\rm coll} =& T_{\rm vib} + T_{\rm rot} + V(\bg), \label{eq:collH} \\
T_{\rm vib} =& \frac{1}{2}D_{\beta\beta}(\bg)\dot{\beta}^2
+ D_{\beta\gamma}(\bg)\dot{\beta}\dot{\gamma} +
 \frac{1}{2}D_{\gamma\gamma}(\bg)\dot{\gamma}^2, \\
T_{\rm rot} =& \frac{1}{2}\sum_{k=1}^3 \Jc_k(\bg) \omega^2_k,
\end{align}
where $V(\bg)$ is the collective potential in the $(\bg)$ plane.
The quantities $T_{\rm vib}$ and $T_{\rm rot}$ are the vibrational and rotational
kinetic energies.
The inertial functions $D_{\beta\beta}, D_{\gamma\gamma}$, 
and $D_{\beta\gamma}$ are the vibrational masses associated with the 
time-derivatives of the 
two quadrupole deformation variables, $\dot{\beta}$ and $\dot{\gamma}$,
and $\Jc_k$ are the rotational moments of inertia associated with the 
three components of the rotational angular velocities $\omega_k$
defined with respect to the principal axes.

The collective potential and the inertial functions in 
the collective Hamiltonian (\ref{eq:collH}) are 
determined microscopically in the CHFB + LQRPA method.
The collective potential is determined by solving the CHFB equation
\begin{align}
 \bra{\phi(\bg)}\Hhat_{\rm CHFB}(\bg)\ket{\phi(\bg)} = 0,
\end{align}
where the CHFB Hamiltonian is given as
\begin{align}
 \Hhat_{\rm CHFB} = \Hhat - \sum_{\tau=n,p} \lambda^{(\tau)}(\bg)
 \Nt^{(\tau)} 
 - \sum_{m=0,2}\mu_{m}(\bg) \Dhat_{2m}^{(+)},
\end{align}
with the constraints on particle numbers and quadrupole deformations.
Here $\Hhat$ is the microscopic Hamiltonian, $\ket{\phi(\bg)}$ is 
the CHFB state, $\lambda^{(\tau)}(\bg)$ and $\mu_m(\bg)$
 are the Lagrange multipliers, 
$\Nt^{(\tau)}\equiv\Nhat^{(\tau)} - N_0^{(\tau)}$
are the particle number operators measured from $N_0^{(\tau)}$
which are the neutron and proton particle numbers of the nucleus.
The operators $\Dhat^{(+)}_{2m}$ are the Hermitian part of the 
quadrupole operators given by $\Dhat^{(+)}_{2m}\equiv (\Dhat_{2m} +
\Dhat_{2-m})/2$.
The collective potential is given by
\begin{align}
 V(\bg) = \bra{\phi(\bg)}\Hhat\ket{\phi(\bg)}.
\end{align}
On top of the CHFB state,
the local normal modes are calculated by solving the LQRPA
equations
\begin{multline}
 \delta \bra{\phi(\bg)}[\Hhat_{\rm CHFB}(\bg), \Qhat^i(\bg)] \\
 -\frac{1}{i} \Phat_i(\bg) \ket{\phi(\bg)} = 0, 
\end{multline}
\begin{multline}
 \delta \bra{\phi(\bg)}\left[\Hhat_{\rm CHFB}(\bg),
 \frac{1}{i}\Phat_i(\bg)\right] \\
 - C_i(\bg) \Qhat^i(\bg) \ket{\phi(\bg)} = 0.
\end{multline}
Here $\Qhat^i(\bg)$ and $\Phat_i(\bg)$ are the infinitesimal generators
locally defined as functions of $(\bg)$.
The quantity $C_i(\bg)=\omega^2_i(\bg)$ is the squared eigen frequency of the normal
mode.
We choose two collective modes from the LQRPA modes, following the 
minimal metric criterion in Ref.~\cite{PhysRevC.82.064313}.
The vibrational masses $D_{\beta\beta}, D_{\gamma\gamma}$, and
$D_{\beta\gamma}$ are determined from the transformation
of the collective coordinates spanned by the two LQRPA modes 
into $(\beta, \gamma)$.

The rotational moments of inertia are calculated by solving the LQRPA 
equations for rotation on top of the CHFB state.
\begin{align}
\delta \bra{\phi(\bg)}[\Hhat_{\rm CHFB}, \hat{\Psi}_k(\bg)] - \frac{1}{i}
 (\Jc_k)^{-1}\Ihat_k \ket{\phi(\bg)} = 0,
\end{align}
\begin{align}
\bra{\phi(\bg)} [\hat{\Psi}_k(\bg), \Ihat_{k'}]\ket{\phi(\bg)} = i \delta_{kk'},
\end{align}
where $\hat{\Psi}_k(\bg)$ and $\Ihat_k$ represent the rotational angles
and  the angular momentum operators with respect to the three principal
axes associated with the CHFB state $\ket{\phi(\bg)}$,
and $\Jc_k$ are the LQRPA moments of inertia.

Pauli's prescription is used to quantize the
classical collective Hamiltonian (\ref{eq:collH}).
From the solution of the collective Schr\"odinger equation
\begin{align}
\left\{\hat{T}_{\rm vib} + \hat{T}_{\rm rot} + V(\bg)\right\} \Psi_{\alpha
 IM}(\bg,\Omega) = E_{\alpha I} \Psi_{\alpha IM}(\bg,\Omega), \label{eq:Schroedinger}
\end{align}
we obtain the collective wave function $\Psi_{\alpha IM}(\bg,\Omega)$
as functions of quadrupole deformations $(\bg)$ and three Euler angles $\Omega$.
The collective wave function is specified by the angular momentum $I$,
and its projection onto the $z$-axis of the laboratory frame, $M$,
and $\alpha$ distinguishes the states which have the same $I$ and $M$.

The collective wave function is written in the following form
\begin{align}
 \Psi_{\alpha IM}(\bg,\Omega) = \sum_{K\ge 0,{\rm even}} \Phi_{\alpha
 IK}(\bg) \langle\Omega|IMK\rangle,
\end{align}
where $\Phi_{\alpha IK}(\bg)$ is the vibrational part of the collective wave function,
and the rotational part is written as
\begin{align}
\langle\Omega|IMK\rangle = \sqrt{\frac{2I+1}{16\pi^2(1+\delta_{k0})}}
[ D^I_{MK}(\Omega) + (-)^I D^I_{M-K}(\Omega)].
\end{align}
Here $D^I_{MK}$ is the Wigner's rotation matrix and $K$ is the
projection of the angular momentum onto the $z$-axis
in the body-fixed frame.

The vibrational wave functions are normalized as
\begin{align}
 \int d\beta d\gamma |\Phi_{\alpha I}(\bg)|^2 |G(\bg)|^{\frac{1}{2}} = 1,
\end{align}
where
\begin{align}
 |\Phi_{\alpha I}(\bg)|^2 \equiv \sum_{K\ge 0, {\rm even}} |\Phi_{\alpha
 IK}(\bg)|^2,
\end{align} 
and the volume element $|G(\bg)|^{\frac{1}{2}}d\beta d\gamma$ is given by
\begin{align}
 |G(\bg)|^{\frac{1}{2}}d\beta d\gamma = 2\beta^4 \sqrt{W(\bg) R(\bg)}
 \sin 3\gamma d\beta d\gamma, \label{eq:volume}
\end{align}
\begin{align}
 W(\bg) =& \{ D_{\beta\beta}(\bg) D_{\gamma\gamma}(\bg) -
 [D_{\beta\gamma}(\bg)]^2\} \beta^{-2}, \label{eq:metric} \\
 R(\bg) =& D_1(\bg)D_2(\bg) D_3(\bg),
\end{align}
where $D_k(\bg)$ are related to the moments of inertia as 
$\Jc_k(\bg) = 4\beta^2 D_k(\bg) \sin^2(\gamma - 2\pi k/3)$.

The requantization form, the symmetries, and boundary conditions of the
collective Hamiltonian are described in Ref.~\cite{Kumar1967608}.

The electric properties are calculated following the discussions in Refs.~\cite{Kumar1967608,Sato201153}.
The value of $B(E2)$ and the spectroscopic quadrupole moment are given by 
\begin{align}
 B(E2; \alpha I \rightarrow \alpha' I') =
(2I+1)^{-1} |\langle \alpha I || \Dhat^{'(E2)} || \alpha' I' \rangle
 |^2, \label{eq:BE2}
\end{align}
and
\begin{align}
 Q(\alpha I) = \sqrt{\frac{16\pi}{5}} 
 \begin{pmatrix}
 I & 2 & I \\ -I & 0 & I
 \end{pmatrix}
 \langle \alpha I || \Dhat^{'(E2)} || \alpha I \rangle. \label{eq:Qmom}
 \end{align}
The reduced matrix element in Eqs.~(\ref{eq:BE2}) and (\ref{eq:Qmom})
is calculated as
\begin{align}
\langle \alpha I|| \Dhat^{'(E2)} || \alpha' I' \rangle =
\int d\beta d\gamma |G(\bg)|^{\frac{1}{2}} 
\rho^{(E2)}_{\alpha I \alpha' I'}(\bg),
\end{align}
\begin{align}
\rho^{(E2)}_{\alpha I \alpha' I'} & (\bg) =
\sqrt{(2I+1)(2I'+1)} (-)^I  \nonumber \\ 
& \sum_{K\ge 0,{\rm even}}
\left\{ 
\begin{pmatrix} I & 2 & I'\\ -K & 0 & K \end{pmatrix}
\Phi_{\alpha,I,K}D^{(E2)}_{0+}\Phi_{\alpha',I',K'}
\right. 
\nonumber \\
+ \sqrt{1+\delta_{K0}} & \left. \left[ \left\{
\begin{pmatrix} I & 2 & I' \\ -K-2 & 2 & K \end{pmatrix}
\Phi_{\alpha,I,K+2}D^{(E2)}_{2+}\Phi_{\alpha',I',K} 
\right. \right. \right.
\nonumber \\
+ (-)^{I+I'} & \left. \left.
\begin{pmatrix} I & 2 & I' \\ K & 2 & -K-2 \end{pmatrix}
\Phi_{\alpha, I, K}D^{(E2)}_{2+}\Phi_{\alpha',I',K+2}
\right] \right\} \label{eq:E2density}
\end{align}
where $\rho^{(E2)}_{\alpha,I,\alpha',I'}(\bg)$
is the $E2$ transition density.
The quantities $D^{(E2)}_{m+}(\bg)$ are the expectation values of the $E2$ operator 
in the intrinsic frame,
\begin{align}
 D^{(E2)}_{m+}(\bg) = \bra{\phi(\bg)} \sum_{\tau=n,p} e_{\rm eff}^{(\tau)} \Dhat_{m+}^{(\tau)} \ket{\phi(\bg)}, \label{eq:qoperator}
\end{align}
where $e_{\rm eff}^{(\tau)}$ are the neutron and proton effective charges, and $\Dhat_{m+}^{(\tau)}\equiv (\Dhat_m^{(\tau)} + \Dhat_{-m}^{(\tau)})/2$
are the neutron and proton parts of the quadrupole operators.

\subsection{Model Hamiltonian and parameters}

The pairing-plus-quadrupole model \cite{Bes-Sorensen,Baranger1968490}
including the quadrupole-pairing interaction is adopted 
as a microscopic Hamiltonian in the present work.
The single-particle model space consists of harmonic oscillator two-major shells ($p$-shell and $sd$-shell)
both for neutrons and protons.
The modified oscillator values are used as the spherical
single-particle energies \cite{Nilsson-Ragnarsson}.
The values of the neutron and proton monopole pairing strengths and the quadrupole
particle-hole interaction strengths are summarized in
Table~\ref{table:int}.
The strength of the quadrupole-pairing interaction is evaluated
at the spherical CHFB state with use of the prescription proposed by
Sakamoto and Kishimoto \cite{Sakamoto1990321}.
The interaction strengths of $^{24}$Mg are adjusted to reproduce the 
quadrupole deformation of the prolate potential minimum and the pairing energy at spherical shape of 
$^{24}$Mg calculated with Skyrme SkM* density functional and the mixed
surface-volume type pairing functional in Ref.~\cite{PhysRevC.81.064307}.
A simple mass number dependence of the interaction strengths
is used to obtain those for $^{26}$Mg, $^{24}$Ne, and $^{28}$Si.
Only the quadrupole strength for $^{28}$Si is increased
in order to adjust the deformation of the oblate HFB minimum.

The Fock term is neglected following the conventional prescription of
the P+Q model. Therefore we call the present framework Hartree-Bogoliubov (HB) instead of
Hartree-Fock-Bogoliubov (HFB).
Following Baranger and Kumar \cite{Baranger1968490},
the reduction factors are multiplied by the quadrupole matrix elements 
between the single-particle states in the upper shells,
and the nuclear radial parameter $R=1.2A^{1/3}$ fm is used
in the calculation of the harmonic oscillator length $b$.
In the calculation of $E2$ transition strengths and quadrupole moments,
the quadrupole operator without reduction factor is used \cite{Baranger1968490},
and the nuclear radial parameter with the higher order $A$-dependence
$R=1.12A^{1/3}(1+3.84A^{-2/3})^{1/2}$ fm \cite{BMvol1} is adopted
to quantitatively evaluate $E2$ matrix elements.

The two-dimensional mesh in the $\beta$ and $\gamma$ directions
is used to express the collective Hamiltonian in the ($\bg$) plane.
The 60 mesh points are taken both in the range 
$0<\beta<\beta_{\rm max}$ and $0^\circ < \gamma < 60^\circ$.
As for $\beta_{\rm max}$, 0.6 is used as common value, except 
$\beta_{\rm max}=0.5$ used for  $^{28}$Si,
because we could not get converged solution which satisfies
the CHB equation with four constraints
at large deformation $\beta > 0.5$ near the prolate region in $^{28}$Si.

\begin{table}[htbp]
\begin{center}
\caption{Neutron and proton monopole pairing interaction strengths
 $G_0^{(\tau)}$ and quadrupole
particle-hole interaction strength $\chi'=\chi b^4.$\label{table:int}}
\begin{tabular}{cccc} \\ \hline \hline
           & $G_0^{(n)}$ (MeV) & $G_0^{(p)}$ (MeV) & $\chi'$ (MeV) \\ \hline
 $^{24}$Mg & 0.79 & 0.83 & 1.56 \\
 $^{26}$Mg & 0.73 & 0.77 & 1.37 \\
 $^{24}$Ne & 0.79 & 0.83 & 1.56 \\
 $^{28}$Si & 0.68 & 0.71 & 1.30 \\ \hline\hline
\end{tabular}
\end{center}
\end{table}

\section{Results}\label{sec:results}

\subsection{Collective potentials}

\begin{figure}[htbp]
\includegraphics[width=70mm]{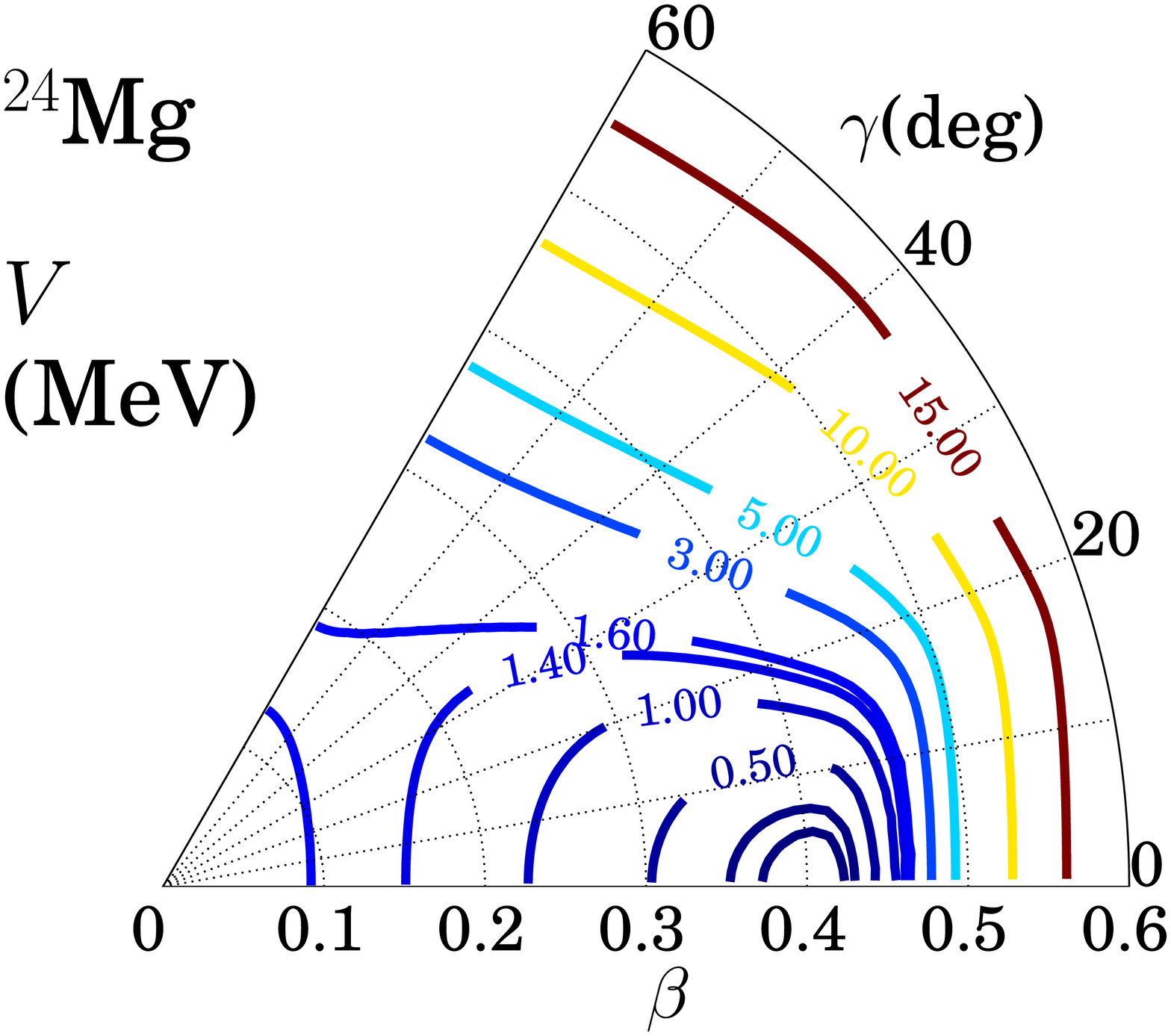}
\caption{\label{fig:24Mg-V} 
(Color online) Collective potential $V(\bg)$ for $^{24}$Mg.}
\end{figure}

\begin{figure}[htbp]
\includegraphics[width=70mm]{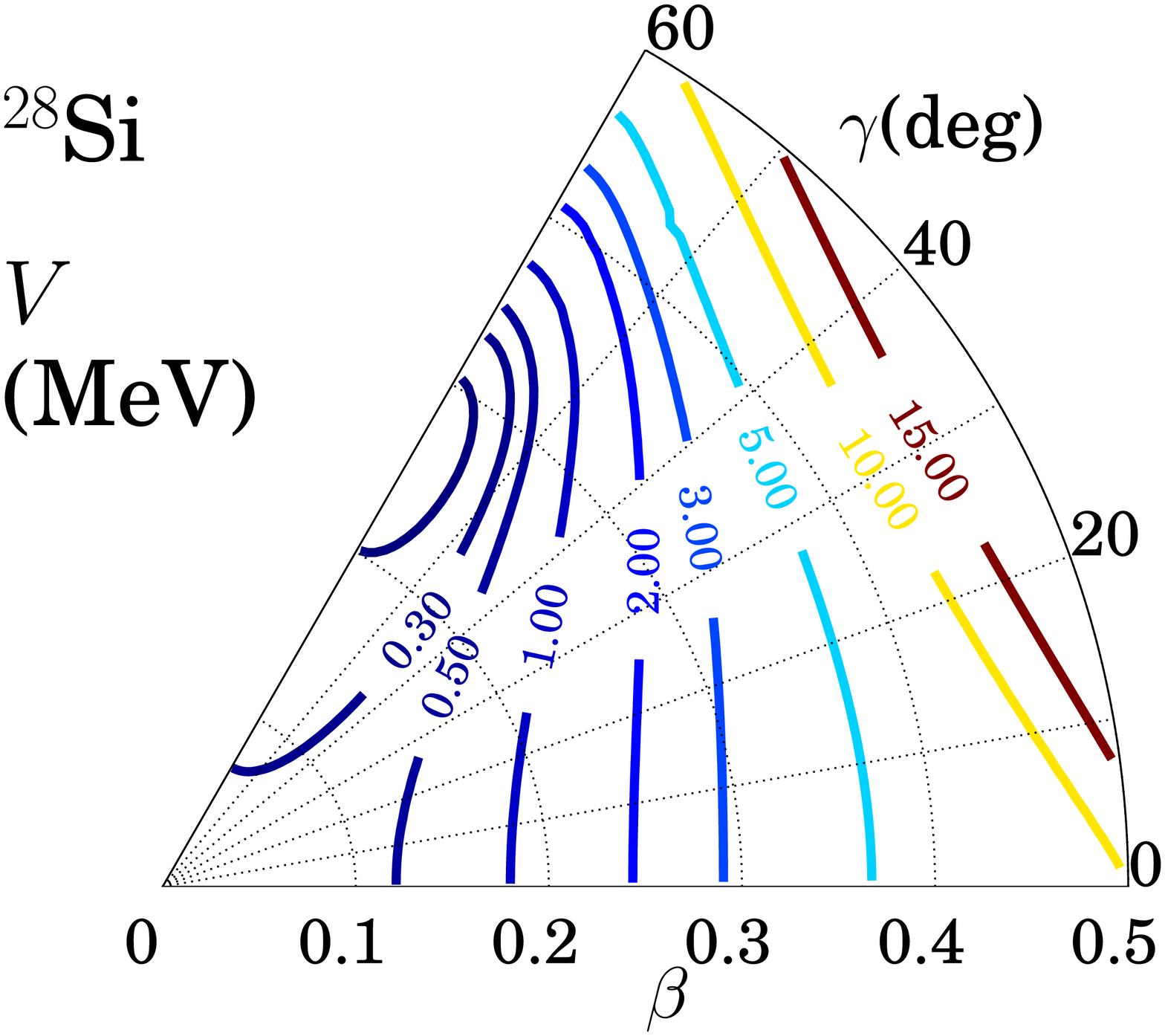}
\caption{\label{fig:28Si-V} 
(Color online) Collective potential $V(\bg)$ for $^{28}$Si.}
\end{figure}

\begin{figure}[htbp]
\includegraphics[width=70mm]{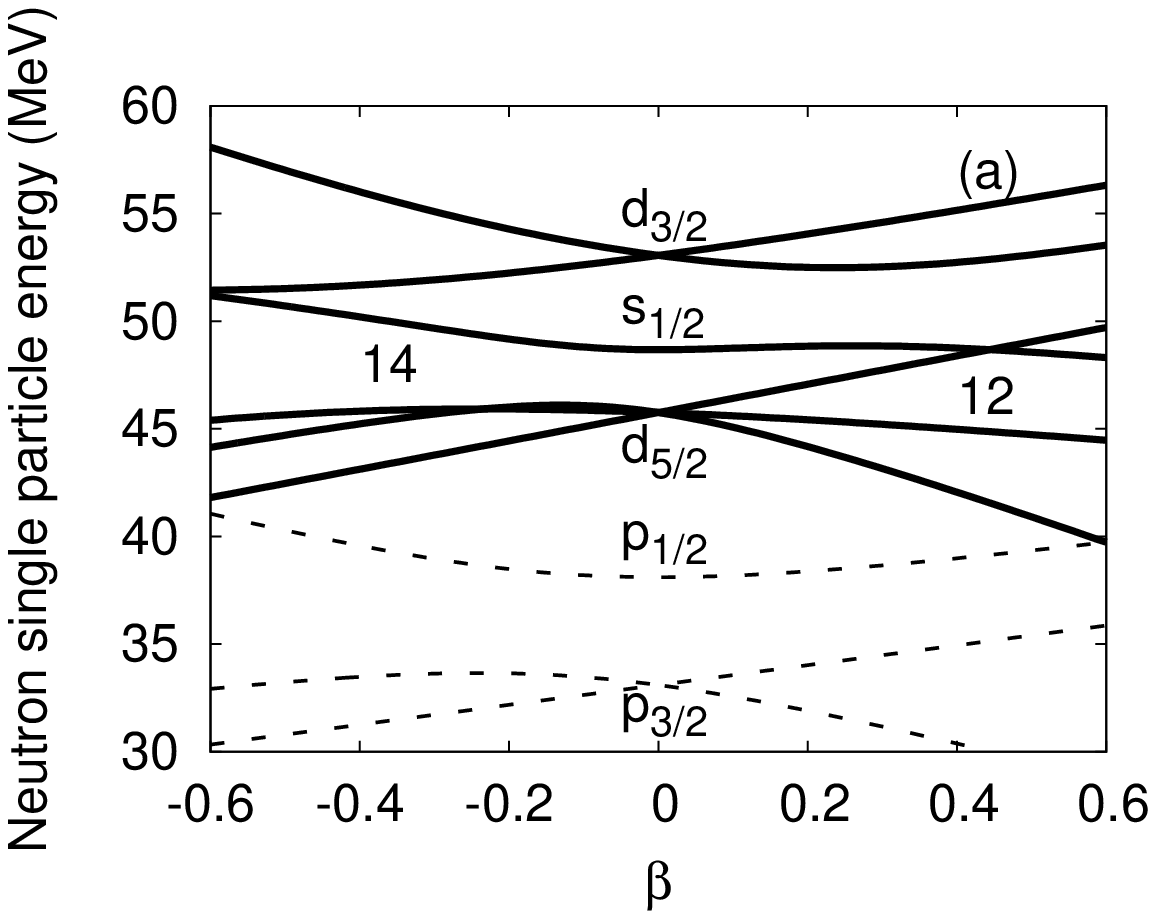} \\
\includegraphics[width=70mm]{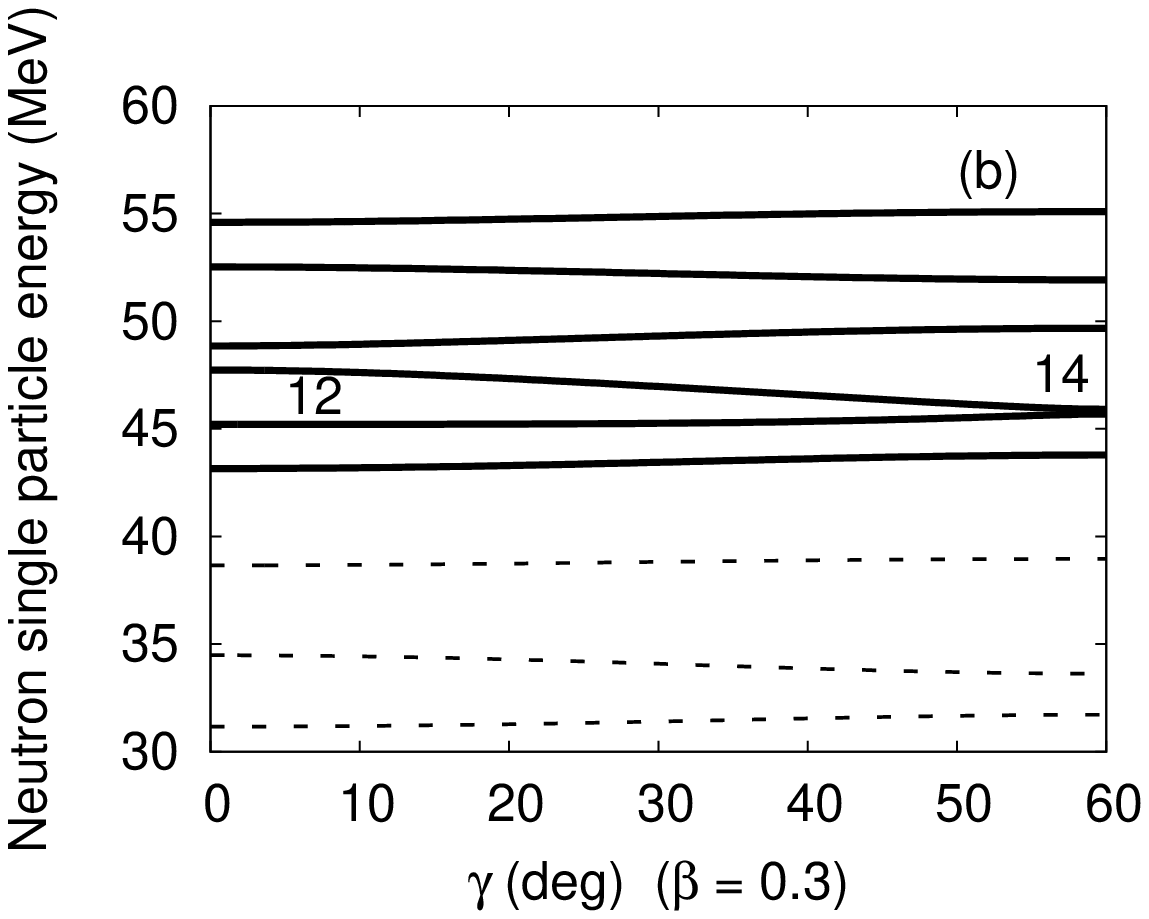} \\
\caption{\label{fig:NilssonDiag}
Neutron Nilsson diagrams calculated within $p$ and $sd$-shell model space
as functions of $\beta$ (a) and $\gamma$ (b).
Even and odd parity orbits are drawn with solid and dashed lines,
 respectively.}
\end{figure}

The collective potentials calculated for $N=Z$ nuclei, $^{24}$Mg and $^{28}$Si,
are shown in Figs.~\ref{fig:24Mg-V} and \ref{fig:28Si-V}.
The collective potential of $^{24}$Mg shows a prolate minimum at
$\beta\sim 0.41$, while that of $^{28}$Si shows an oblate minimum 
at $\beta\sim 0.26$.
As shown in Fig.~\ref{fig:NilssonDiag}, deformed shell gaps 
are found at prolately and oblately deformed regions
in the present model. It corresponds to the appearance of the 
deformed minima in the collective potentials of $^{24}$Mg and $^{28}$Si.
In $^{24}$Mg, 
the triaxial potential valley from the prolate minimum to the oblate region
exists in the small deformed region with $\beta\sim 0.2$.
In $^{28}$Si, the collective potential curve around the oblate minimum is steep against the triaxial
deformation.
A prolate local minimum suggested by other microscopic calculations
for $^{28}$Si \cite{Ragnarsson1982387,PhysRevC.71.014303,PhysRevC.80.044316} does not
appear in the present calculation.
Note that a potential energy surface similar to the present work 
is reported in the Skyrme HF + BCS calculation \cite{Kaneko2009214}.

\begin{figure}[htbp]
\includegraphics[width=70mm]{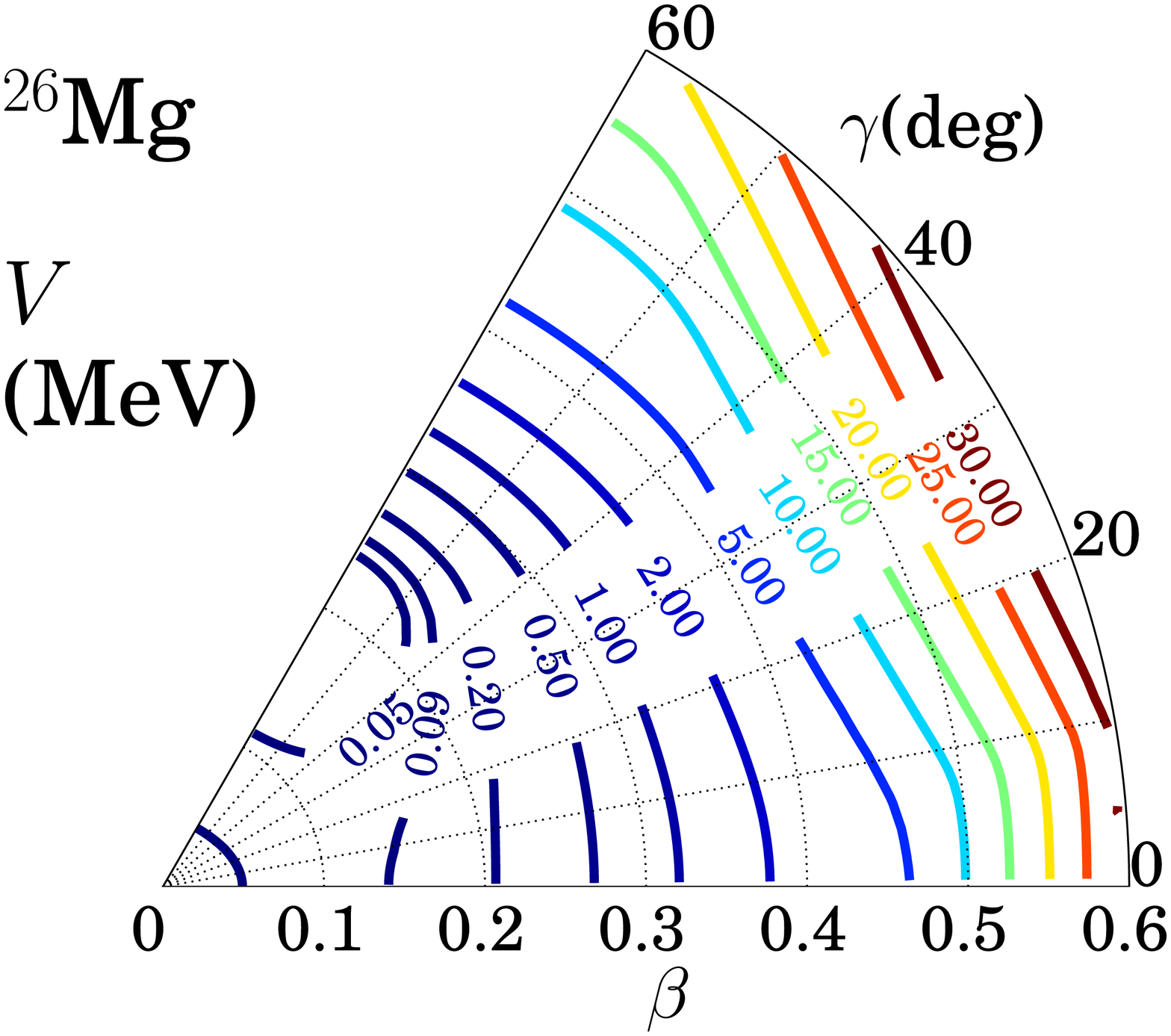}
\caption{\label{fig:26Mg-V} 
(Color online) Collective potential $V(\bg)$ for $^{26}$Mg.}
\end{figure}

\begin{figure}[htbp]
\includegraphics[width=70mm]{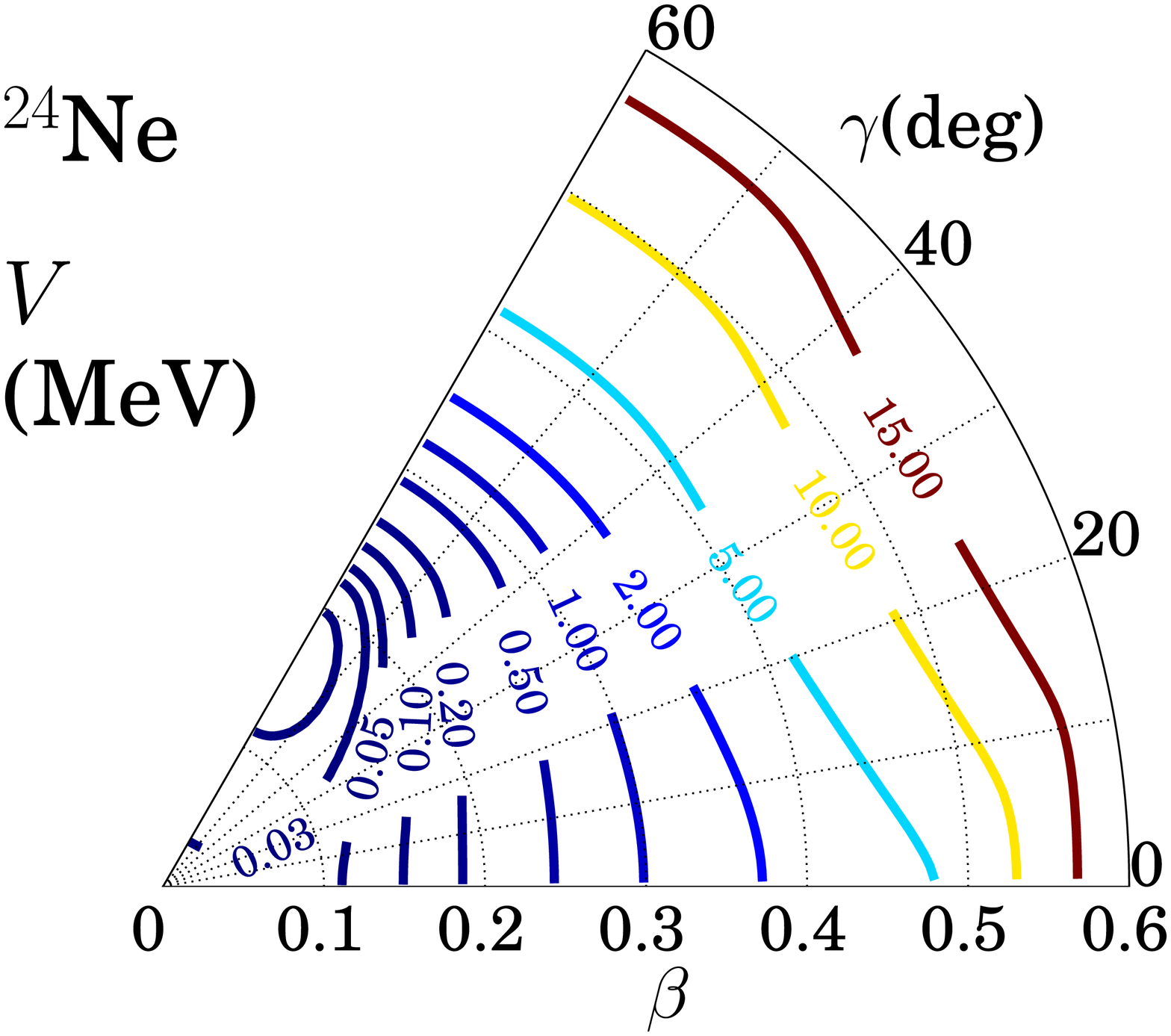}
\caption{\label{fig:24Ne-V} 
(Color online) Collective potential $V(\bg)$ for $^{24}$Ne.}
\end{figure}

In contrast to the deep oblate and prolate minima in $N=Z$ nuclei,
the collective potentials of $^{26}$Mg and $^{24}$Ne presented in
Figs.~\ref{fig:26Mg-V} and \ref{fig:24Ne-V} show $\beta$ and $\gamma$ soft situations.
The potential minima in $^{26}$Mg and $^{24}$Ne show small oblate
deformations with $\beta = 0.16$ and 0.20, respectively. 
Around the potential minima, the collective potentials are soft against both axial and triaxial
quadrupole deformations, suggesting the anharmonic situations in these nuclei.



\begin{figure}[htbp]
\begin{tabular}{cc}
\includegraphics[width=45mm]{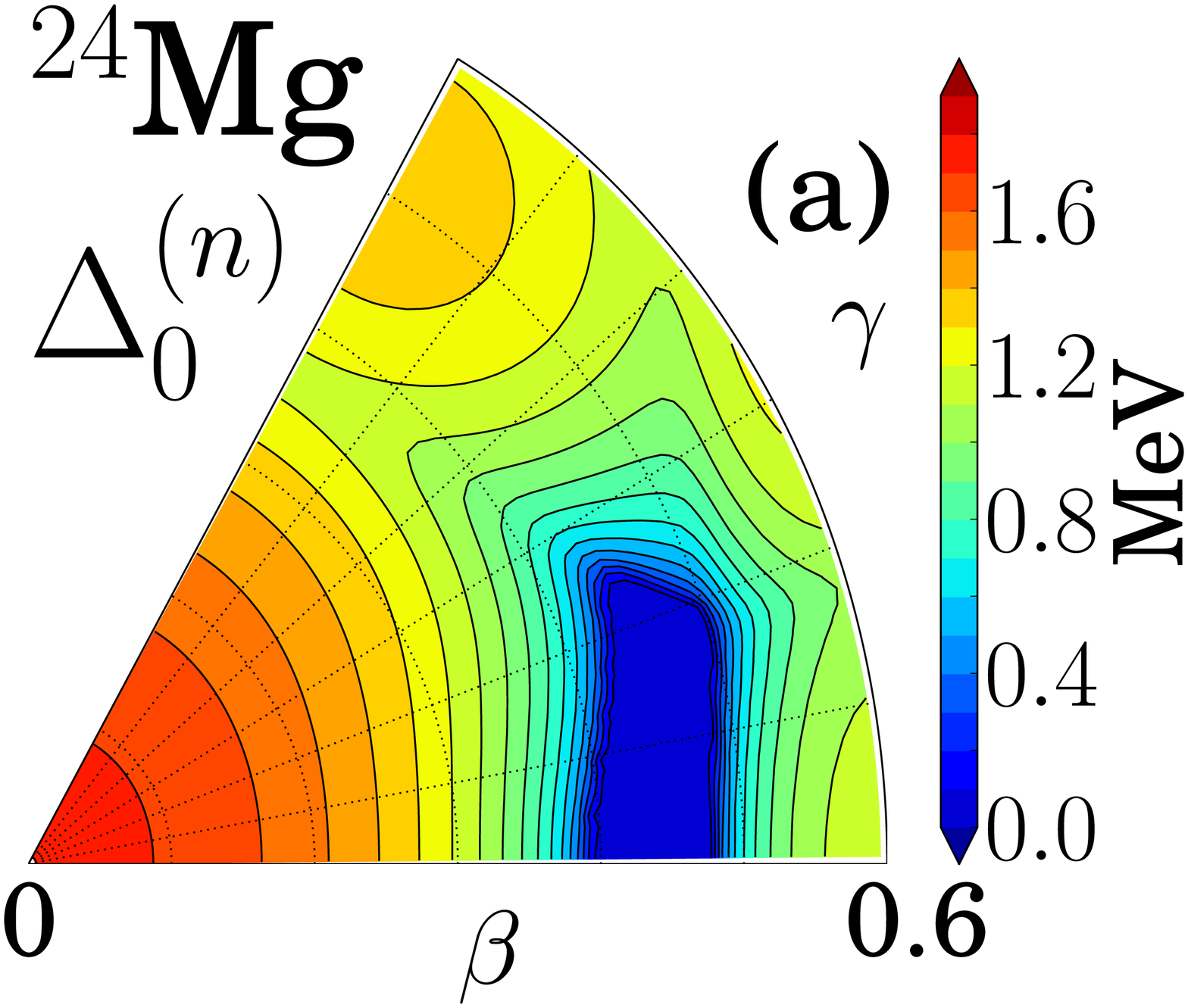} &
\includegraphics[width=45mm]{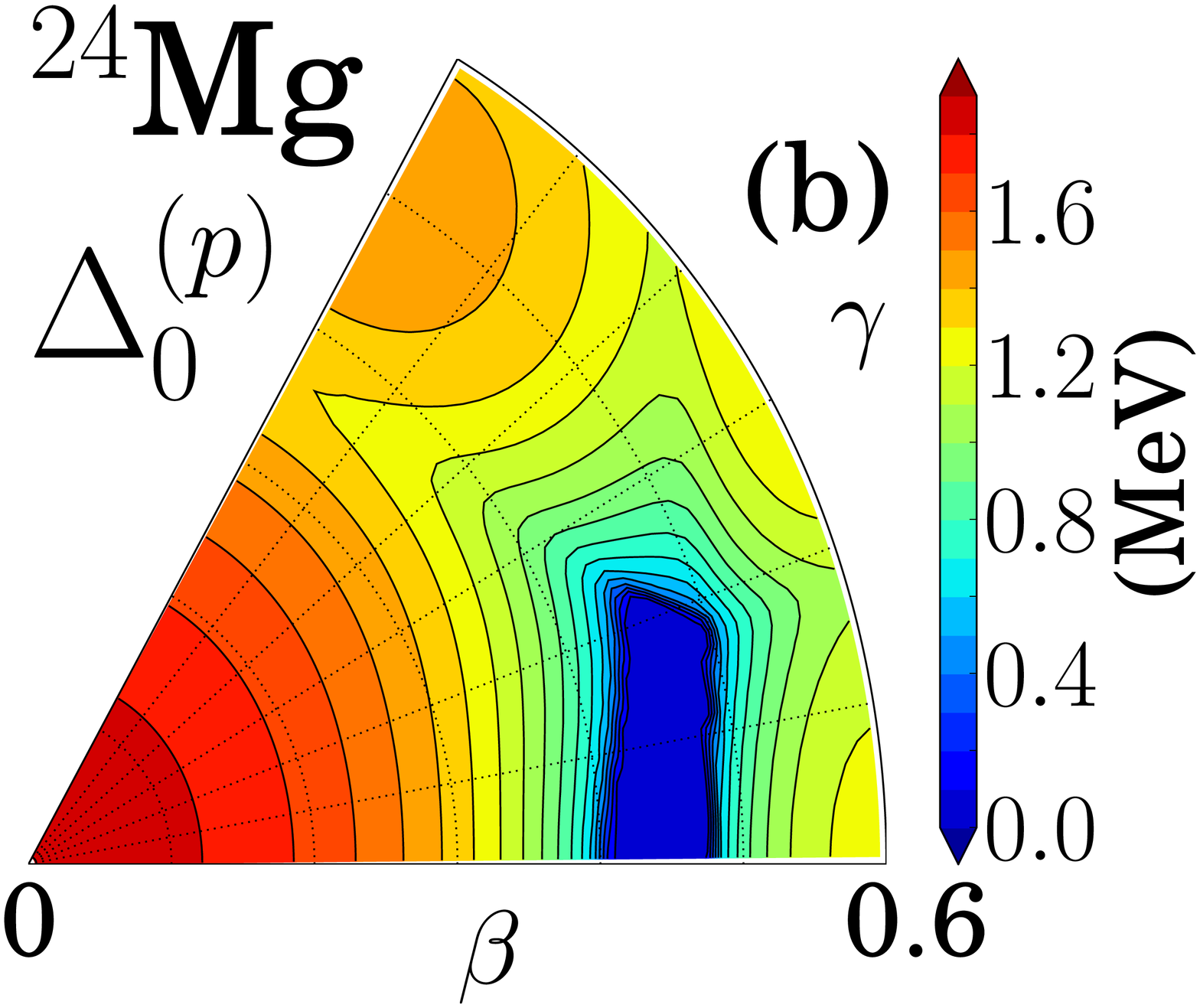} \\
\includegraphics[width=45mm]{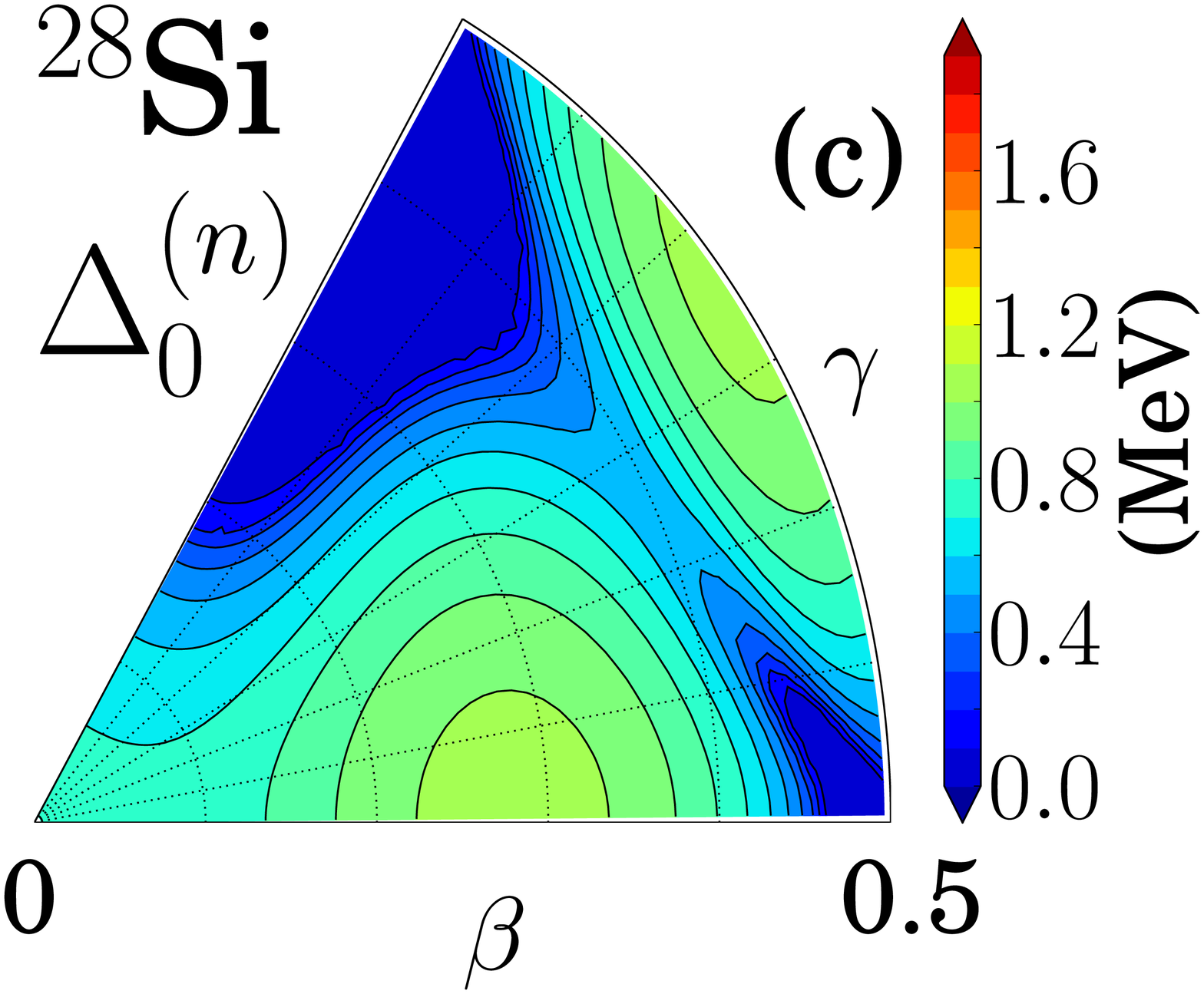} &
\includegraphics[width=45mm]{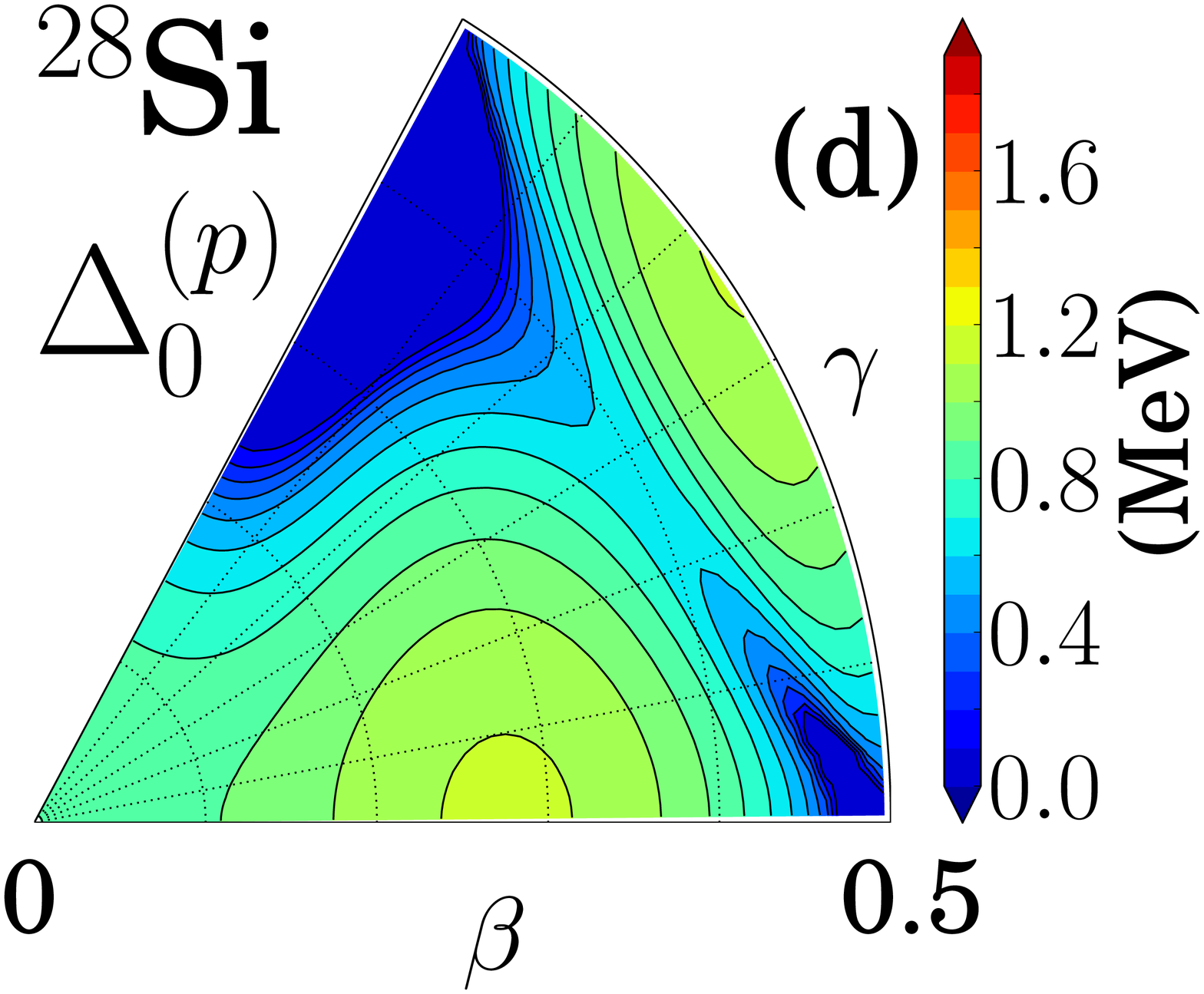} \\
\includegraphics[width=45mm]{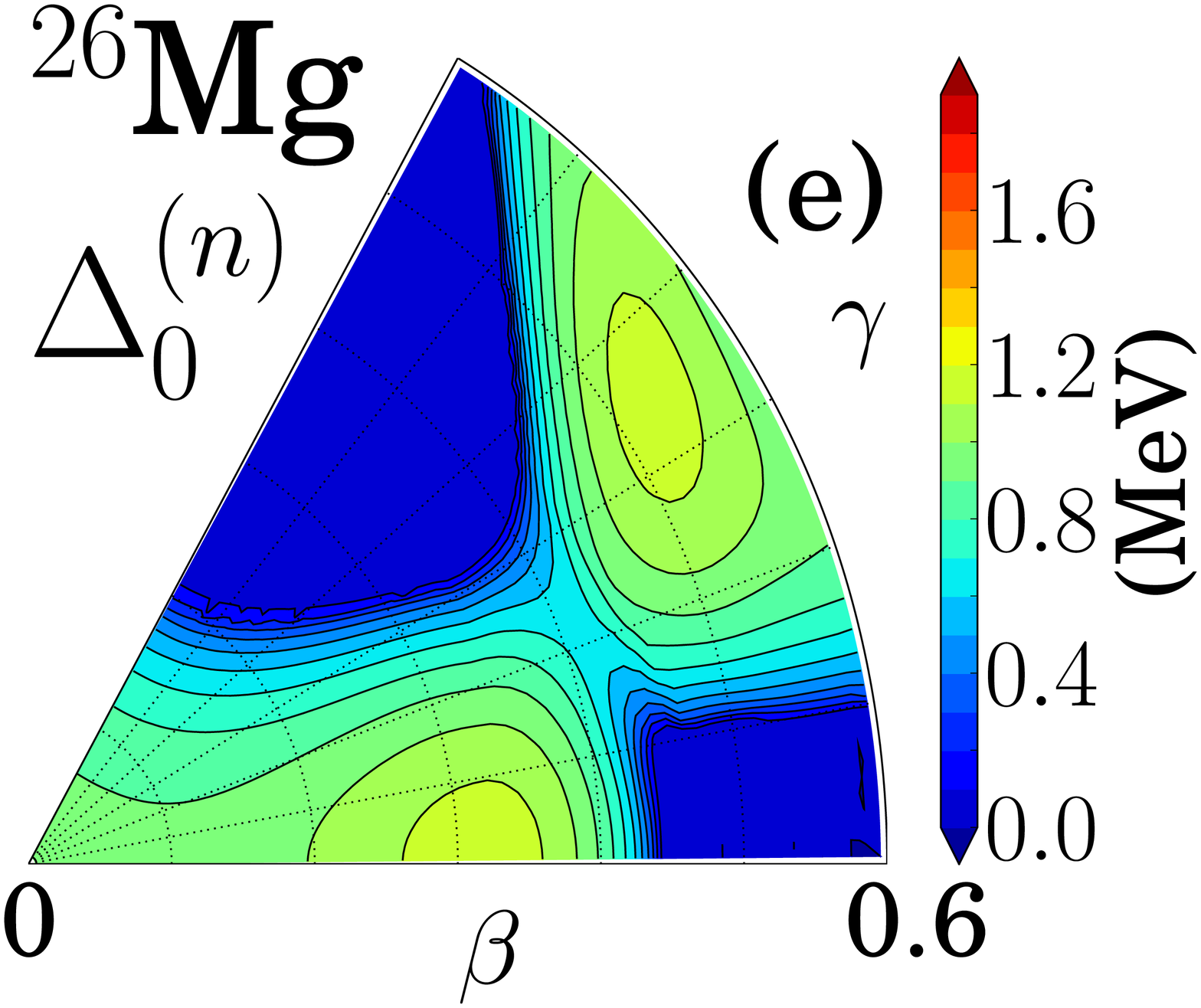} &
\includegraphics[width=45mm]{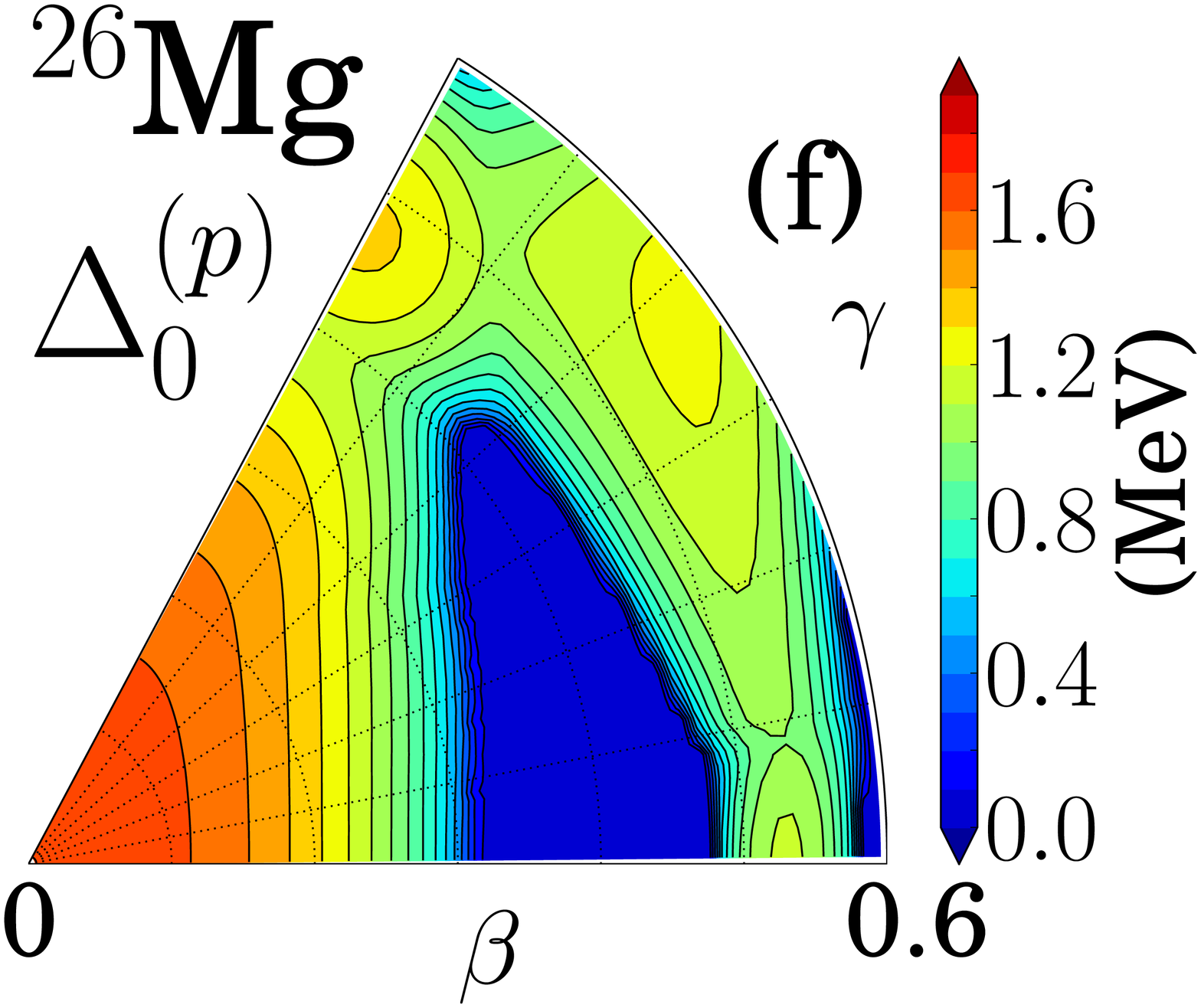} \\
\includegraphics[width=45mm]{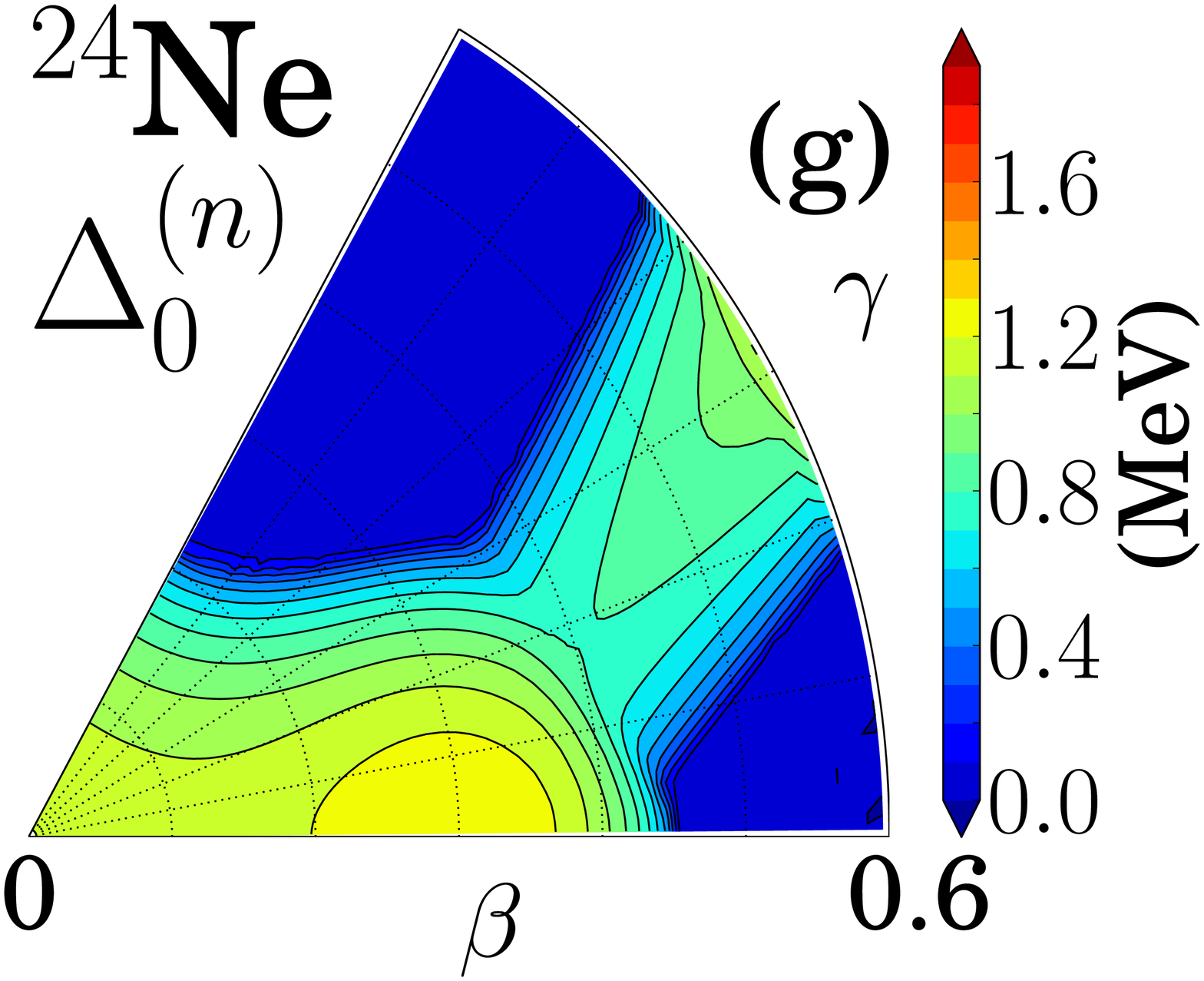} &
\includegraphics[width=45mm]{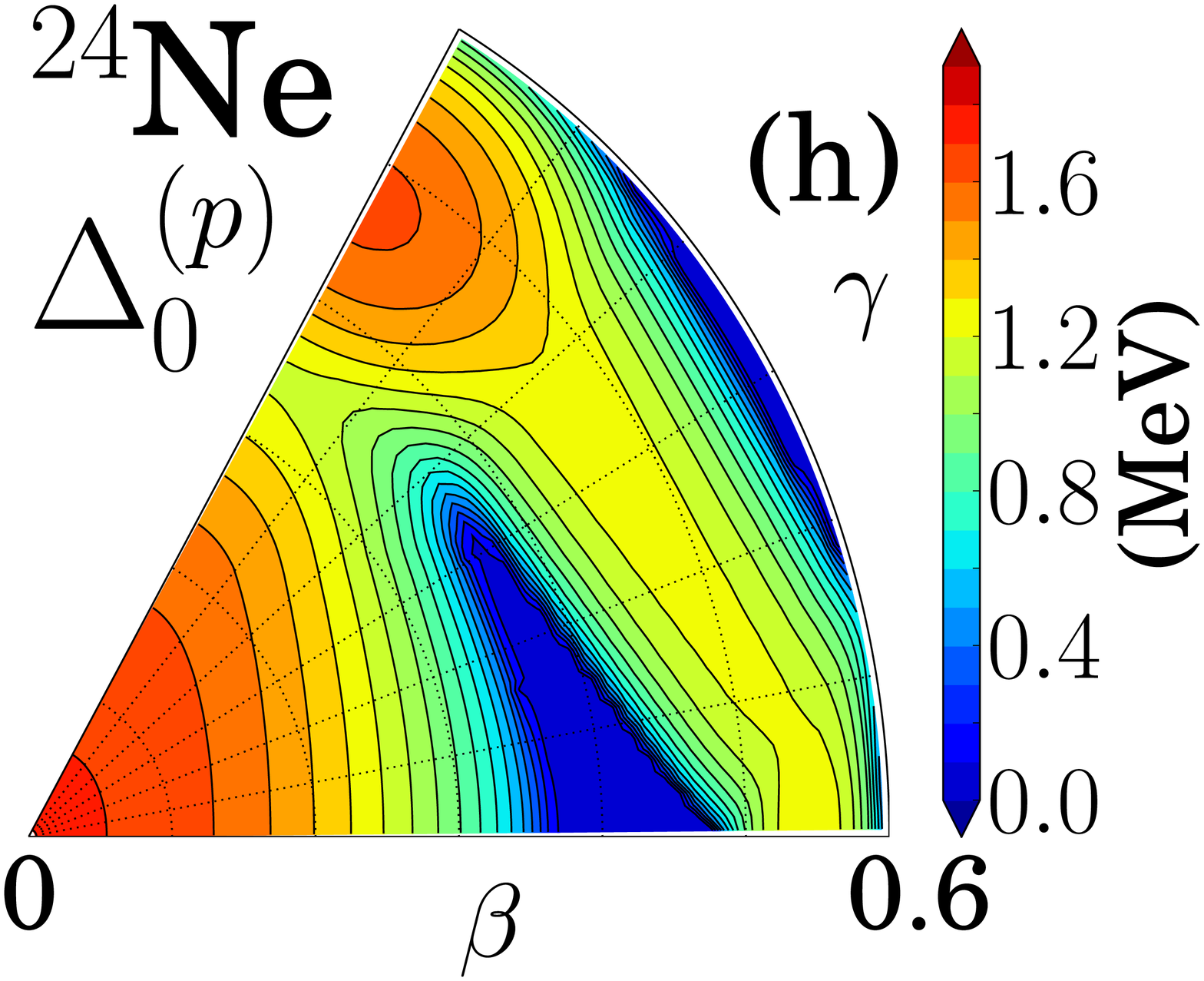} 
\end{tabular}
\caption{\label{fig:gap}
(Color online) 
Neutron and proton pairing gaps $\Delta_{0}^{(n)}$ and $\Delta_{0}^{(p)}$
for $^{24}$Mg, $^{28}$Si, $^{26}$Mg, and $^{24}$Ne.}
\end{figure}

In Fig.~\ref{fig:gap}, 
proton and neutron monopole pairing gaps are plotted as functions of $(\bg)$.
Basic characters of the proton pairing gaps in $^{26}$Mg and $^{24}$Ne are 
same as that in $^{24}$Mg, and those of neutron pairing gaps in $^{26}$Mg and $^{24}$Ne are same as that in $^{28}$Si.
The proton monopole pairing gaps in $^{24}$Mg, $^{26}$Mg, and $^{24}$Ne show strong ($\bg$)
dependence; they have the maximum at the spherical point, 
and they become zero around $\beta\sim0.4$ in the prolate region.
The neutron pairing gaps in $^{28}$Si, $^{26}$Mg, and $^{24}$Ne becomes zero
in the oblate region.
These zero-gap regions correspond to the prolate $Z=12$ and 10
shell gap and the oblate $N=14$ shell gap in Fig.~\ref{fig:NilssonDiag}, respectively.
Comparing the proton pairing gap for $^{26}$Mg with that for $^{24}$Mg,
and the neutron pairing gap for $^{26}$Mg with that for $^{28}$Si,
an interesting feature is seen in $^{26}$Mg;
the zero-gap region extends 
widely in the ($\bg$) plane in the case of $^{26}$Mg.

In the lower panel of Fig.~\ref{fig:NilssonDiag}, the neutron Nilsson diagram
is shown as functions of $\gamma$ at $\beta=0.3$.
The energies of the Nilsson orbits change gradually depending on $\gamma$.
Especially, the oblate shell gap at 14 and the prolate shell gap at 12 more smoothly
vanish in the $\gamma$ direction than in the $\beta$ direction.

\subsection{Properties of the LQRPA modes}
\begin{figure}[htbp]
\includegraphics[width=80mm]{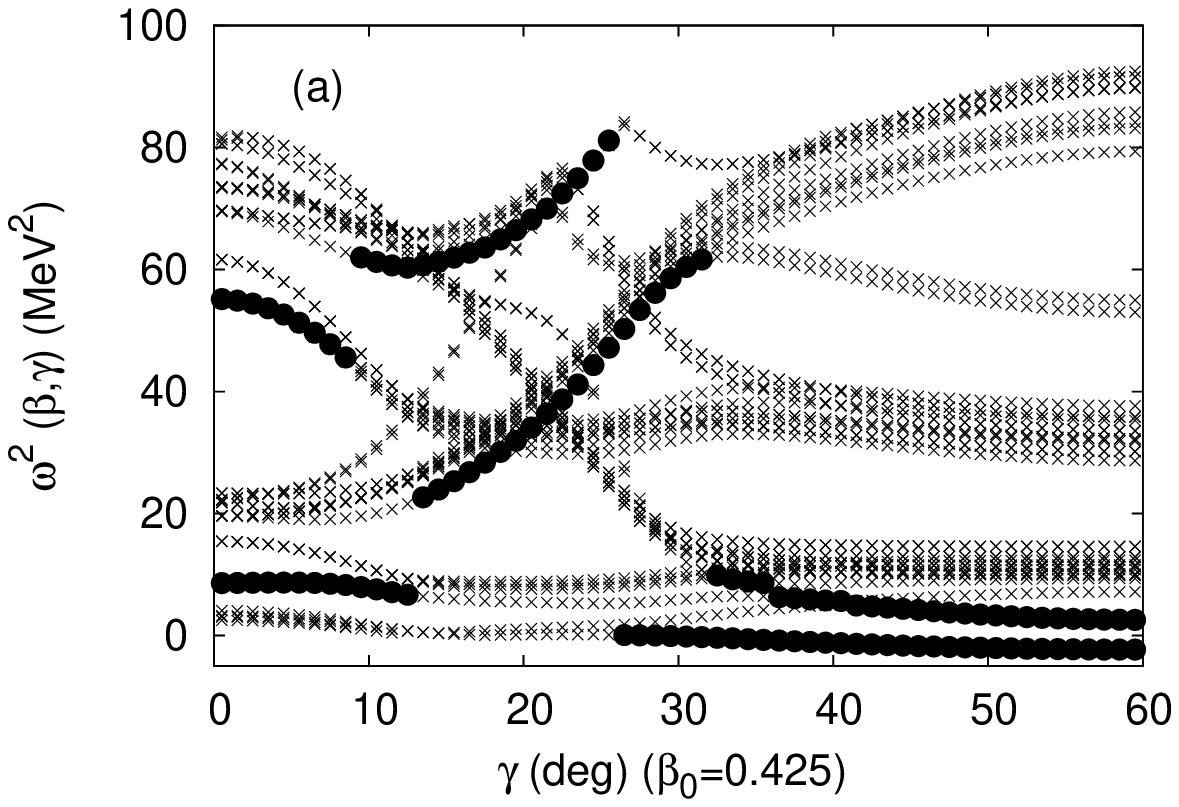} \\
\includegraphics[width=80mm]{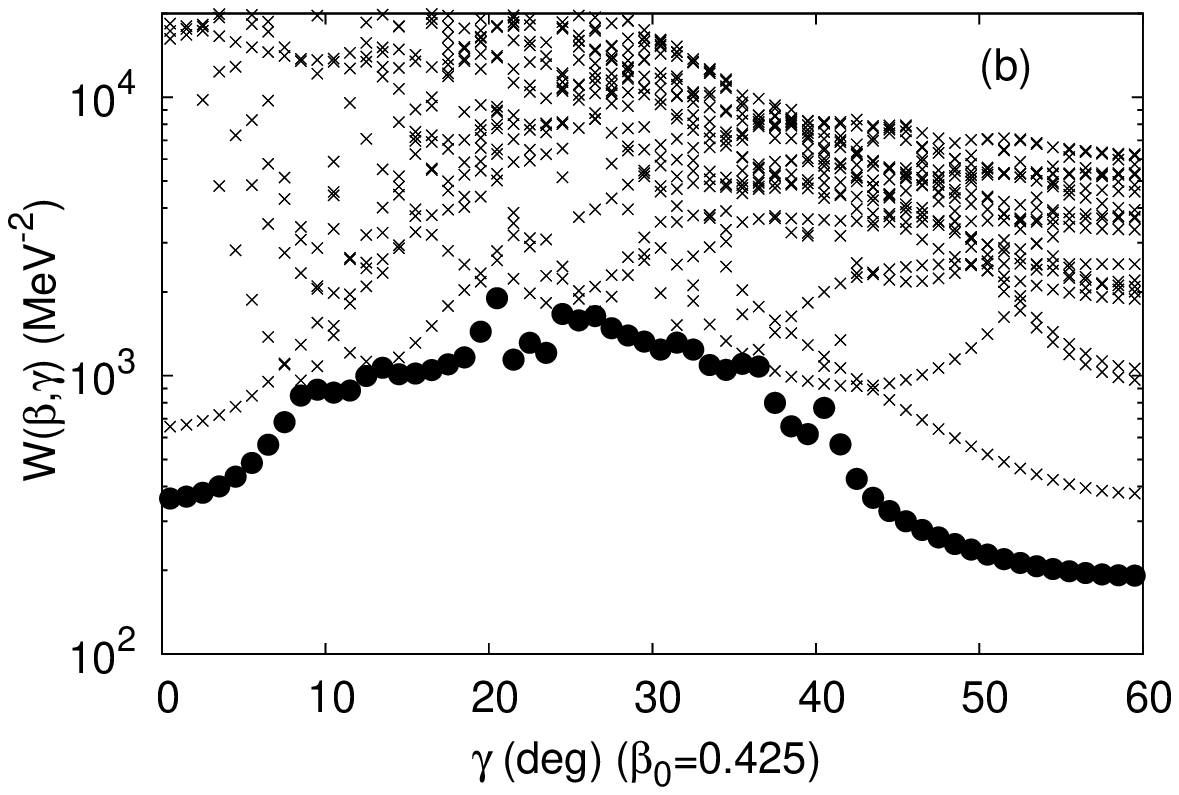} \\
\includegraphics[width=80mm]{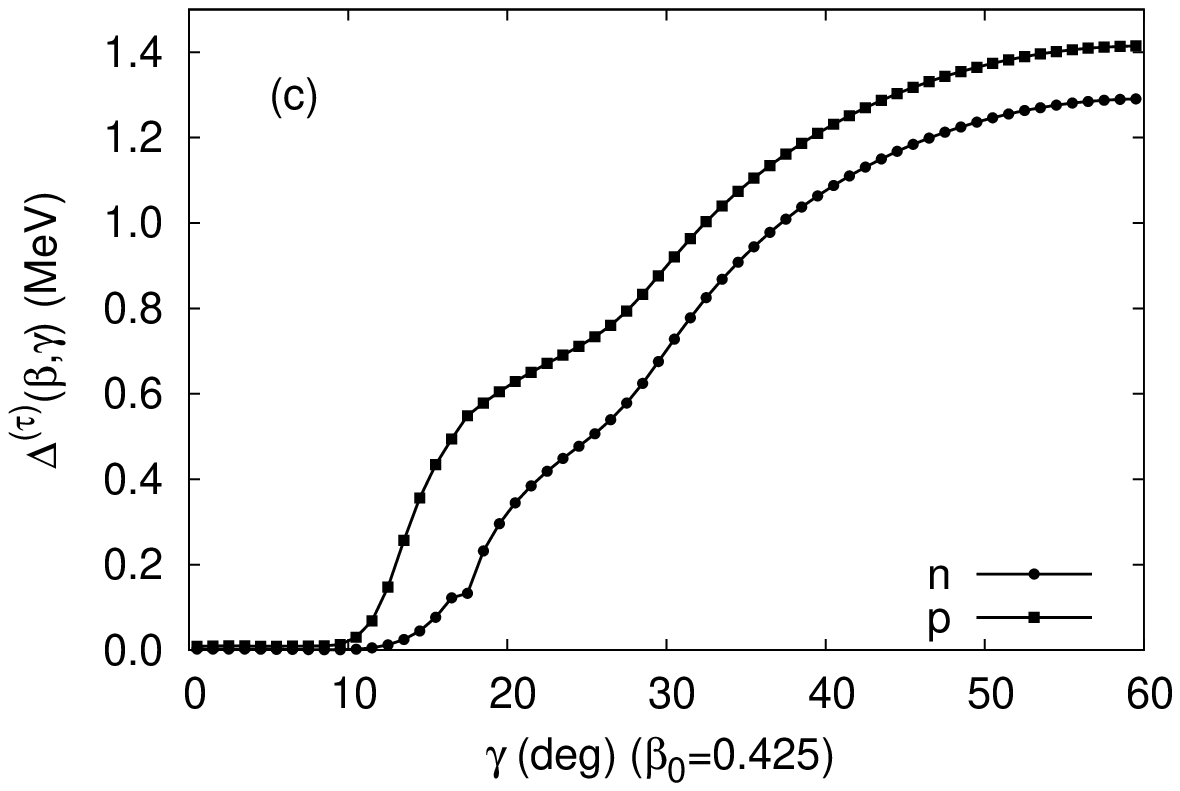}
\caption{\label{fig:LQRPAmodes}
(a) The LQRPA eigen frequencies $\omega^2$, (b) the vibrational part of the metric $W$,
and (c) the neutron and proton monopole pairing gaps $\Delta_0^{(\tau)}$ 
in $^{24}$Mg plotted
as a function of $\gamma$ for a constant value $\beta=0.425$.
In (a), the frequencies squared of low-lying LQRPA eigen modes are denoted with the cross symbols, while the two collective modes chosen at each deformation to calculate the vibrational collective masses are denoted by the filled circles.
In (b), the vibrational metrics are denoted by the filled circles,
while the vibrational metrics calculated with other possible combinations of the LQRPA modes are denoted by the cross symbols.}
\end{figure}

As discussed in the previous subsection,
one of the pairing features of the $sd$-shell nuclei is the collapse
of the pairing gaps around the deformed shell gaps (Fig.~\ref{fig:gap}).
In the former applications of the LQRPA equations to
Se and Kr isotopes \cite{PhysRevC.82.064313,Sato201153},
where the systems are in the superconducting phase in all over the $(\beta,\gamma)$
region considered,
the properties and choices of the
LQRPA modes in the normal phase has not
been analyzed though they are interesting theoretical
issues to be clarified.
In Fig.~\ref{fig:LQRPAmodes},
the eigen frequencies squared of the LQRPA modes $\omega^2(\beta,\gamma)$,
the vibrational part of the metric $W(\beta,\gamma)$,
and the neutron and proton monopole pairing gaps $\Delta_0^{(n)}$
and $\Delta_0^{(p)}$ in $^{24}$Mg
are plotted as functions of $\gamma$
along the $\beta=0.425$ line.

Let us label the chosen two collective modes in
Fig.~\ref{fig:LQRPAmodes} (a) as `mode A' and `mode B'.
Mode A denotes the collective mode lower in energy at $\gamma=0^\circ$, and 
this mode jumps around $\gamma=12^\circ, 32^\circ, 36^\circ$ and $41^\circ$.
It becomes the second lowest mode at $\gamma=60^\circ$.
Mode B denotes the collective mode higher in energy at $\gamma=0^\circ$, and 
this mode jumps around $\gamma=8^\circ$ and $25^\circ$.
 It becomes the lowest mode at $\gamma=60^\circ$.

As seen in Fig.~\ref{fig:LQRPAmodes} (c), in the region with $0^\circ< \gamma < 8^\circ$, both the neutron and proton pairing gaps vanish.
In this region, lowest four eigen modes are close in energy around 1-2 MeV, and they correspond to the neutron and proton pairing vibrational modes at a normal phase (pair addition and pair annihilation).
The fifth mode is chosen as a collective mode (mode A). It has the $\gamma$-vibrational character at the axial limit, and this mode is chosen continuously in the small $\gamma$ where the neutron is normal.
One important character in the normal system is that
the collectivity of the $\beta$ vibration is weakened \cite{BMvol2}.
In the present case, the $\beta$ vibrational mode is found at around 7 MeV.
In this energy region, 
the $\beta$-vibrational collective mode is sometimes embedded in other 
non-collective modes.
However, the figure shows that one can always find such modes using
the minimal metric criterion to select the two collective modes.

Around $8^\circ < \gamma < 12^\circ$, the proton becomes superconducting,
and the character of the low-lying modes changes to the neutron
pair addition and annihilation, proton pairing vibration, and proton
pairing rotation (Nambu-Goldstone mode). 
Note that in Fig.~\ref{fig:LQRPAmodes} (a), the zero-energy pairing rotational mode is not shown.

In $12^\circ < \gamma < 25^\circ$, 
neutrons also become superconducting.
As the pairing gaps significantly change as functions of deformation in
this region, the low-lying vibrational modes have the pairing
vibrational characters, and thus they are not chosen to evaluate the
quadrupole collective masses. In this region, the vibrational part of
the metric increases.

This situation changes in $25^\circ < \gamma < 41^\circ$.
The pairing vibration and quadrupole vibration mix in the lowest LQRPA mode,
and the pairing-vibrational character of the lowest mode decreases.
Therefore, the lowest LQRPA mode is continuously chosen as a collective mode (mode B) to oblate limit.
On the other hand, 
the other collective mode (mode A) changes in several LQRPA modes in this region,
because large quadrupole collectivity appears in these LQRPA modes.
The quadrupole collectivities of these modes are similar, and
the vibrational part of the metric $W$ is continuous around the jump.

Near the oblate region with $41^\circ < \gamma < 60^\circ$, 
the lowest two modes are chosen as the collective modes, 
and are decoupled
with other LQRPA modes in energy.
At the oblate axial limit, the lowest mode (mode B) becomes the $\gamma$ vibration,
and the second lowest mode (mode A) corresponds to the $\beta$ vibration.

This analysis shows that the minimal metric criterion for selecting the  two collective modes among low-lying LQRPA modes also works
for the situation where the pairing gap vanishes.

\subsection{Collective levels}

In this subsection, we present the excitation spectra,
electric transition properties, and collective wave functions
that are obtained by solving the collective Schr\"{o}dinger equation (\ref{eq:Schroedinger})
for $I\le 6$ states.

\subsubsection{$^{24}$Mg}

\begin{figure}
\includegraphics[width=80mm]{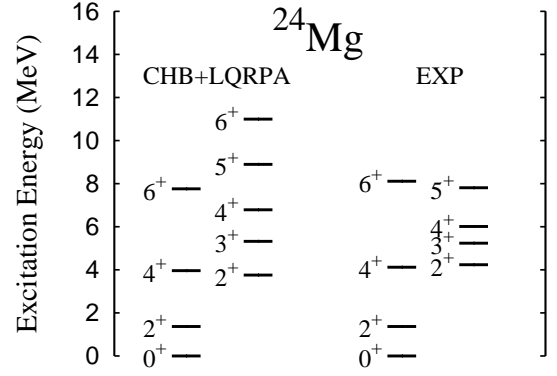}
\caption{\label{fig:24Mg-energy}
Excitation spectra calculated for $^{24}$Mg
by means of the CHB+LQRPA method and experimental data~\cite{ENSDF}.}
\end{figure}

\begin{figure}[htbp]
\begin{tabular}{cc}
\includegraphics[width=25mm]{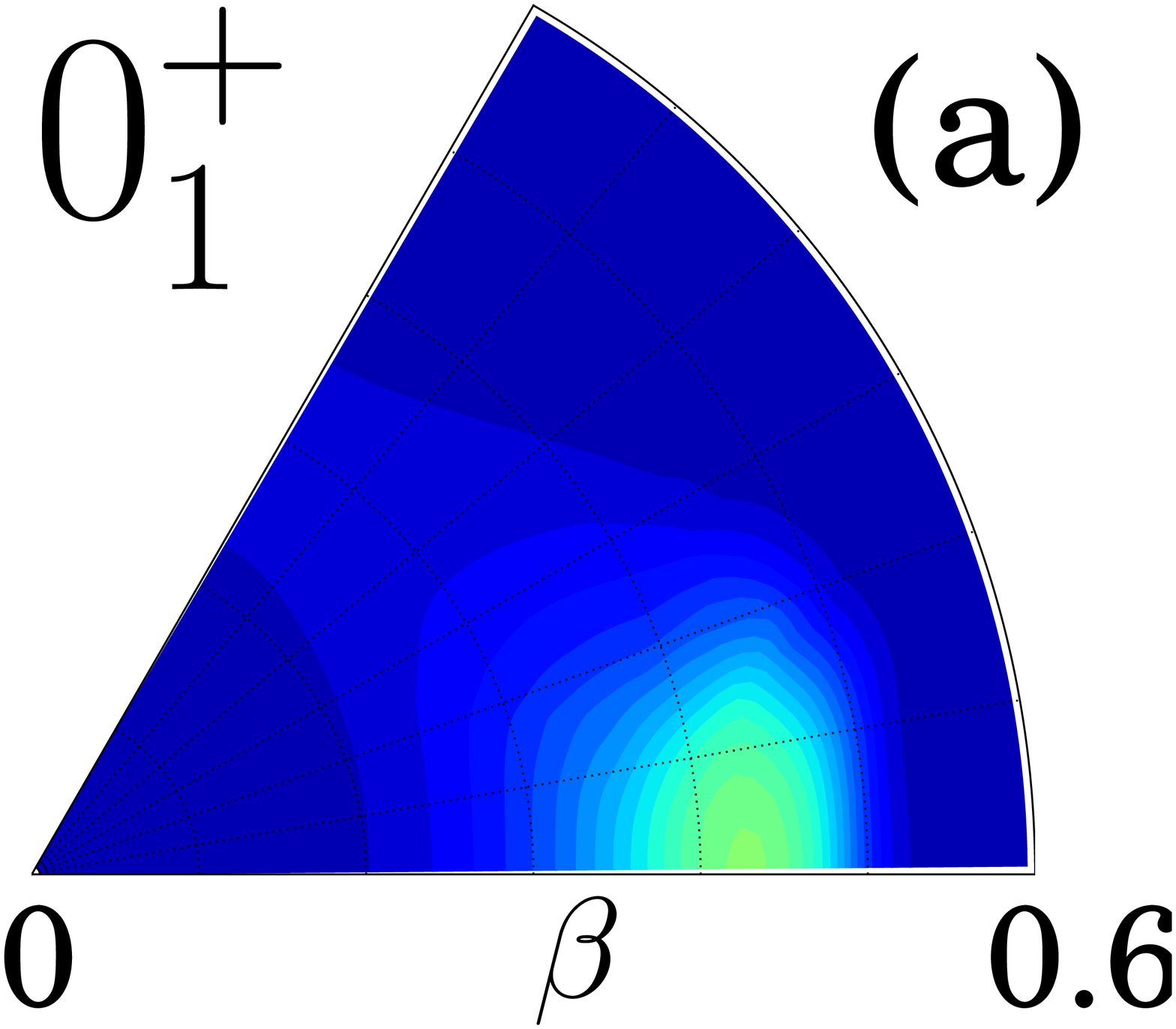} &
\includegraphics[width=25mm]{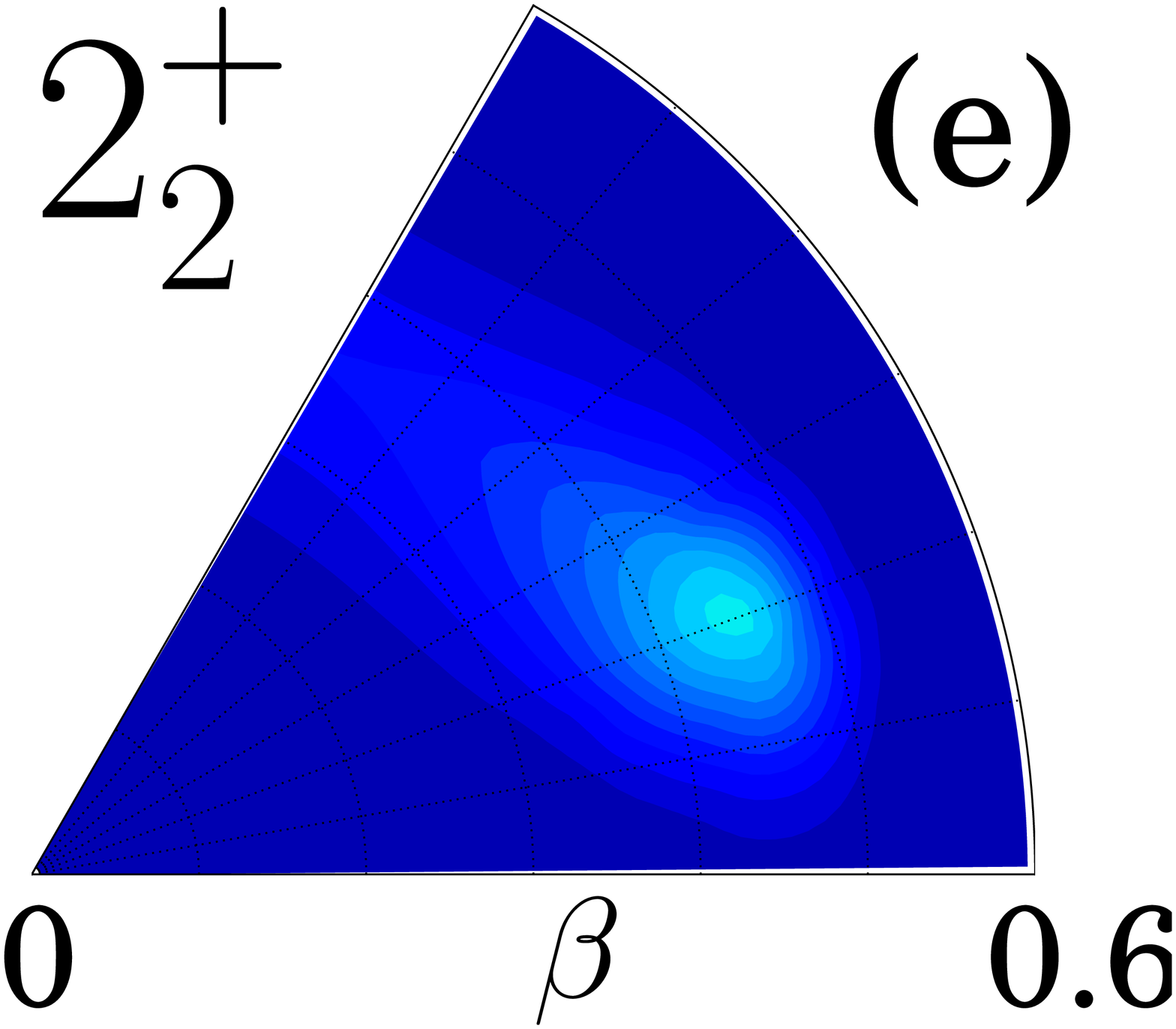} \\
\includegraphics[width=25mm]{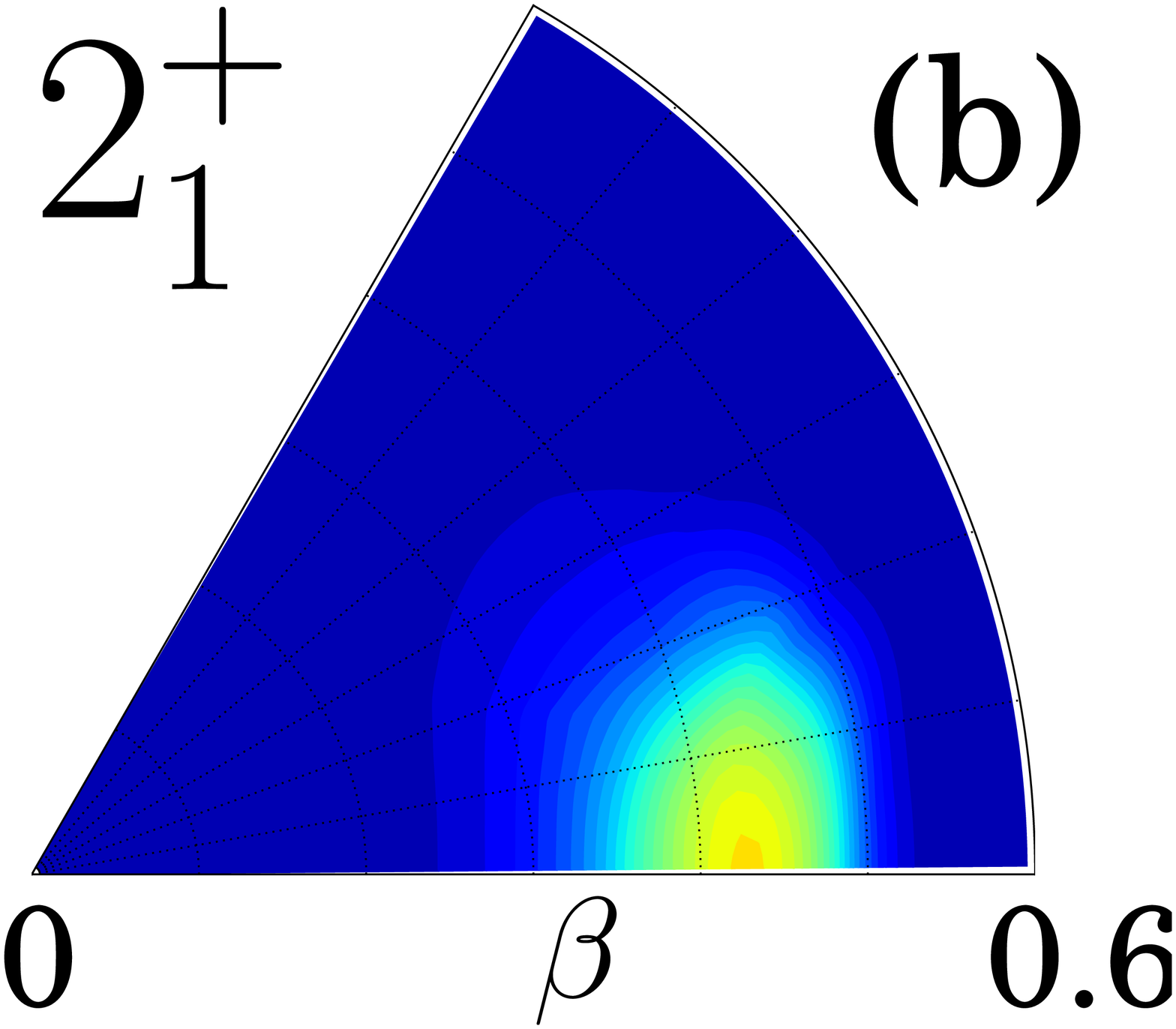} &
\includegraphics[width=25mm]{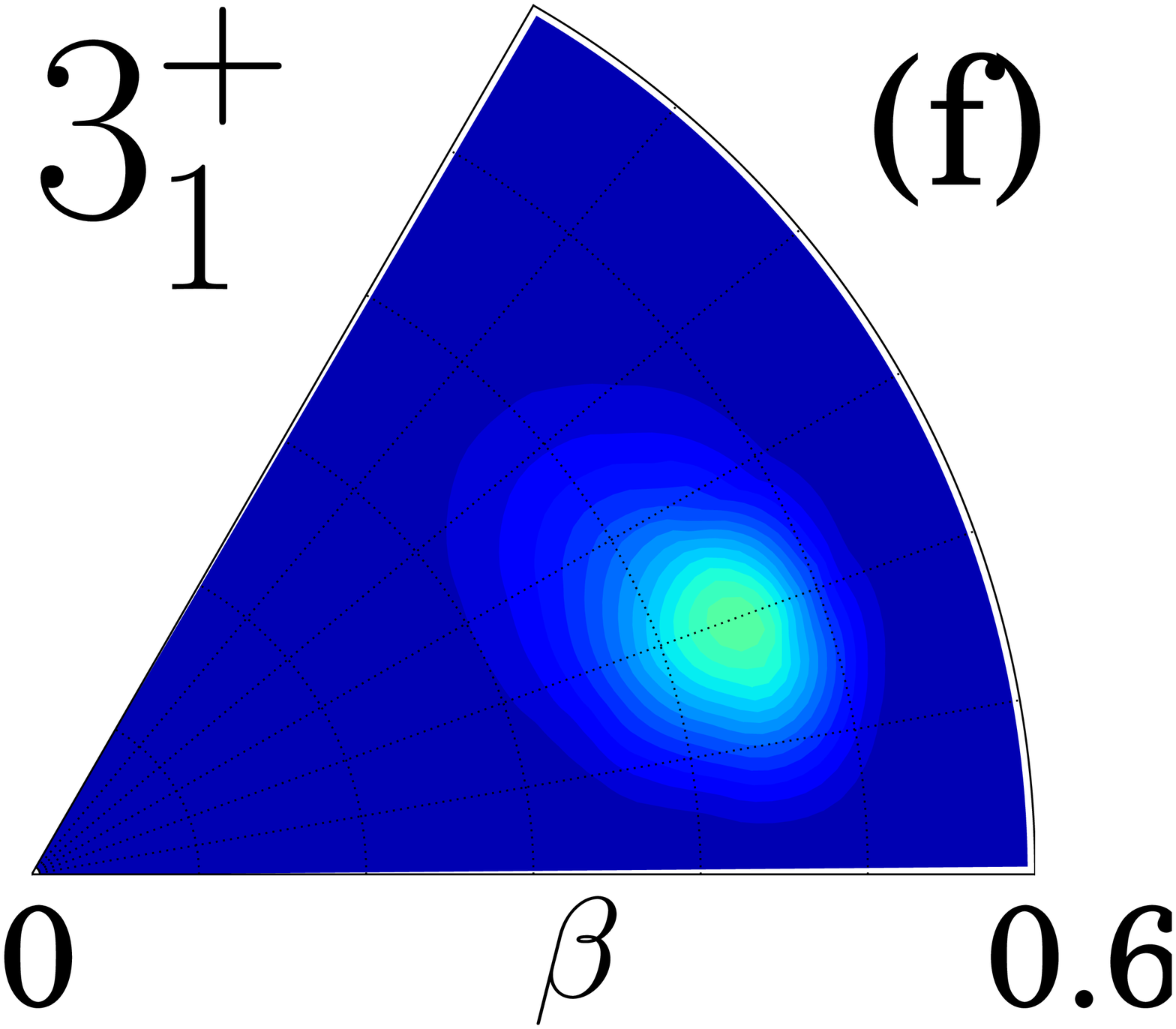} \\
\includegraphics[width=25mm]{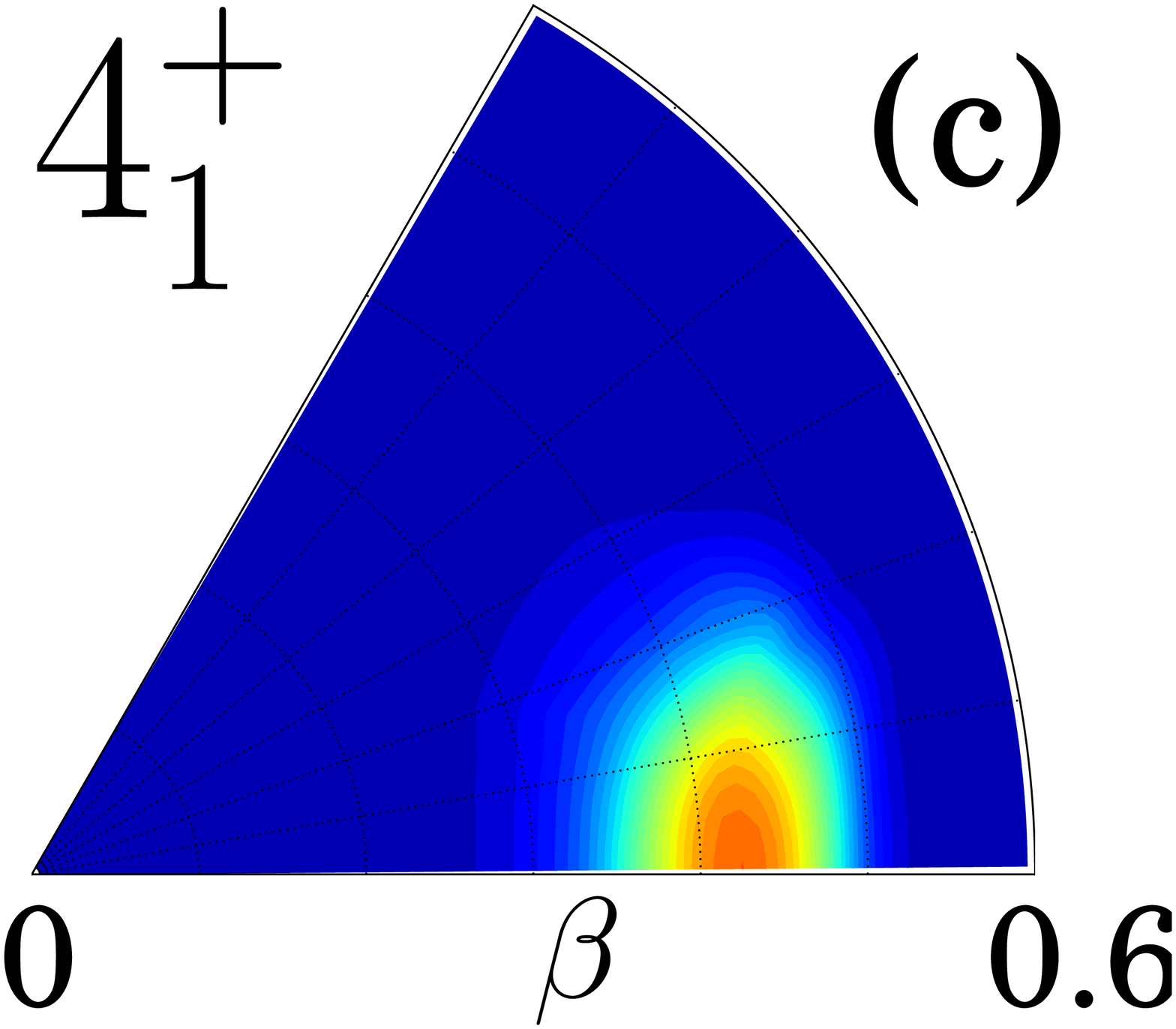} &
\includegraphics[width=25mm]{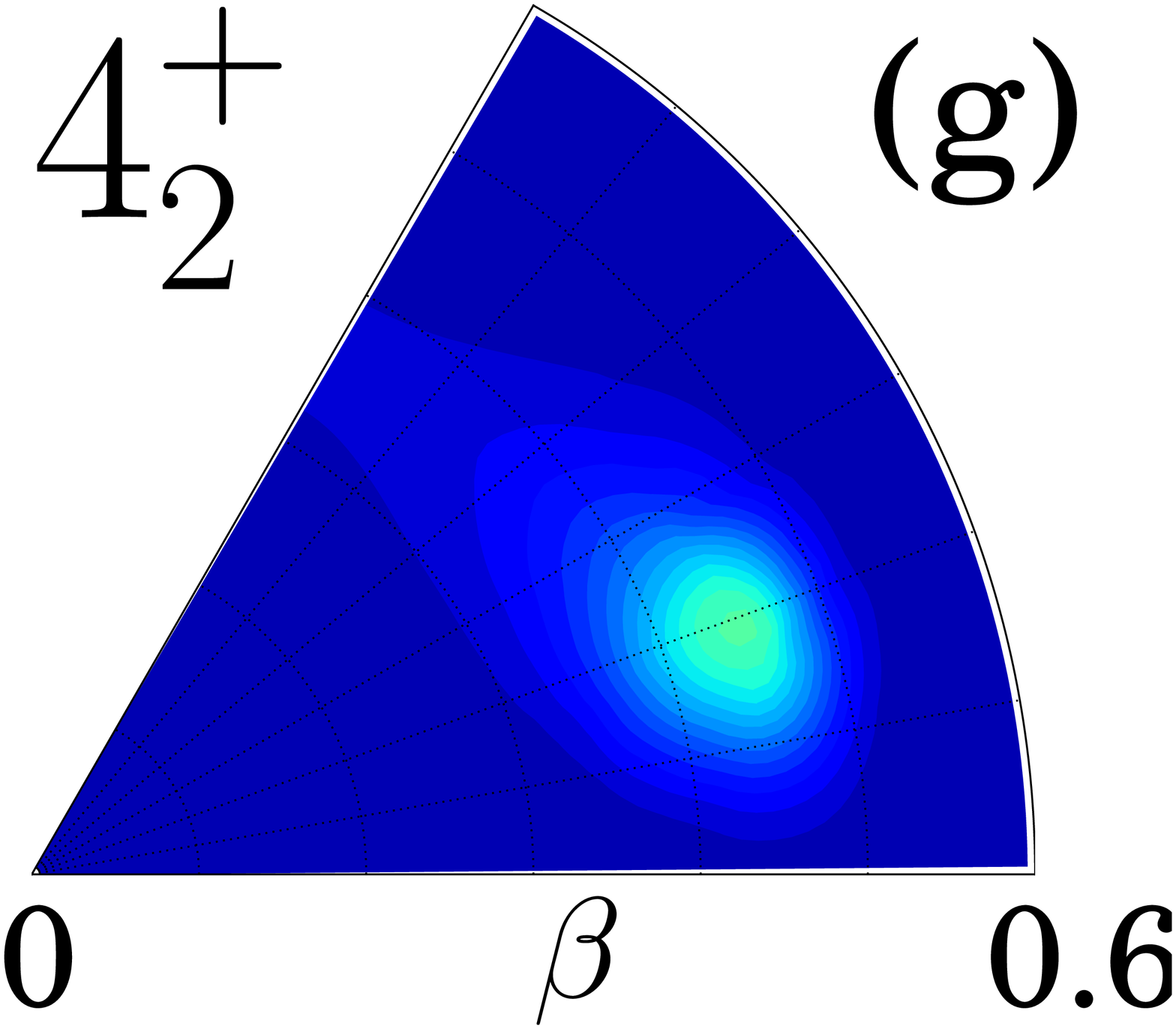} \\
\includegraphics[width=25mm]{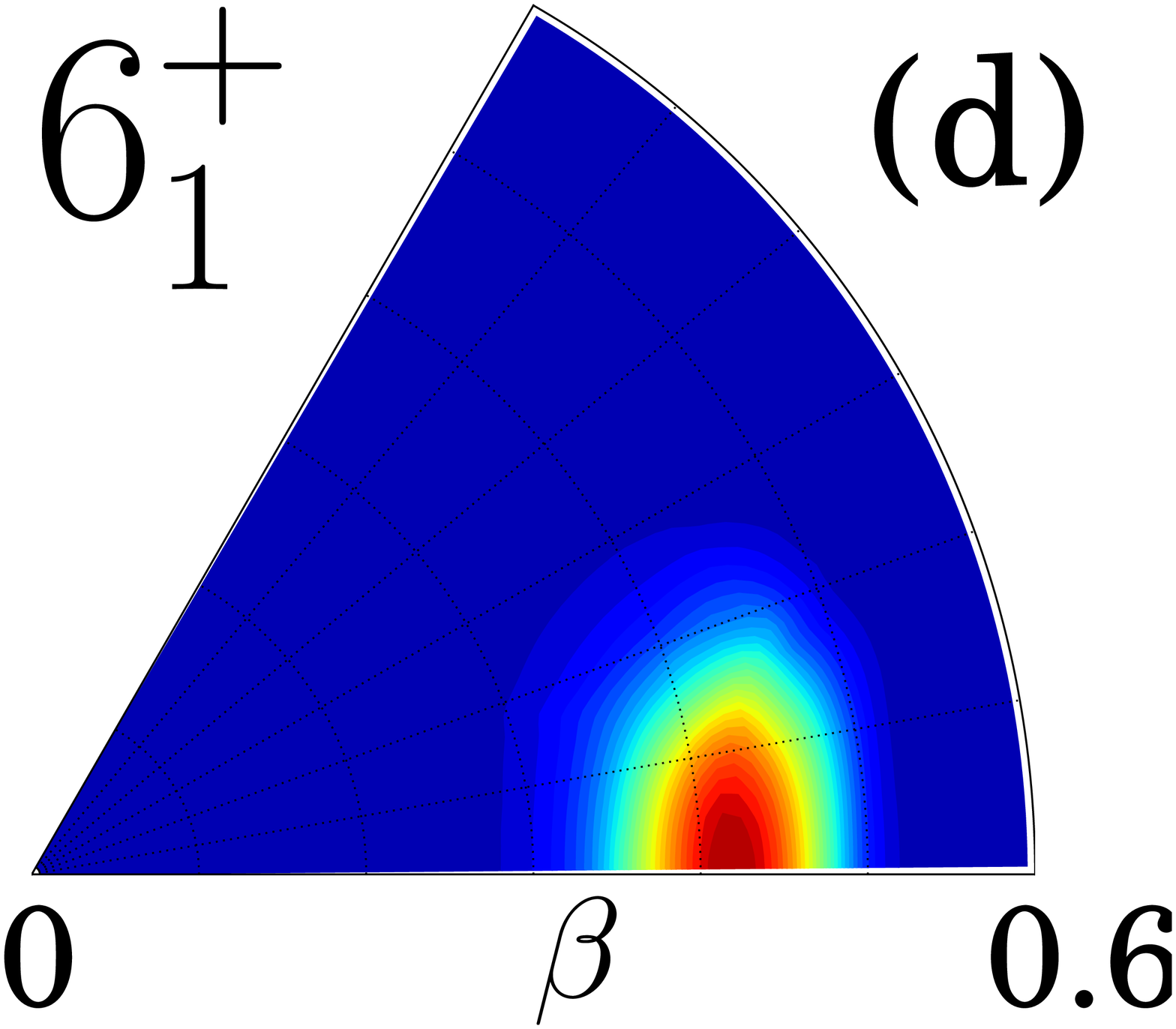} & 
\includegraphics[width=25mm]{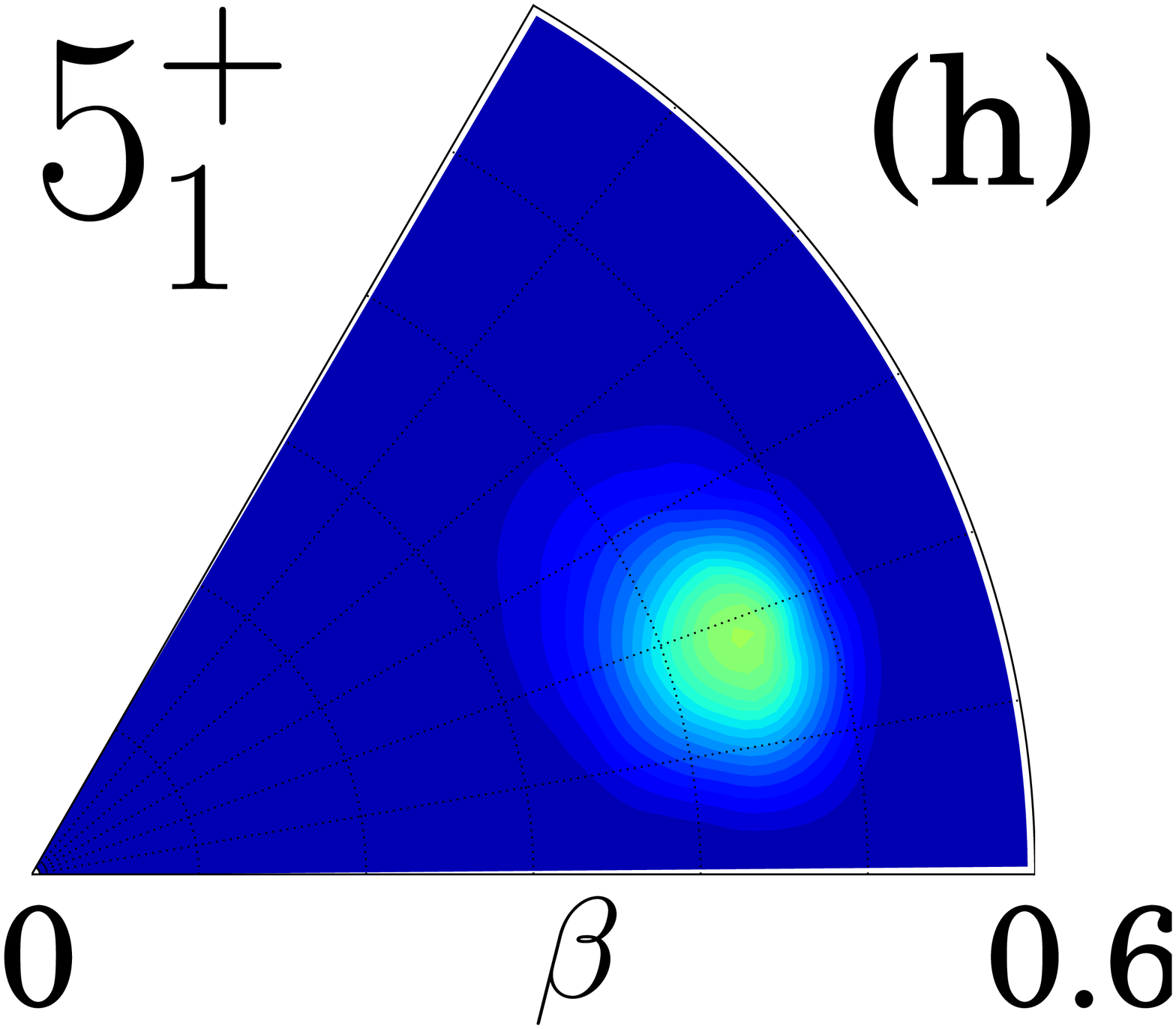} \\ &
\includegraphics[width=25mm]{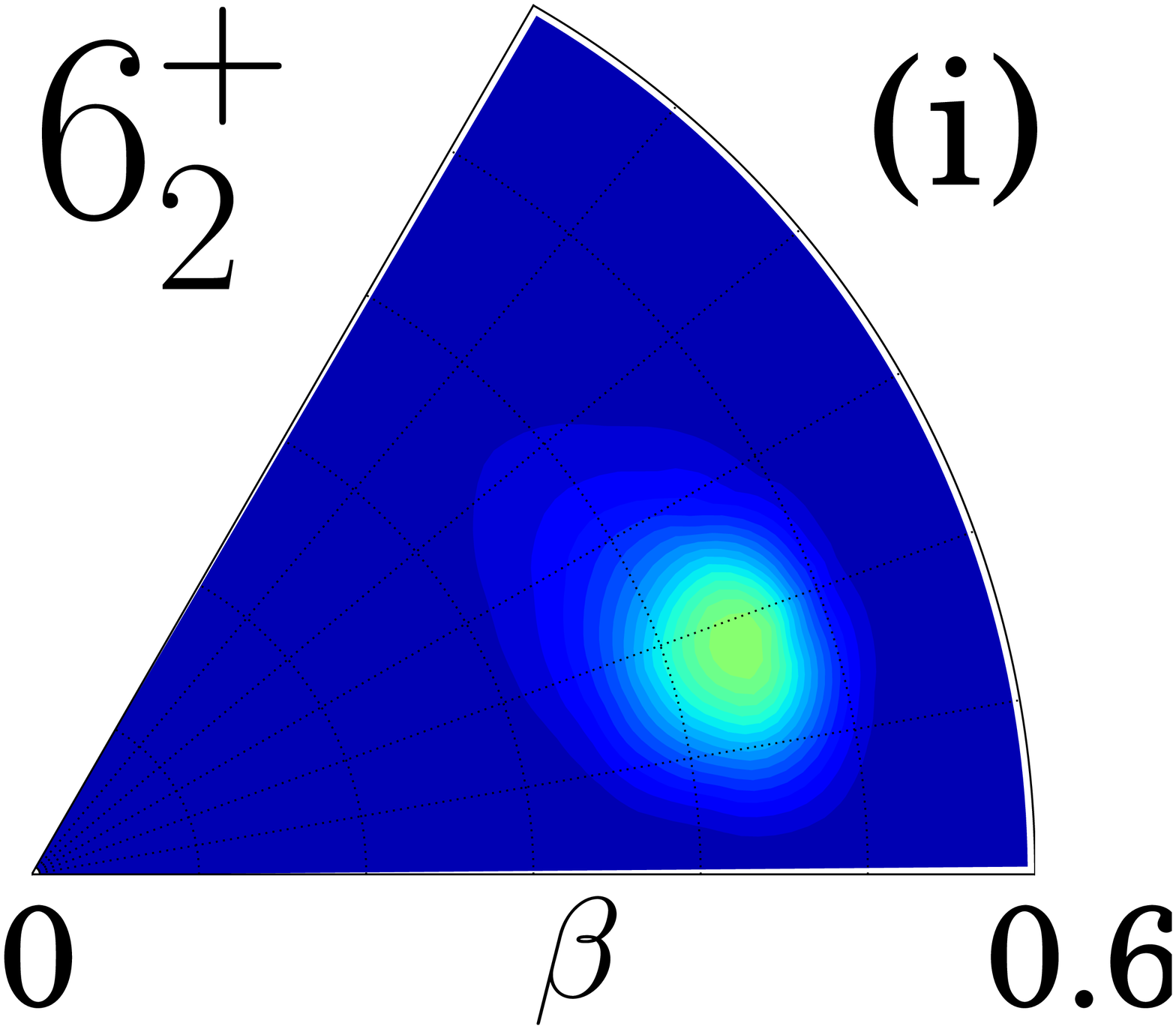} 
\end{tabular}
\caption{\label{fig:24Mg-wave}
(Color online) Vibrational wave functions squared
 $\beta^4 |\Phi_{\alpha I}(\beta,\gamma)|^2$ calculated for $^{24}$Mg.}
\end{figure}

The excitation spectra for $^{24}$Mg are shown in
Fig.~\ref{fig:24Mg-energy}.
The calculation yields an yrast rotational band composed of $0_1^+, 2_1^+, 4_1^+$,
and $6_1^+$ states, and an excited side band composed of $2_2^+, 3_1^+, 4_2^+, 5_1^+$,  and $6_2^+$
states.
These two bands are in a very good agreement with the experimental energy levels.
To analyze the structure of each state, 
the vibrational wave functions squared are displayed in
Fig.~\ref{fig:24Mg-wave}, where
the $\beta^4$ factor is multiplied, which carries the main $\beta$
dependence from the volume element $|G(\bg)|^{1/2}d\beta d\gamma$ in Eq.~(\ref{eq:volume}).
The members of the yrast band are localized around the prolate
minimum, showing the prolate character of the yrast rotational band.
The vibrational wave functions of the excited band
are concentrated in the triaxial region around the prolate
minimum with $\gamma\sim 20^\circ$.
This can be interpreted as the $\gamma$ vibrational band of the prolate yrast state.
We also analyze the $K$-component fraction for these states in
Table~\ref{table:24Mg-Kprob}. 
The table shows that the $K$-mixing in these states are very small,
and the results support the $K=0$ ground band and $K=2$ excited band.
These features are almost unchanged even with the increase of angular momentum.

In Tables \ref{table:24Mg-E2} and \ref{table:24Mg-Q}, 
the electric properties of the low-lying states are summarized.
In addition to the theoretical results calculated with the effective charges
($e_{\rm eff}^{(n)},e_{\rm eff}^{(p)}$)=(0.5,1.5),
we also list the values for pure neutron and proton contributions
obtained by using ($e_{\rm eff}^{(n)},e_{\rm eff}^{(p)}$)=(1,0) and (0,1).
The calculated results reasonably reproduce the trend of the $E2$ transition
strengths though they tend to underestimate the absolute values of experimental data.
Using a different set of effective charges improves the
systematical underestimation of the theoretical values.
However, still there are some disagreement, for example, in $4_1^+ \rightarrow 2_1^+$, $3_1^+\rightarrow 2_2^+$ and $5_1^+ \rightarrow 3_1^+$ transitions.
One can see that the neutron and proton matrix elements are almost equal
for all the transitions listed in the table.
The prolate feature is shown also in the spectroscopic quadrupole
moments.
The calculated values for the yrast and the side bands are well described
with the estimated values from the rotational collective model of a
prolate deformation with
$K=0$ and $2$,  respectively.
This shows the rigid deformation feature in $^{24}$Mg.

As mentioned in Introduction, several 
triaxial GCM calculations are performed for the study of 
low-lying states in $^{24}$Mg employing modern density functionals
\cite{PhysRevC.81.044311,bender:024309,PhysRevC.81.064323}.
In comparison with them, the present calculation 
gives a remarkable agreement with the experimental data
for the ground $K=0$ band and the excited $K=2$ band
despite the schematic effective interaction and restricted model space.
Especially, the agreement in the excitation energies are 
better than the GCM calculations,
while that in the $B(E2)$ values are worse.
The small $K$-mixing properties in the ground and excited bands shown in Table
\ref{table:24Mg-Kprob} are consistent with them \cite{PhysRevC.81.044311,PhysRevC.81.064323}.

\begin{table}[htbp]
\caption{\label{table:24Mg-Kprob}
Probability amplitude of quantum states of $^{24}$Mg for each $K$-component.}
\begin{tabular}{ccccc} \hline\hline
$I^\pi_\alpha$ & $K=0$ & $K=2$ & $K=4$ & $K=6$\\ \hline
$0_1^+$ & 1.00 & -    & -    & -   \\
$2_1^+$ & 0.99 & 0.01 & -    & -   \\
$4_1^+$ & 1.00 & 0.00 & 0.00 & -   \\
$6_1^+$ & 1.00 & 0.00 & 0.00 & 0.00\\ \hline
$2_2^+$ & 0.03 & 0.97 &    - & -   \\
$3_1^+$ &  -   & 1.00 &   -  & -   \\
$4_2^+$ & 0.03 & 0.95 & 0.02 & -   \\
$5_1^+$ &  -   & 1.00 & 0.00 & -   \\
$6_2^+$ & 0.02 & 0.97 & 0.01 & 0.00\\ \hline\hline
\end{tabular}
\end{table}

\begin{table}[htbp]
\caption{\label{table:24Mg-E2}
The values of $B(E2)$ for $^{24}$Mg listed in units of $e^2$fm$^4$. 
The theoretical values are calculated with the effective charges 
($e_{\rm eff}^{(n)}, e_{\rm eff}^{(p)}$) = (0.5,1.5).
The values calculated from pure neutron contribution ($e_{\rm eff}^{(n)},e_{\rm eff}^{(p)}$) = (1,0)
and proton contribution ($e_{\rm eff}^{(n)},e_{\rm eff}^{(p)}$) = (0,1) 
are also listed. Experimental data are taken from Ref.~\cite{ENSDF}.}
\begin{tabular}{ccccc} \hline \hline
 & EXP & CHB+LQRPA & neutron & proton \\ \hline
$2_1^+ \rightarrow 0_1^+$ & 88 &  63.026 &  16.189 &  15.614 \\
$4_1^+ \rightarrow 2_1^+$ &160 &  96.171 &  24.663 &  23.838 \\
$6_1^+ \rightarrow 4_1^+$ &155 & 108.032 &  27.684 &  26.784 \\ \hline
$3_1^+ \rightarrow 2_2^+$ &239 & 103.484 &  26.686 &  25.602 \\
$4_2^+ \rightarrow 3_1^+$ & - &  80.216 &  20.674 &  19.849 \\
$5_1^+ \rightarrow 4_2^+$ & - &  57.085 &  14.713 &  14.125 \\
$6_2^+ \rightarrow 5_1^+$ & - &  47.575 &  12.220 &  11.786 \\
$4_2^+ \rightarrow 2_2^+$ & 64 &  44.673 &  11.448 &  11.076 \\
$5_1^+ \rightarrow 3_1^+$ & 149 &  65.981 &  16.944 &  16.347 \\
$6_2^+ \rightarrow 4_2^+$ & - &  83.500 &  21.415 &  20.697 \\ \hline
$4_1^+ \rightarrow 2_2^+$ & - &   0.011 &   0.003 &   0.003 \\
$2_2^+ \rightarrow 2_1^+$ & 15 &  17.197 &   4.216 &   4.327 \\
$2_2^+ \rightarrow 0_1^+$ & 8 &   4.911 &   1.179 &   1.244 \\
$3_1^+ \rightarrow 4_1^+$ & - &   5.091 &   1.241 &   1.284 \\
$3_1^+ \rightarrow 2_1^+$ & 10 &   8.180 &   1.948 &   2.078 \\
$4_2^+ \rightarrow 4_1^+$ & - &  12.100 &   2.925 &   3.059 \\
$6_1^+ \rightarrow 4_2^+$ & - &   0.018 &   0.004 &   0.005 \\
$4_2^+ \rightarrow 2_1^+$ & 5 &   3.493 &   0.849 &   0.882 \\
$5_1^+ \rightarrow 6_1^+$ & - &   4.217 &   1.018 &   1.066 \\
$5_1^+ \rightarrow 4_1^+$ & - &   7.618 &   1.825 &   1.931 \\
$6_2^+ \rightarrow 6_1^+$ & - &   9.756 &   2.348 &   2.470 \\
$6_2^+ \rightarrow 4_1^+$ & - &   3.534 &   0.867 &   0.889 \\ \hline\hline
\end{tabular} 
\end{table}

\begin{table}
\caption{\label{table:24Mg-Q} Spectroscopic quadrupole moments for
 $^{24}$Mg listed in units of $e$ fm$^2$.
See also caption in Table~\ref{table:24Mg-E2}.
Experimental data are taken from Ref.~\cite{ENSDF}.}
\begin{tabular}{ccccc} \hline\hline
 & EXP & CHB+LQPRA & neutron & proton \\ \hline
$Q(2_1^+)$ & $-$16.6 & $-$15.7 & $-$7.97 & $-$7.81 \\
$Q(4_1^+)$ & -       & $-$20.8 & $-$10.5 & $-$10.4 \\
$Q(6_1^+)$ & -       & $-$23.7 & $-$12.0 & $-$11.8 \\ \hline
$Q(2_2^+)$ & -       &    15.5 &    7.87 &    7.72 \\
$Q(3_1^+)$ & -       &    0    &     0   &    0    \\
$Q(4_2^+)$ & -       & $-$5.85 & $-$3.00 & $-$2.90 \\
$Q(5_1^+)$ & -       & $-$13.0 & $-$6.58 & $-$6.45 \\
$Q(6_2^+)$ & -       & $-$14.4 & $-$7.34 & $-$7.16 \\ \hline\hline
\end{tabular}
\end{table}

\subsubsection{$^{28}$Si}

\begin{figure}[htbp]
\includegraphics[width=80mm]{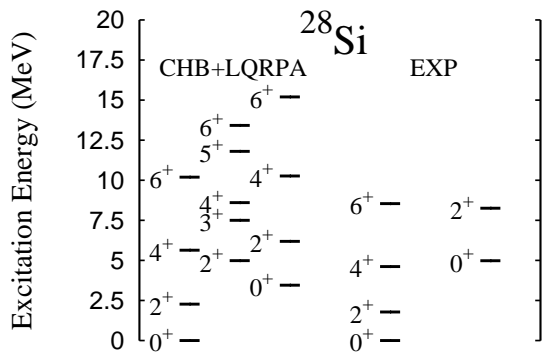}
\caption{\label{fig:28Si-energy}
Excitation spectra calculated for $^{28}$Si
by means of the CHB+LQRPA method and experimental data~\cite{ENSDF}.}
\end{figure}

\begin{figure}[htbp]
\begin{tabular}{ccc}
\includegraphics[width=25mm]{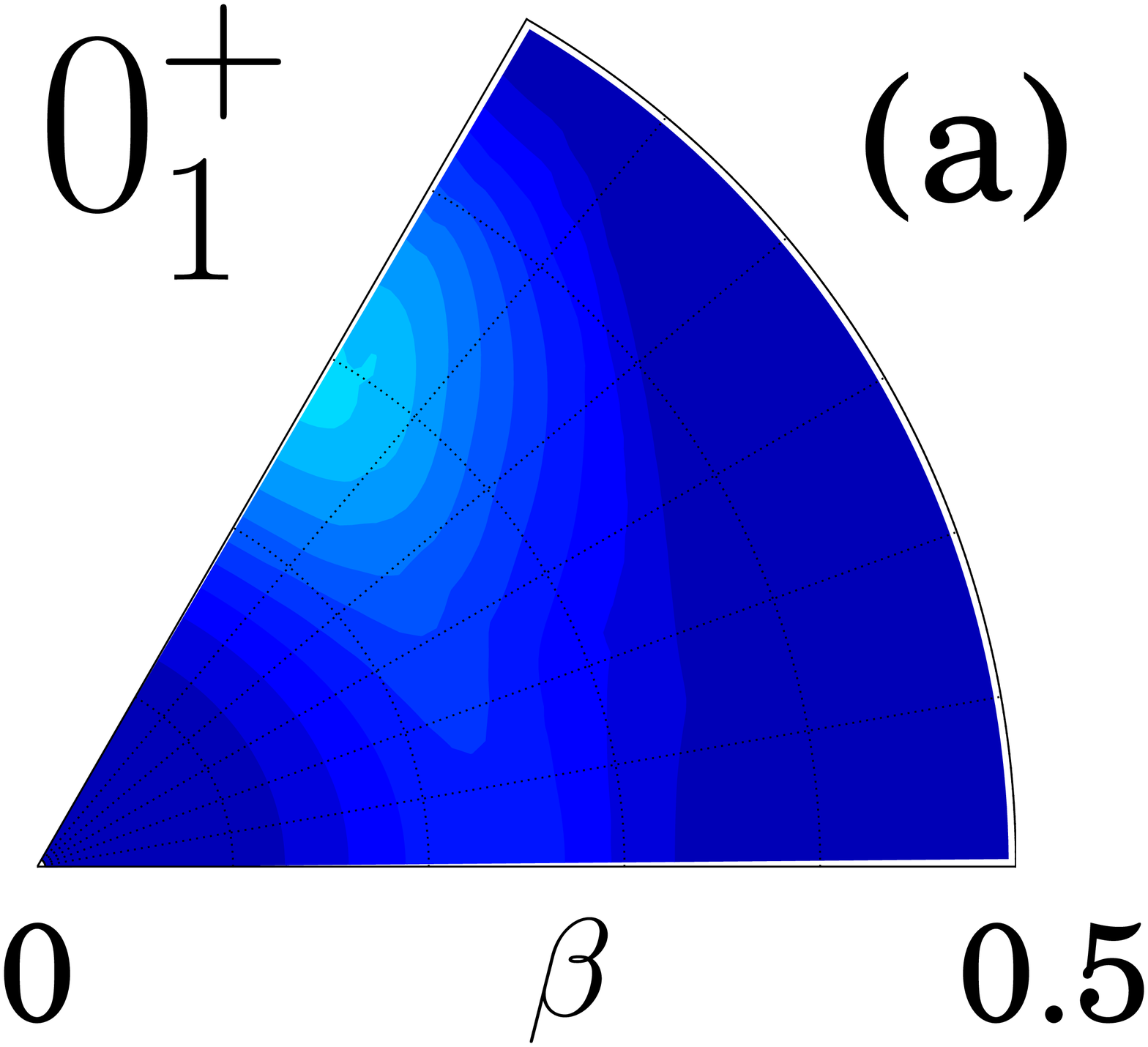} &
\includegraphics[width=25mm]{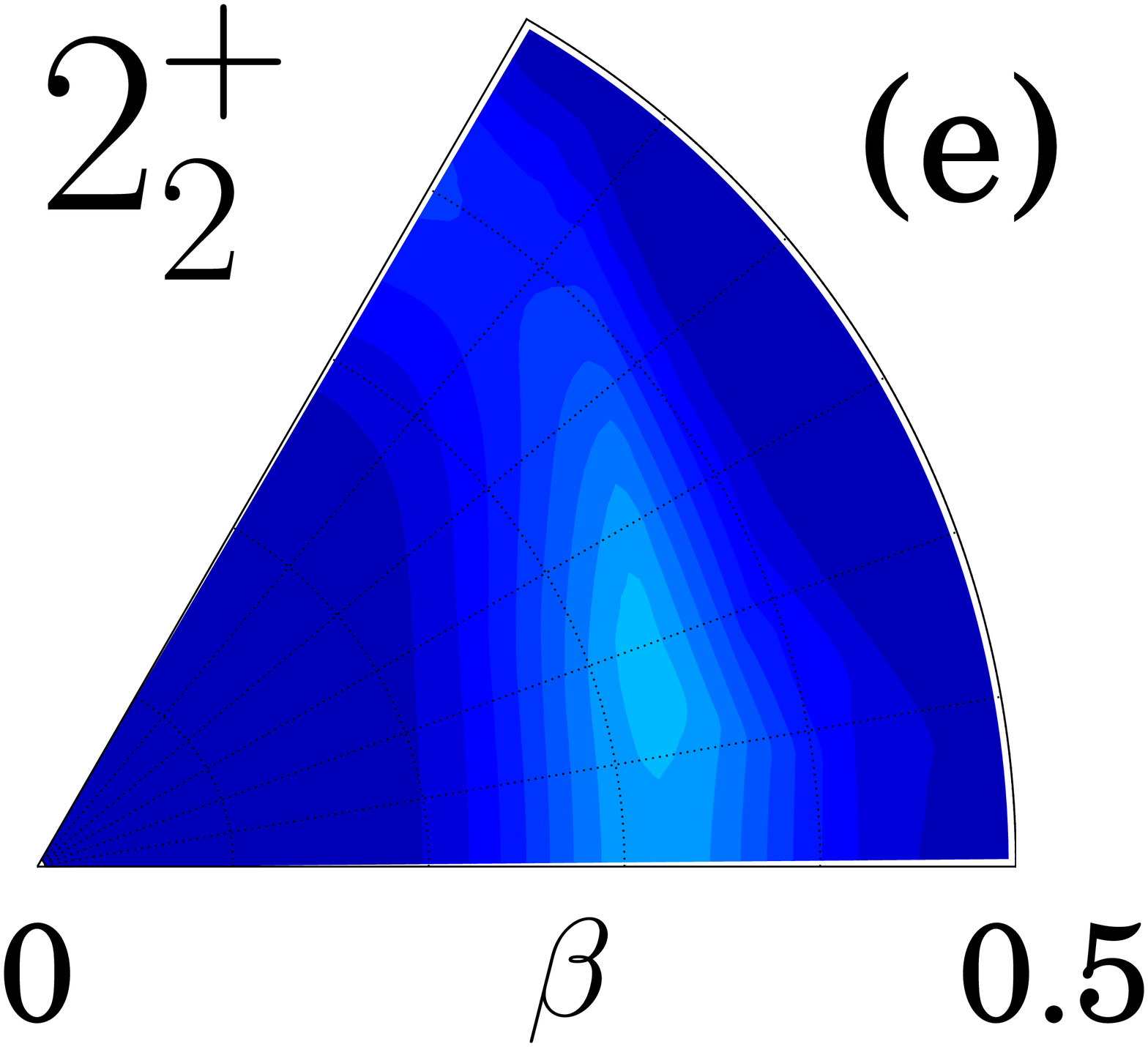} &
\includegraphics[width=25mm]{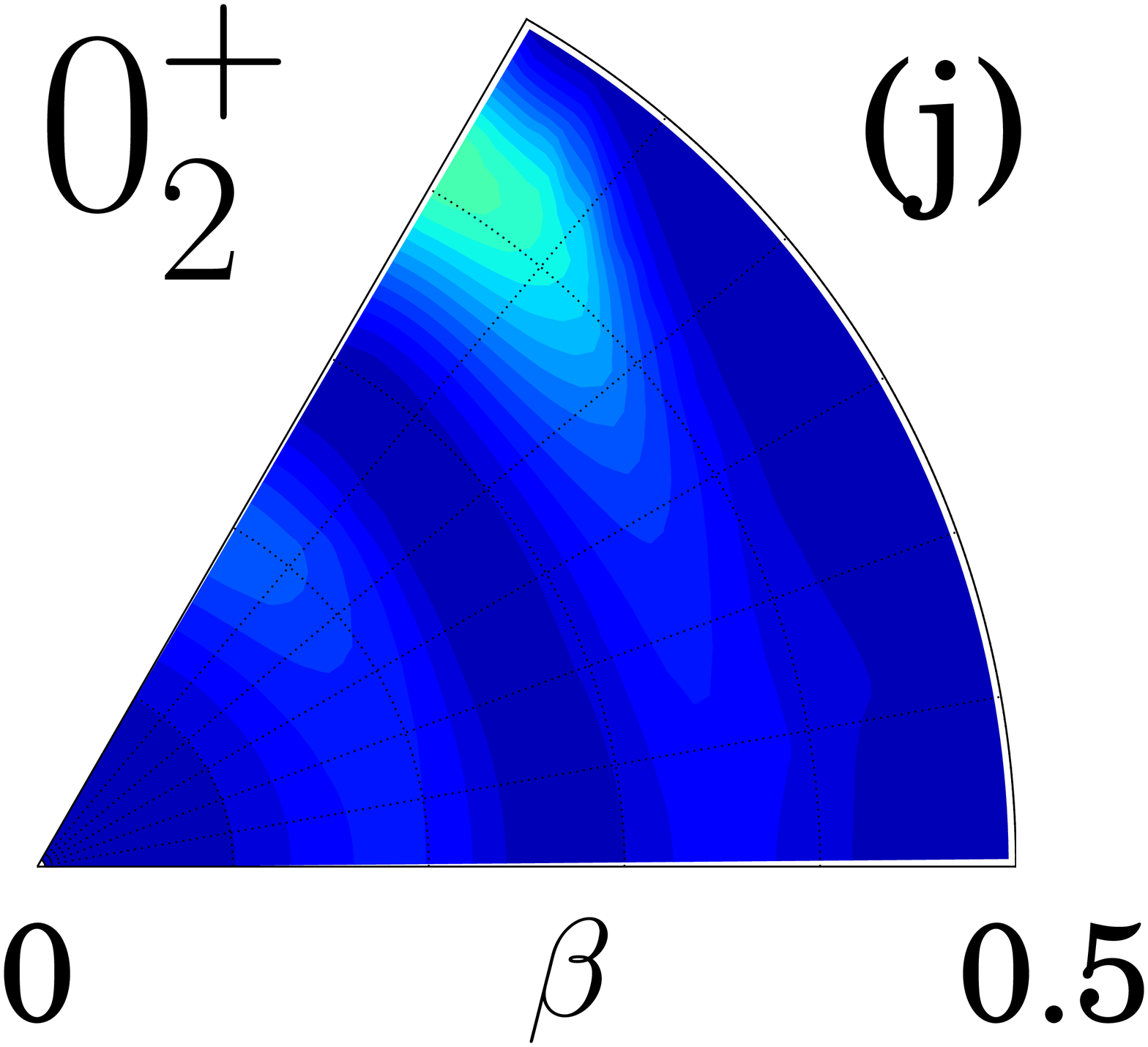} \\
\includegraphics[width=25mm]{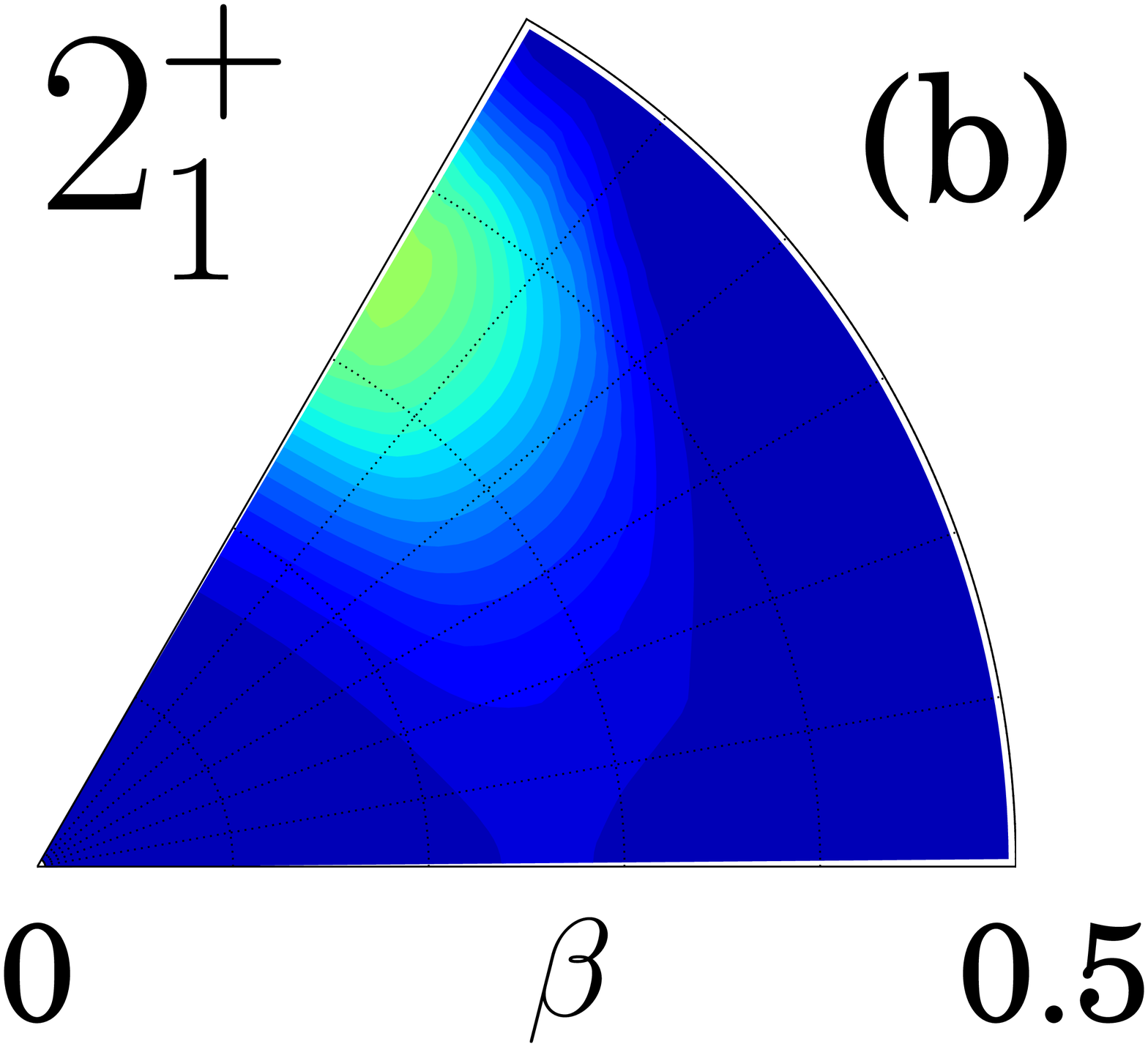} &
\includegraphics[width=25mm]{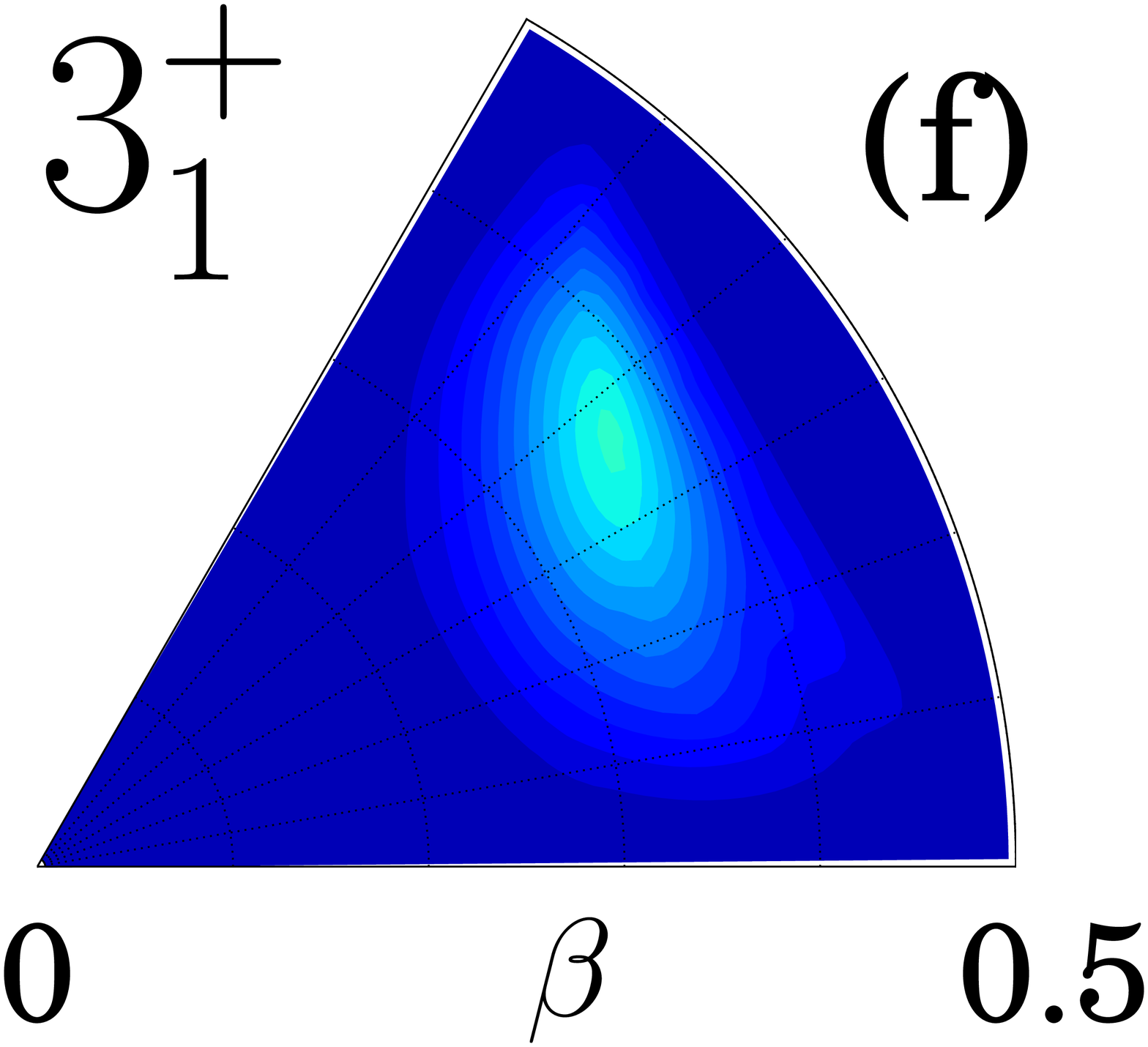} &
\includegraphics[width=25mm]{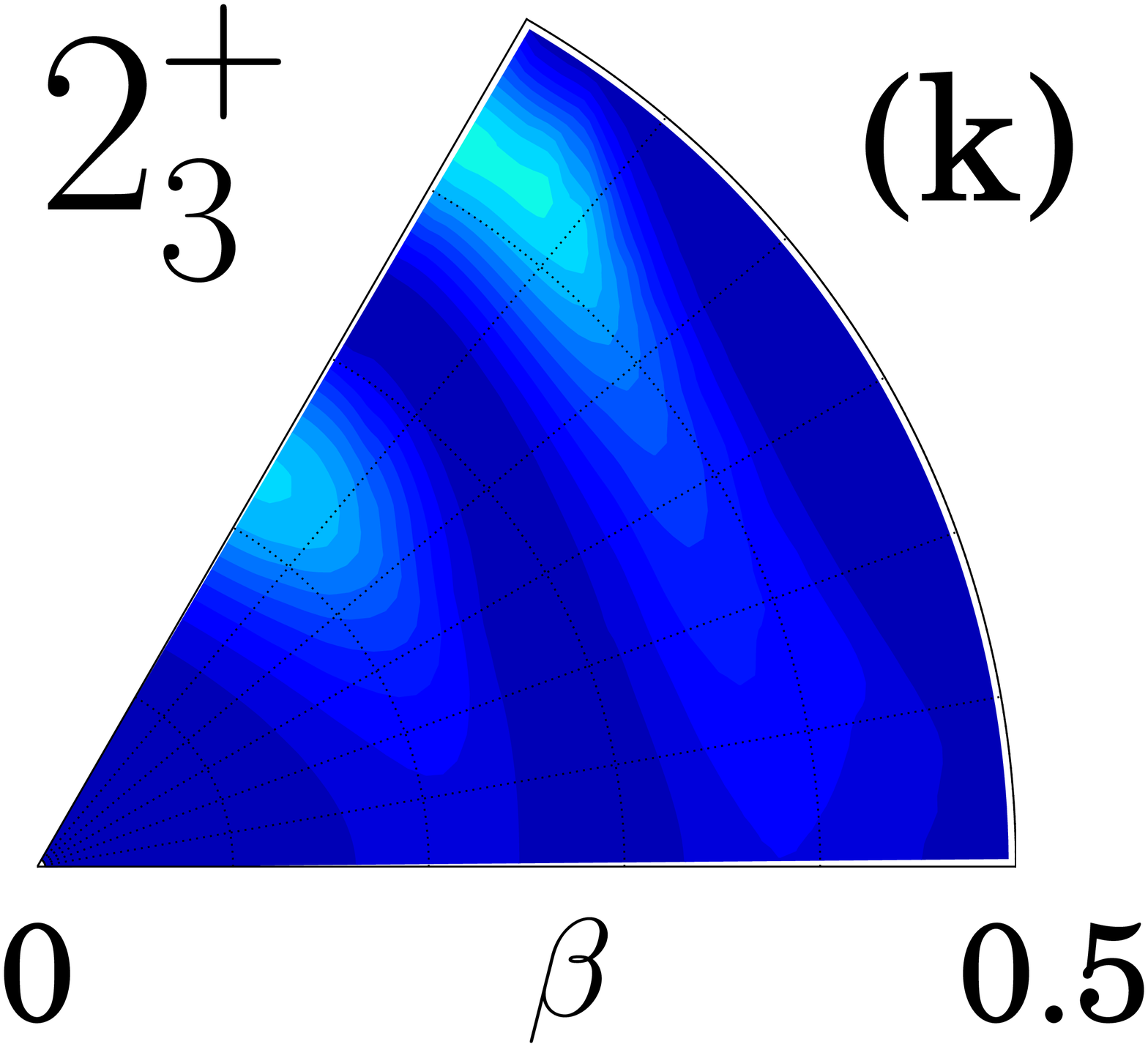} \\
\includegraphics[width=25mm]{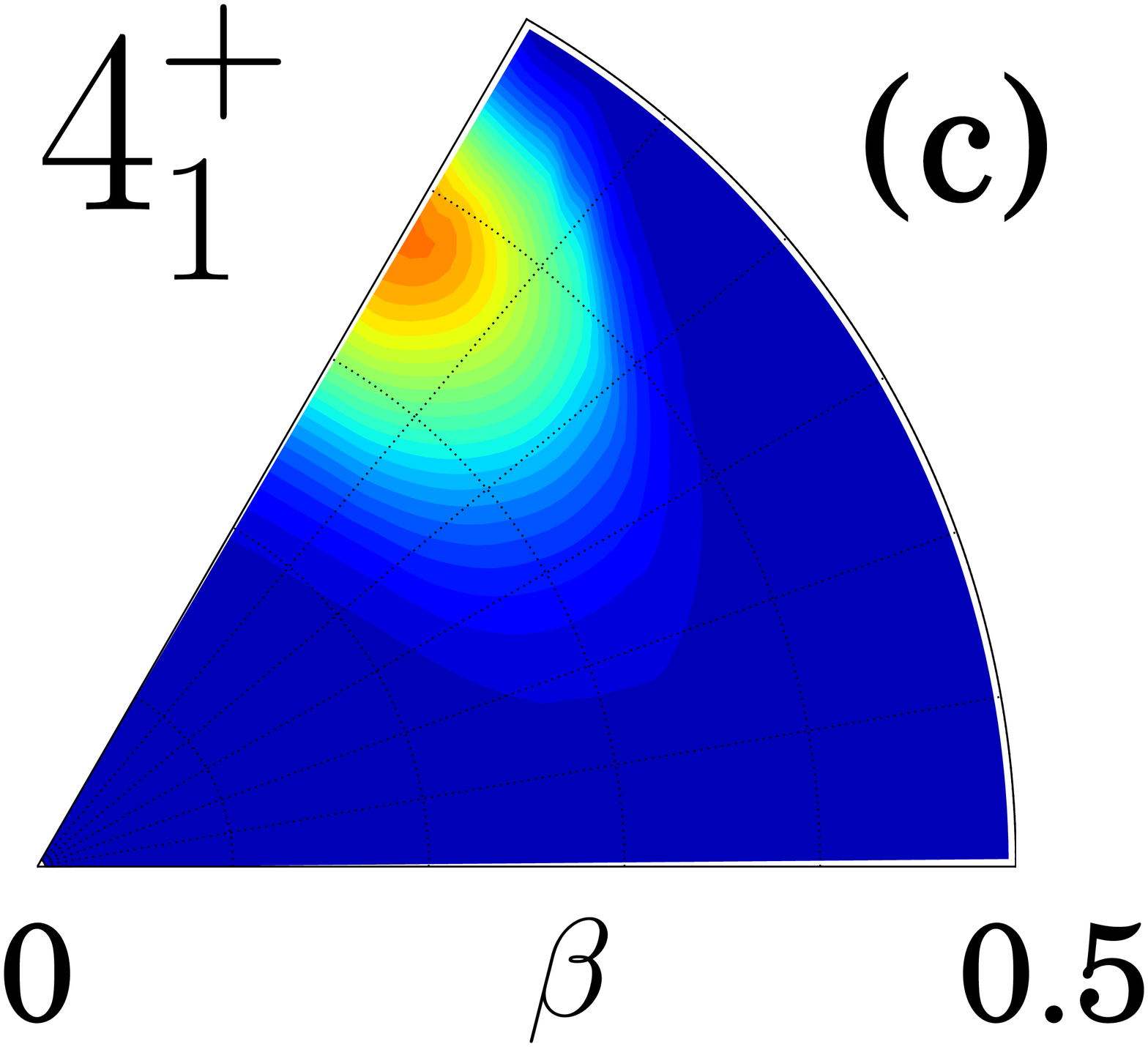} &
\includegraphics[width=25mm]{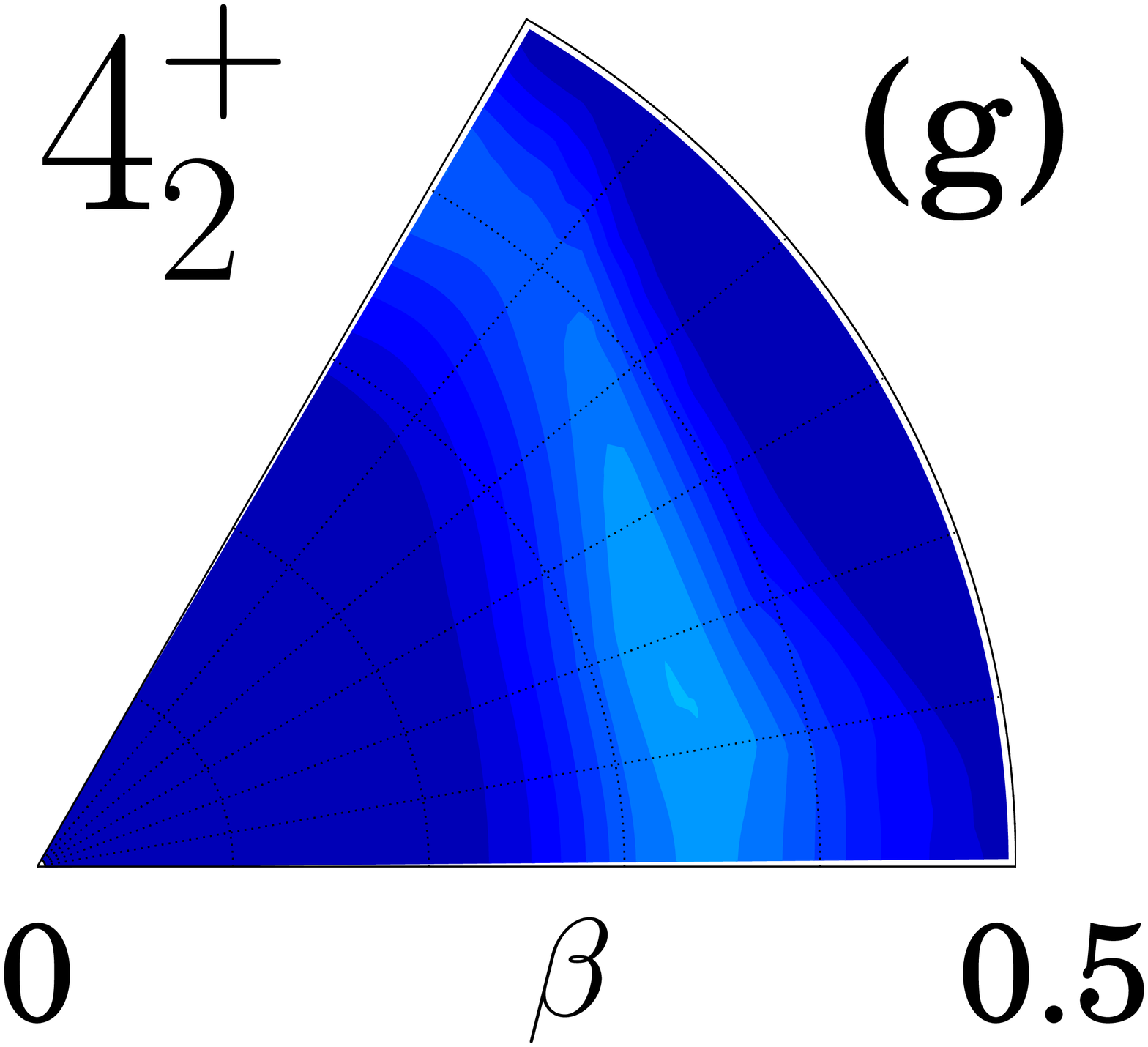} &
\includegraphics[width=25mm]{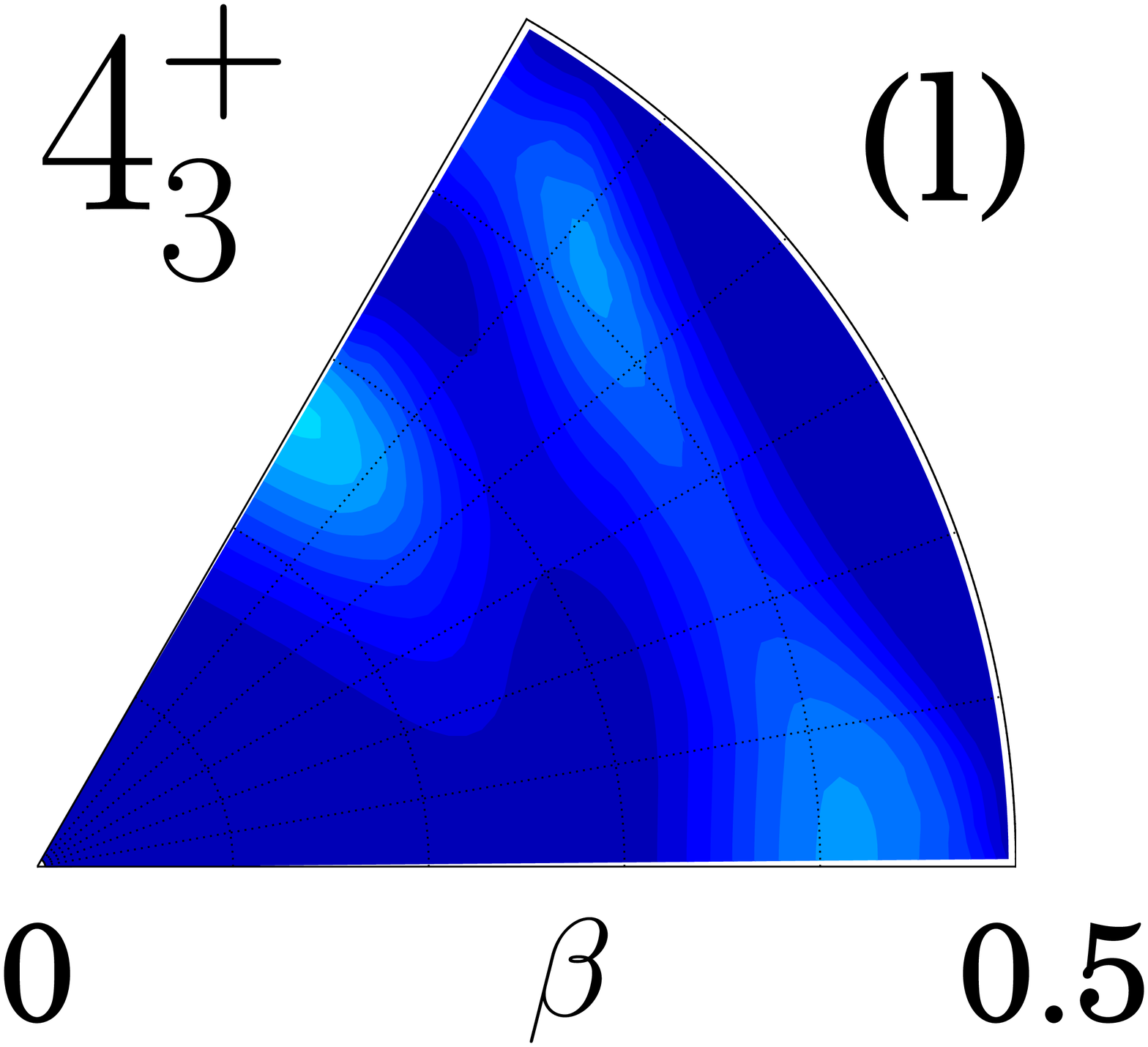} \\
\includegraphics[width=25mm]{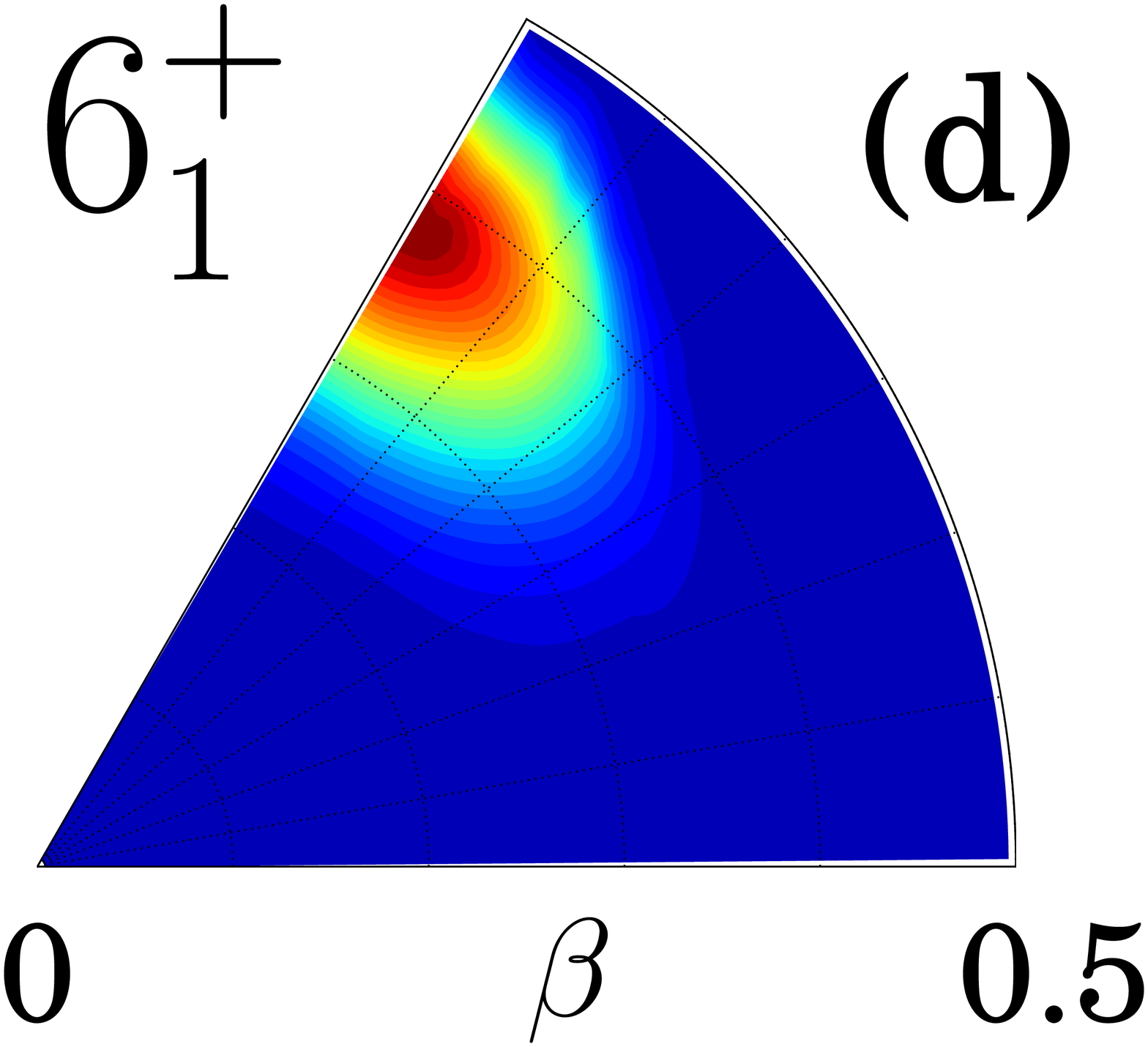} & 
\includegraphics[width=25mm]{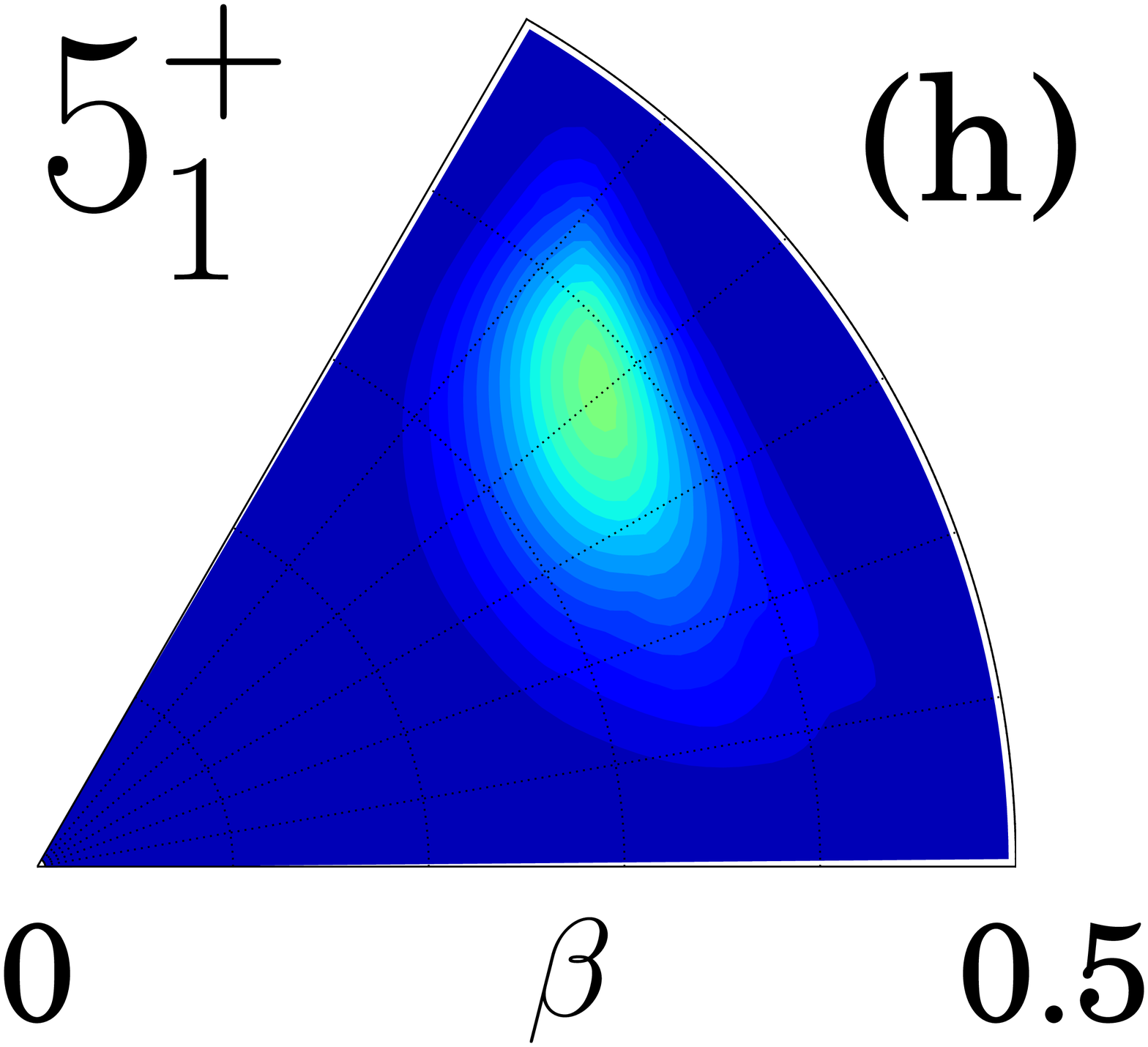} & \\ &
\includegraphics[width=25mm]{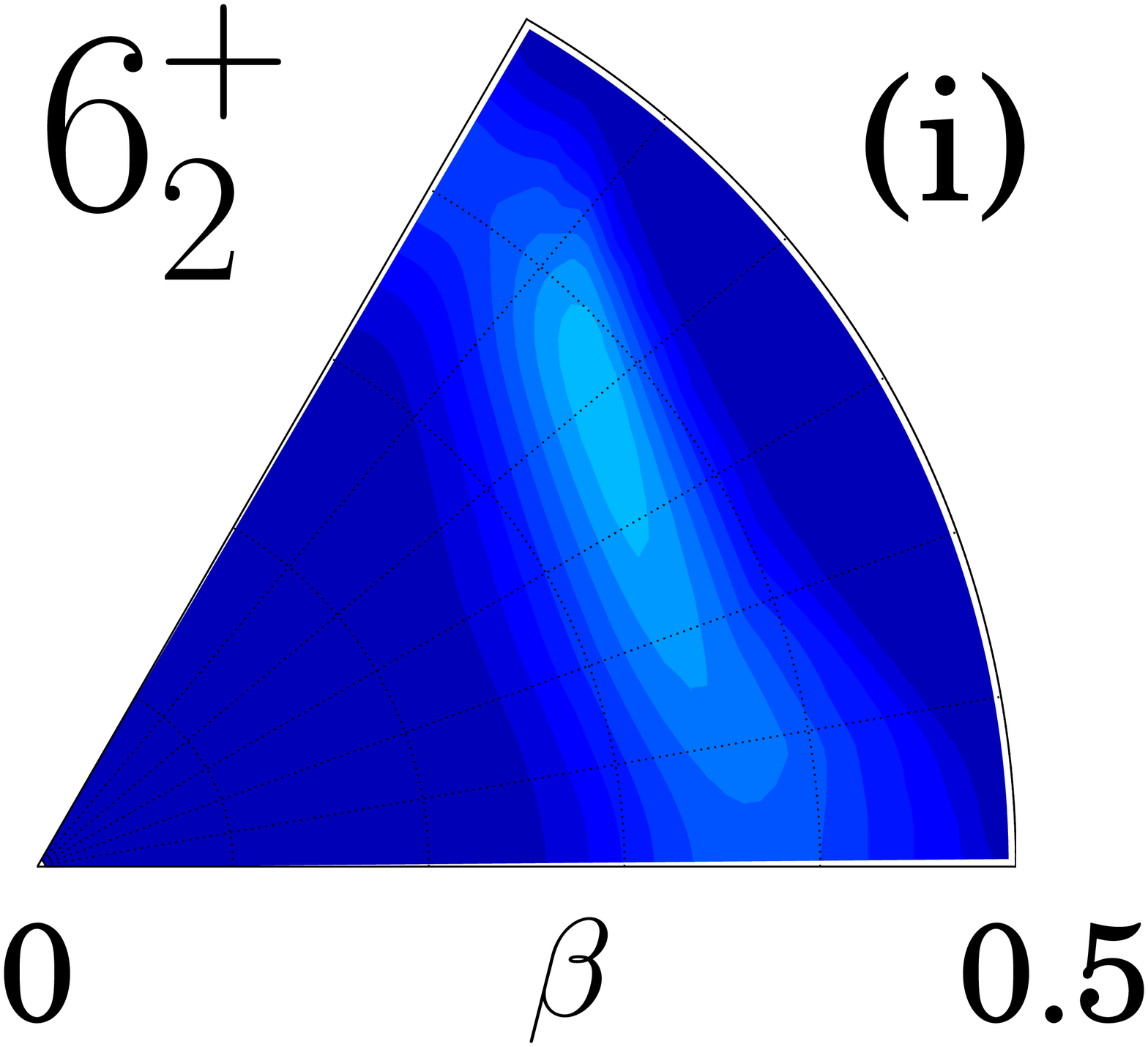} &
\end{tabular}
\caption{\label{fig:28Si-wave}
(Color online) Vibrational wave functions squared
$\beta^4|\Phi_{\alpha I}(\beta,\gamma)|^2$ calculated for $^{28}$Si.}
\end{figure}

In addition to the ground rotational band, we show
two rotational bands built on  
the  $2_2^+$ and $0_2^+$ states in Fig.~\ref{fig:28Si-energy}.
The values of $B(E2)$ and spectroscopic quadrupole moments are 
summarized in Tables~\ref{table:28Si-E2} and \ref{table:28Si-Q},
and the vibrational wave functions squared in $(\bg)$ plane are shown in
Fig.~\ref{fig:28Si-wave}.
The vibrational wave functions and the quadrupole moments indicate the
oblate deformation of the yrast rotational band.
In Table~\ref{table:28Si-Kprob}, 
the $K$ mixing in each state is listed.
The yrast rotational band is consistent with $K=0$ oblate band.
It is seen that the spectroscopic quadrupole moment significantly increases
with the increase of the angular momentum.
The values of $Q(4_1^+)$ and $Q(6_1^+)$ are larger than those estimated from 
the rotational collective model using the calculated $Q(2_1^+)$ value.
(In the case of the $6_1^+$ state, 16.8 $e$ fm$^2$ is the estimated
value for $K=0$ rotational band.)
This indicates that the $\beta$ deformation of the intrinsic state grows
as the angular momentum increases in the yrast band
as seen in Fig.~\ref{fig:28Si-wave}.

We then compare the theoretical results for the ground band 
with the experimental data.
The theoretical results overestimate both of the excitation energies
and $B(E2)$ values for $4_1^+$ and $6_1^+$ states.
The experimental values of $B(E2)$ do not follow the trend of the collective model.
This indicates that the spin alignment of the single-particle states plays a role
in the high angular momentum states, and this reduces the $B(E2)$ values.
In the quadrupole collective Hamiltonian which we derive in the present calculation,
the moments of inertia are evaluated at zero angular momentum,
and the alignment of the single-particle states is not taken into account.
To include such effect by using the cranked mean field approach is an interesting extension of the model, but is beyond the scope of this paper.

The vibrational wave functions of the theoretical $0_2^+$, $2_3^+$, and $4_3^+$
states have nodes in the $\beta$ direction, and this
can be interpreted as the $\beta$-vibrational behavior on top of
the oblate yrast states.
The fact that in-band transitions in the $\beta$-vibrational band
$B(E2;2_3^+\rightarrow 0_2^+)$ is comparable with those of the
ground rotational band $B(E2;2_1^+\rightarrow 0_1^+)$,
and the positive sign of the quadrupole moment in $2_3^+$ and $4_3^+$
supports this interpretation.
However, they contain $K$ mixing and shape mixing with the prolate region in the
vibrational wave functions.

The excited band composed of $2_2^+$, $3_1^+$, $4_2^+$, $5_1^+$, and $6_2^+$
states are also found in the calculation.
The main component of the vibrational wave functions 
is $K=2$, and lies in the triaxially deformed region. 
However, there is no experimental information
corresponding to 
this triaxial band.
Since the prolate local minimum is not found in the collective
potential, the prolate rotational band does not obtained
in the energy spectrum.

\begin{table}[htbp]
\caption{\label{table:28Si-E2}
The values of $B(E2)$ for $^{28}$Si listed in units of $e^2$fm$^4$. 
See also caption in Table~\ref{table:24Mg-E2}.
Experimental data are taken from Ref.~\cite{ENSDF}.}
\begin{tabular}{ccccc} \hline \hline
 & EXP & CHB+LQRPA & neutron & proton \\ \hline
$2_1^+ \rightarrow 0_1^+$ & 66.7 &  41.558 &  10.497 &  10.354 \\
$4_1^+ \rightarrow 2_1^+$ & 69.7 &  77.816 &  19.635 &  19.394 \\
$6_1^+ \rightarrow 4_1^+$ & 50.0 &  98.715 &  24.893 &  24.608 \\ \hline
$3_1^+ \rightarrow 2_2^+$ & - &  65.783 &  16.517 &  16.422 \\
$4_2^+ \rightarrow 3_1^+$ & - &  41.612 &  10.515 &  10.366 \\
$5_1^+ \rightarrow 4_2^+$ & - &  27.185 &   6.840 &   6.782 \\
$6_2^+ \rightarrow 5_1^+$ & - &  30.972 &   7.815 &   7.719 \\
$4_2^+ \rightarrow 2_2^+$ & - &  59.403 &  14.918 &  14.828 \\
$5_1^+ \rightarrow 3_1^+$ & - &  62.666 &  15.768 &  15.633 \\
$6_2^+ \rightarrow 4_2^+$ & - &  98.322 &  24.709 &  24.538 \\ \hline
$2_3^+ \rightarrow 0_2^+$ & 27.8\footnote{$2_5^+ \rightarrow 0_2^+$ transition} &  44.347 &  11.187 &  11.053 \\
$4_3^+ \rightarrow 2_3^+$ & - &  55.761 &  14.108 &  13.884 \\
$4_1^+ \rightarrow 2_2^+$ & - &   0.450 &   0.118 &   0.111 \\
$2_2^+ \rightarrow 2_1^+$ & - &  30.834 &   7.592 &   7.748 \\
$2_2^+ \rightarrow 0_1^+$ & - &   2.618 &   0.617 &   0.667 \\
$3_1^+ \rightarrow 4_1^+$ & - &   8.124 &   1.970 &   2.051 \\
$3_1^+ \rightarrow 2_1^+$ & - &   5.910 &   1.387 &   1.508 \\
$4_2^+ \rightarrow 4_1^+$ & - &  19.928 &   4.864 &   5.022 \\
$6_1^+ \rightarrow 4_2^+$ & - &   0.432 &   0.115 &   0.106 \\
$4_2^+ \rightarrow 2_1^+$ & - &   2.135 &   0.508 &   0.542 \\
$5_1^+ \rightarrow 6_1^+$ & - &   5.897 &   1.408 &   1.497 \\
$5_1^+ \rightarrow 4_1^+$ & - &   6.951 &   1.638 &   1.772 \\
$6_2^+ \rightarrow 6_1^+$ & - &  15.752 &   3.813 &   3.980 \\
$6_2^+ \rightarrow 4_1^+$ & - &   2.836 &   0.683 &   0.718 \\ \hline
$0_2^+ \rightarrow 2_1^+$ & - &  56.831 &  14.217 &  14.205 \\
$2_3^+ \rightarrow 0_1^+$ & - &   0.078 &   0.021 &   0.019 \\
$2_3^+ \rightarrow 2_1^+$ & - &   2.571 &   0.623 &   0.649 \\
$2_3^+ \rightarrow 4_1^+$ & - &  16.159 &   4.033 &   4.042 \\ \hline\hline
\end{tabular}
\end{table}

\begin{table}
\caption{ \label{table:28Si-Q}Spectroscopic quadrupole moments for
 $^{28}$Si listed in units of $e$ fm$^2$.
See also captions in Table~\ref{table:24Mg-E2}.
Experimental data are taken from Ref.~\cite{ENSDF}.}
\begin{tabular}{ccccc}\hline\hline
 & EXP & CHB+LQRPA & neutron & proton \\ \hline 
$Q(2_1^+)$ & 16    &  12.0 &  6.05 &  5.98 \\
$Q(4_1^+)$ & -     &  17.8 &  8.99 &  8.90 \\
$Q(6_1^+)$ & -     &  22.2 &  11.2 &  11.1 \\ \hline
$Q(2_2^+)$ & -     & $-$12.6 & $-$6.33 & $-$6.29 \\
$Q(3_1^+)$ & -     &  0    &  0    & 0     \\
$Q(4_2^+)$ & -     & $-$3.10 & $-$1.55 & $-$1.55  \\
$Q(5_1^+)$ & -     &  10.3 &  5.16 & 5.12  \\
$Q(6_2^+)$ & -     &  1.61 &  0.83 & 0.80 \\ \hline
$Q(2_3^+)$ & -     &  9.69 &  4.89 & 4.83  \\
$Q(4_3^+)$ & -     &  7.35 &  3.69 & 3.67  \\ \hline\hline
\end{tabular}
\end{table}

\begin{table}[htbp]
\caption{\label{table:28Si-Kprob}
Probability amplitude of quantum states of $^{28}$Si for each
 $K_2$-component, where $K_2$ is the projection of the angular momentum 
onto the symmetric axis at $\gamma=60^\circ$.}
\begin{tabular}{ccccc} \hline\hline
$I^\pi_\alpha$ & $K_2=0$ & $K_2=2$ & $K_2=4$ & $K_2=6$\\ \hline
$0_1^+$ & 1.00 & -    & -    & -   \\
$2_1^+$ & 0.97 & 0.03 & -    & -   \\
$4_1^+$ & 0.98 & 0.02 & 0.00 & -   \\
$6_1^+$ & 0.99 & 0.01 & 0.00 & 0.00\\ \hline
$2_2^+$ & 0.11 & 0.89 &    - & -   \\
$3_1^+$ &  -   & 1.00 &   -  & -   \\
$4_2^+$ & 0.16 & 0.71 & 0.14 & -   \\
$5_1^+$ &  -   & 0.97 & 0.03 & -   \\
$6_2^+$ & 0.15 & 0.71 & 0.09 & 0.04\\ \hline
$0_2^+$ & 1.00 & -    & -    & -   \\
$2_3^+$ & 0.94 & 0.06 & -    & -   \\
$4_3^+$ & 0.67 & 0.20 & 0.14 & -   \\ \hline\hline
\end{tabular}
\end{table}

\subsubsection{$^{26}$Mg}

\begin{figure}[htbp]
\includegraphics[width=80mm]{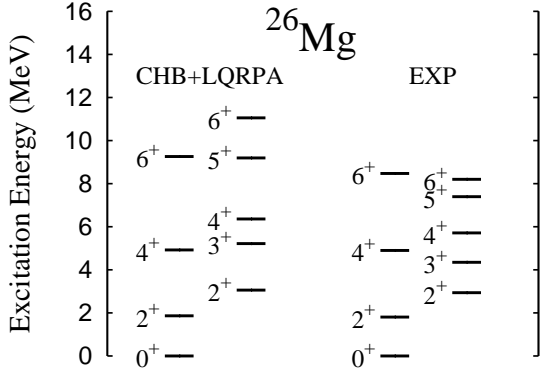}
\caption{\label{fig:26Mg-energy}
Excitation spectra calculated for $^{26}$Mg
by means of the CHB+LQRPA method and experimental data~\cite{ENSDF}.}
\end{figure}

By adding two neutrons to the prolately deformed $^{24}$Mg,
the character of collective dynamics in the low-lying levels drastically
changes from that of $^{24}$Mg.
Figure \ref{fig:26Mg-energy} compares the 
theoretical and experimental low-lying energy levels in $^{26}$Mg.
In addition to the yrast band ($0_1^+, 2_1^+, 4_1^+$ and $6_1^+$), we show a side band
composed of $2_2^+, 3_1^+, 4_2^+, 5_1^+$, and $6_2^+$ states,
which are connected by relatively large $B(E2)$ values (see Table~\ref{table:26Mg-E2}).

\begin{figure}[htbp]
\begin{tabular}{cc}
\includegraphics[width=25mm]{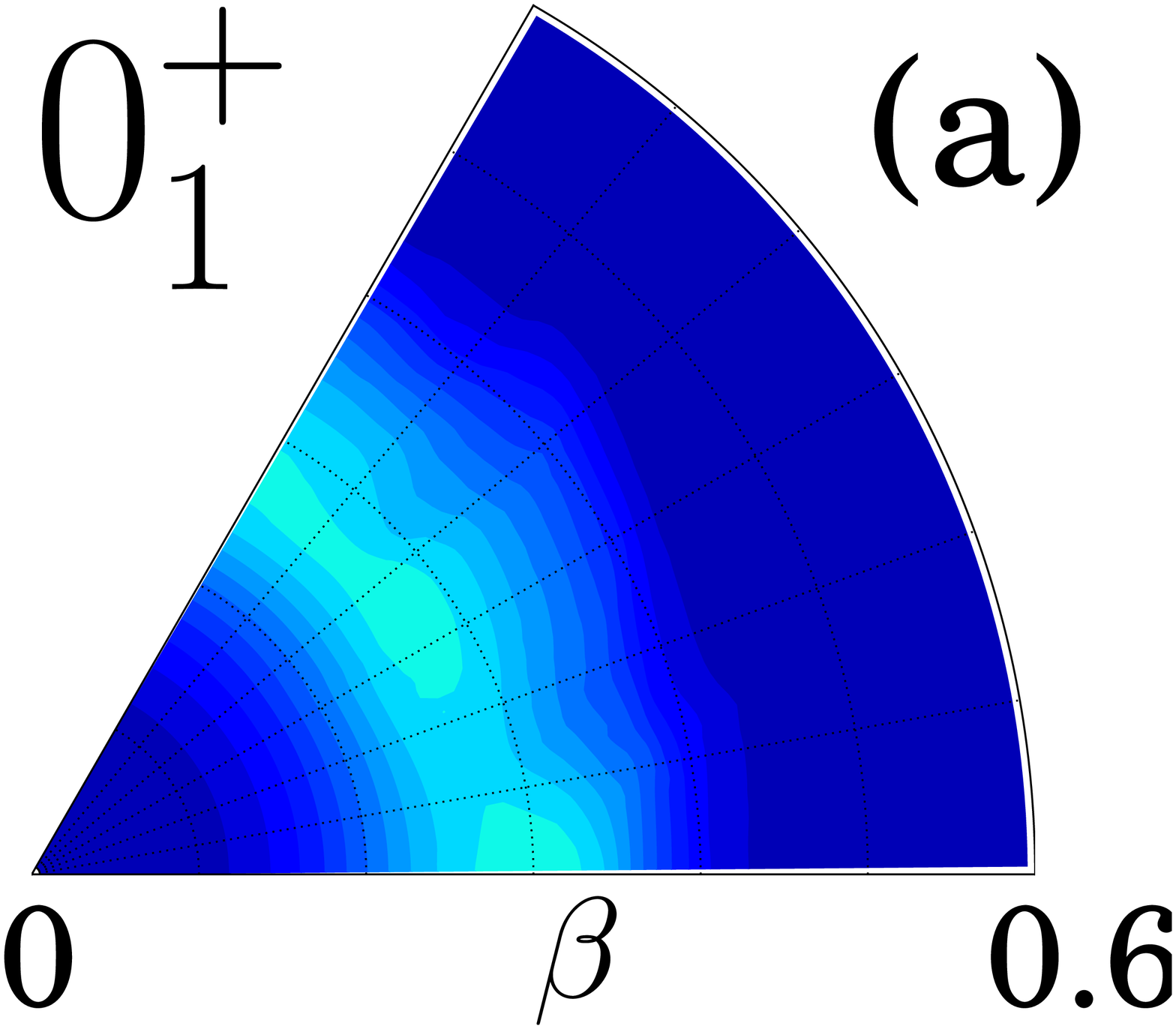} &
\includegraphics[width=25mm]{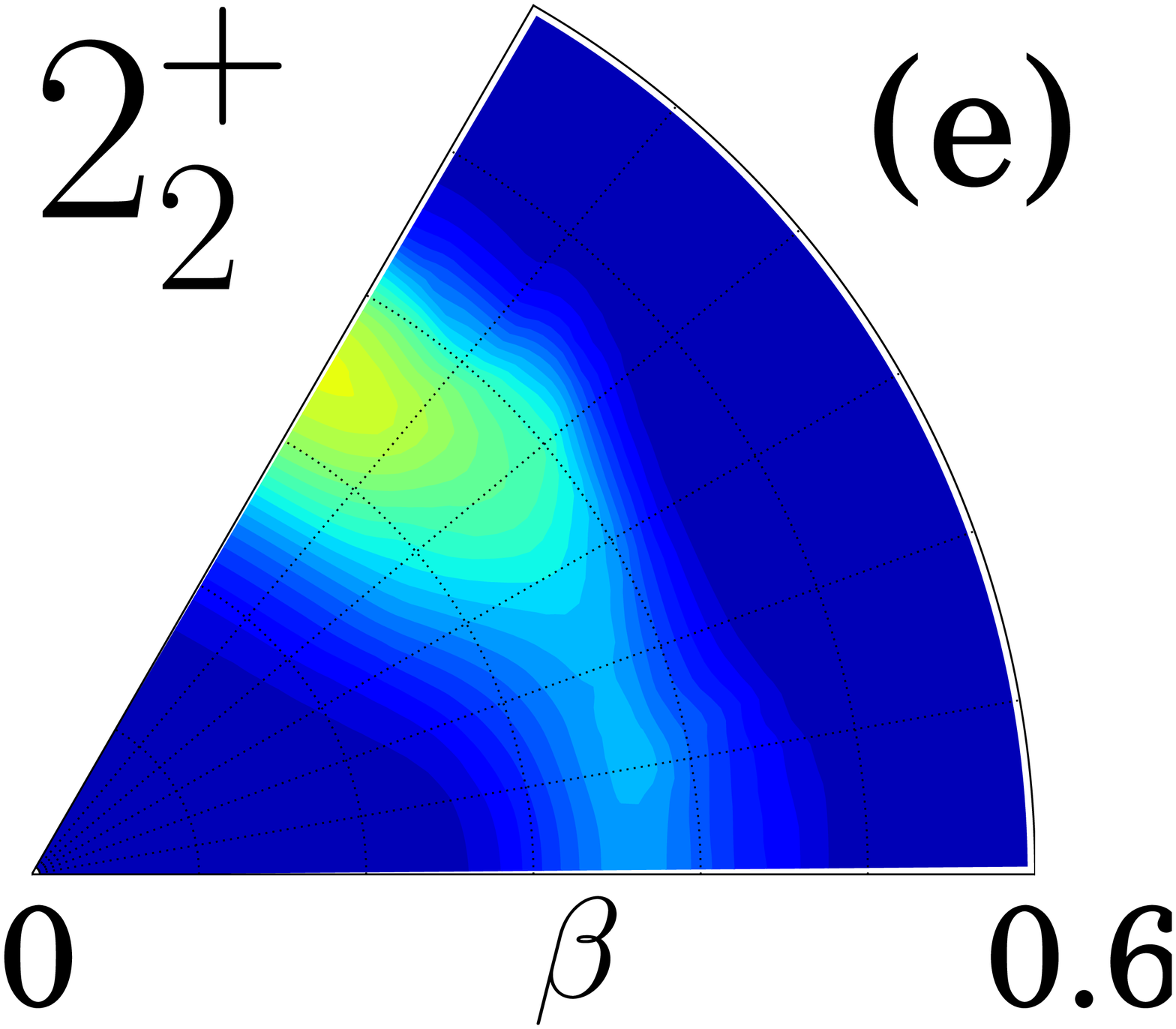} \\
\includegraphics[width=25mm]{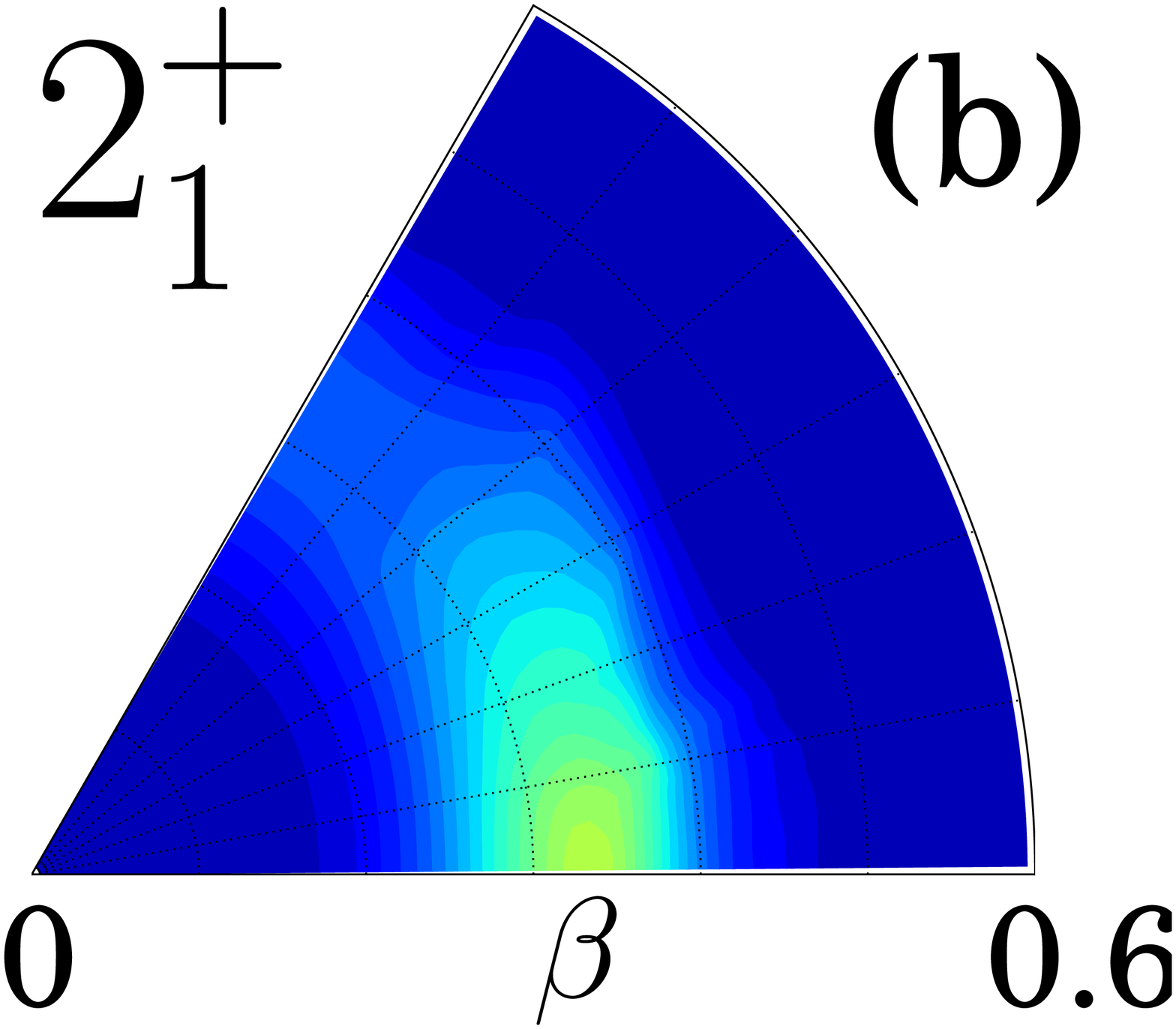} &
\includegraphics[width=25mm]{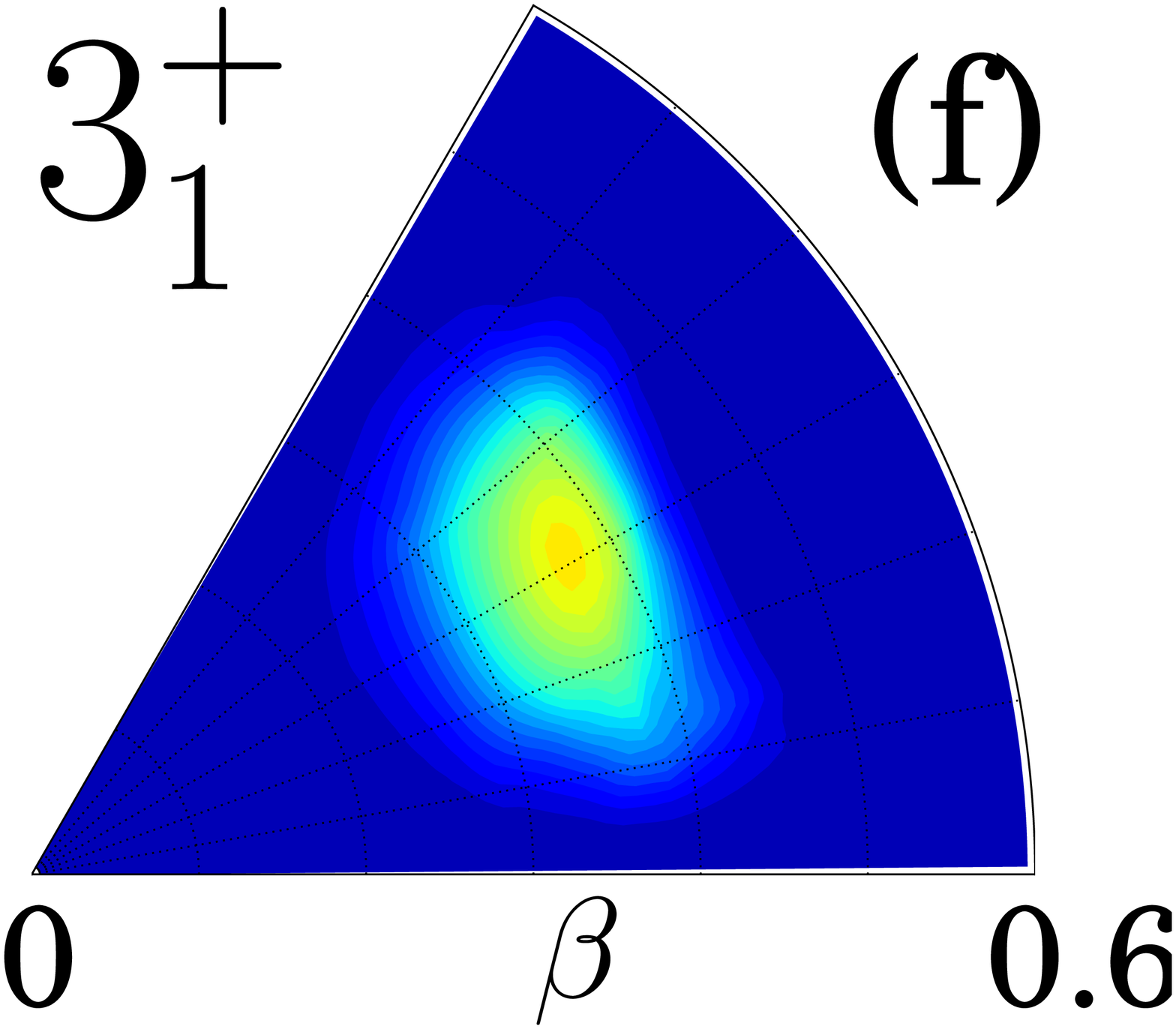} \\
\includegraphics[width=25mm]{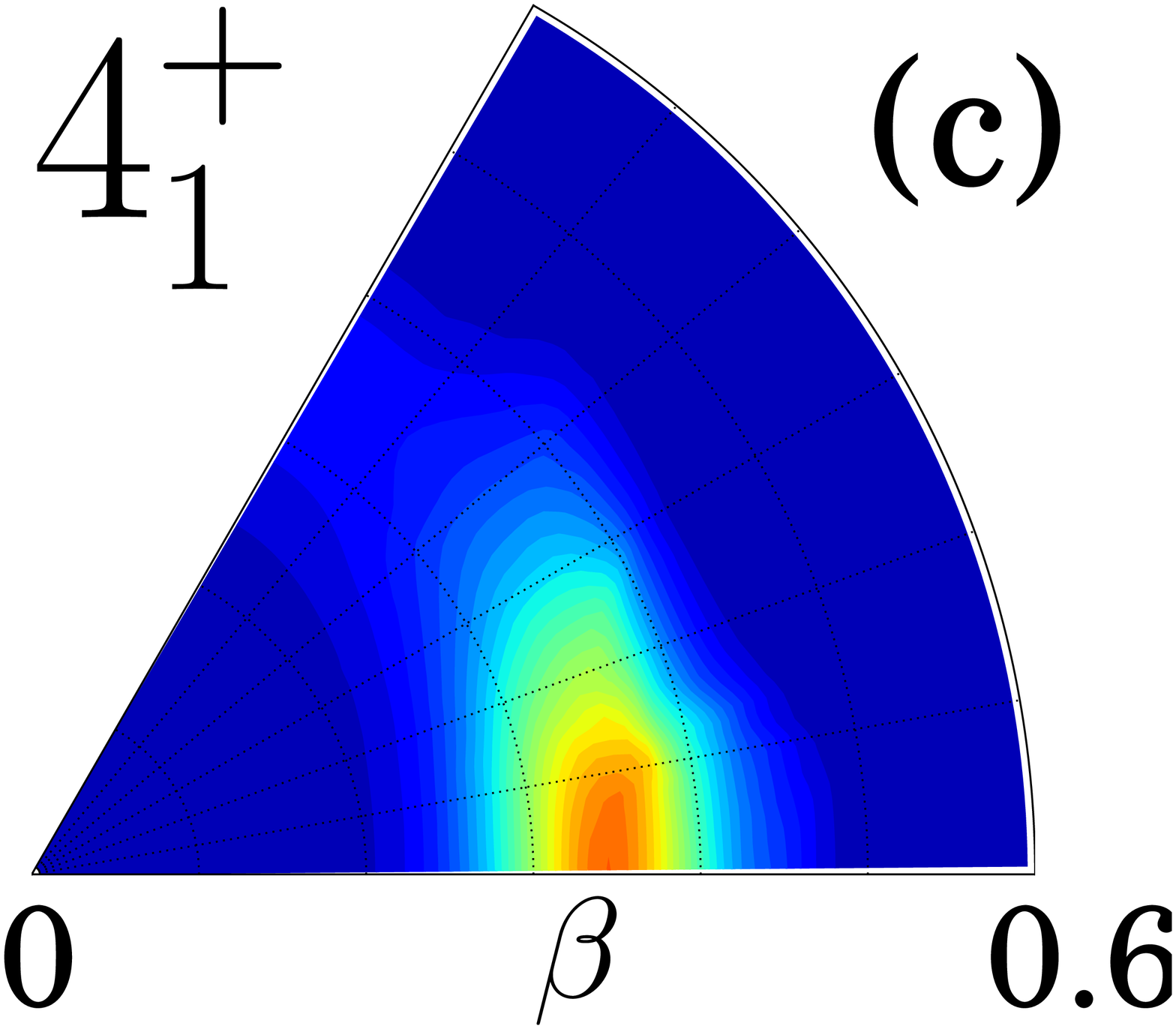} &
\includegraphics[width=25mm]{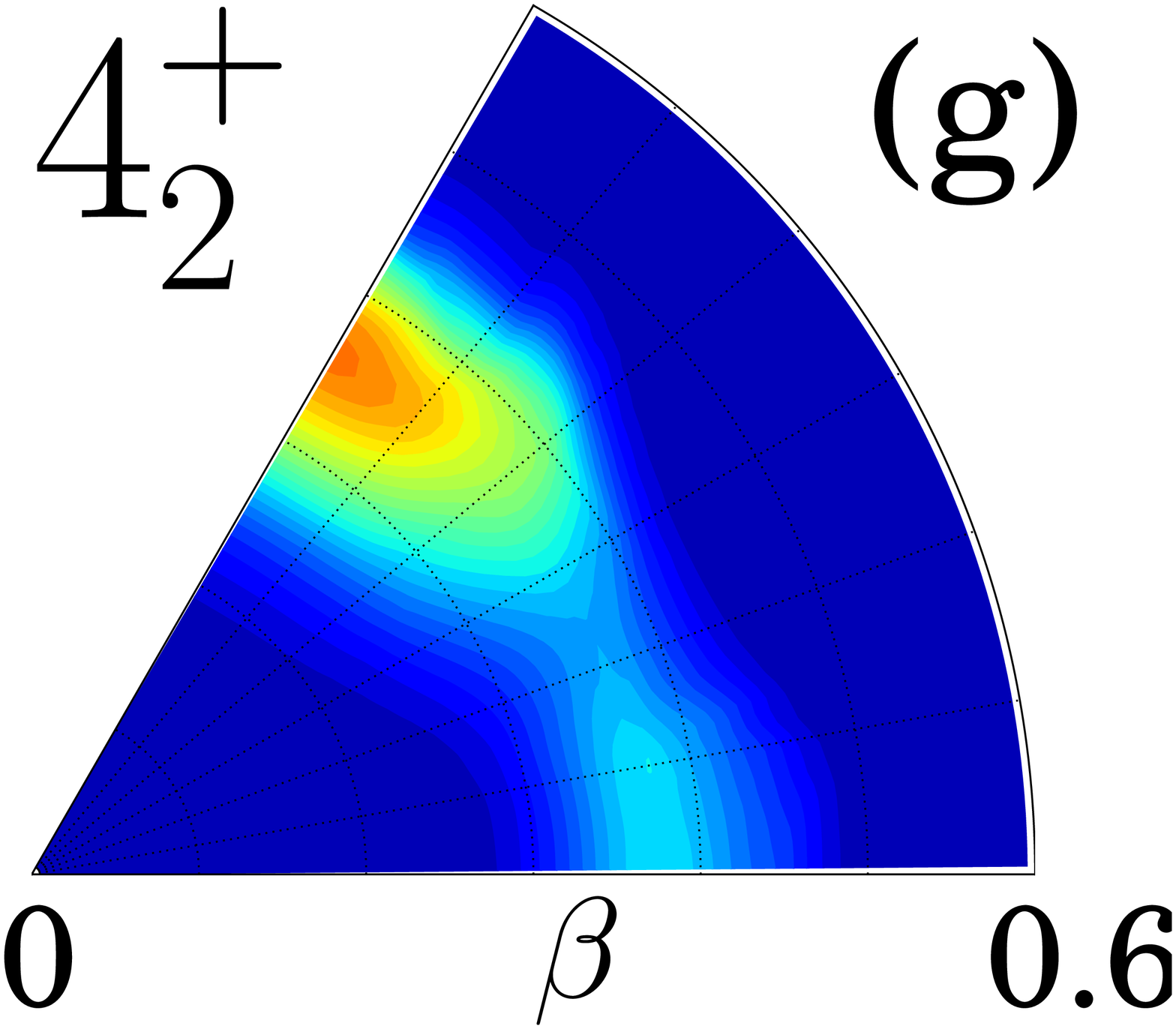} \\
\includegraphics[width=25mm]{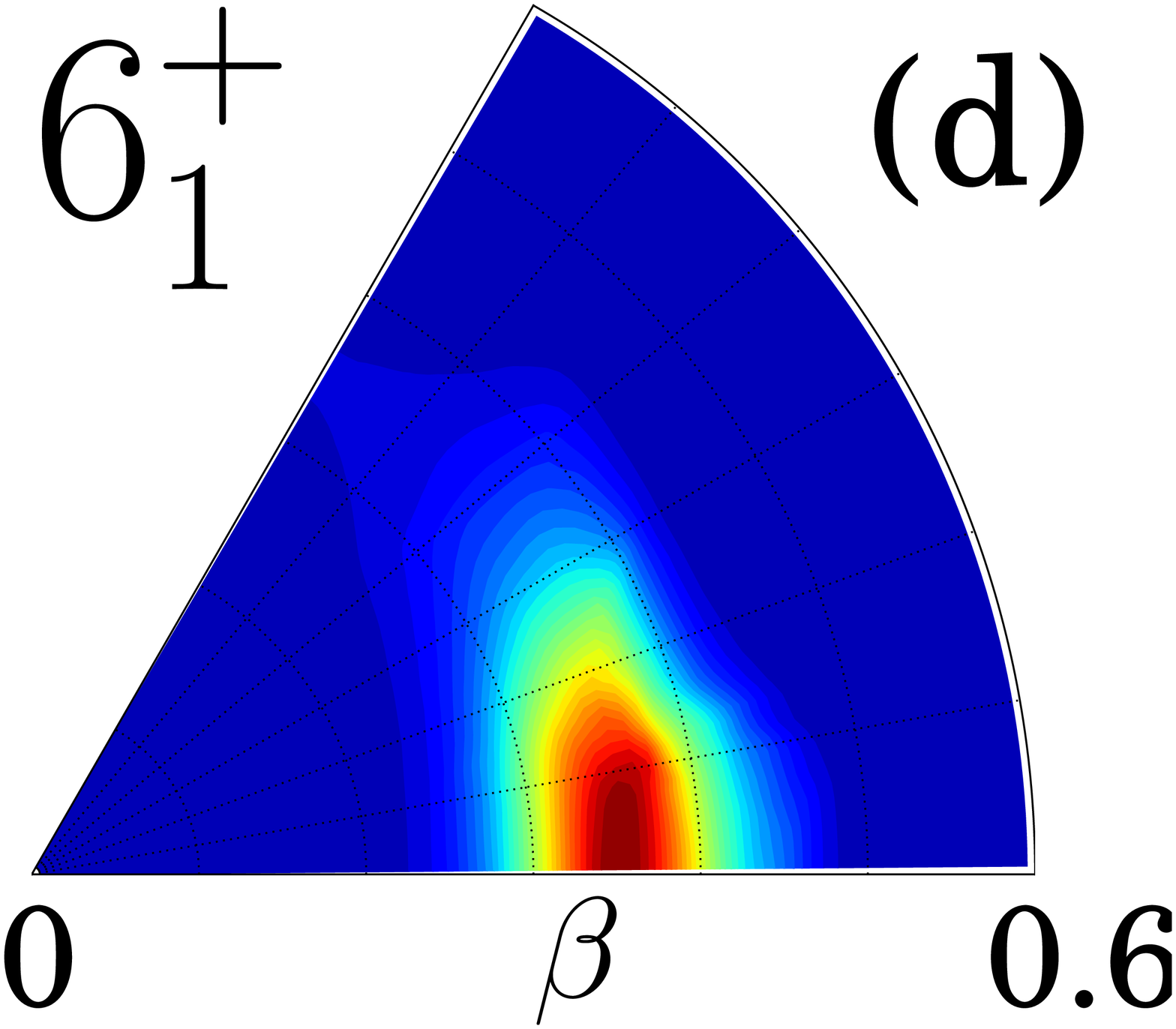} &
\includegraphics[width=25mm]{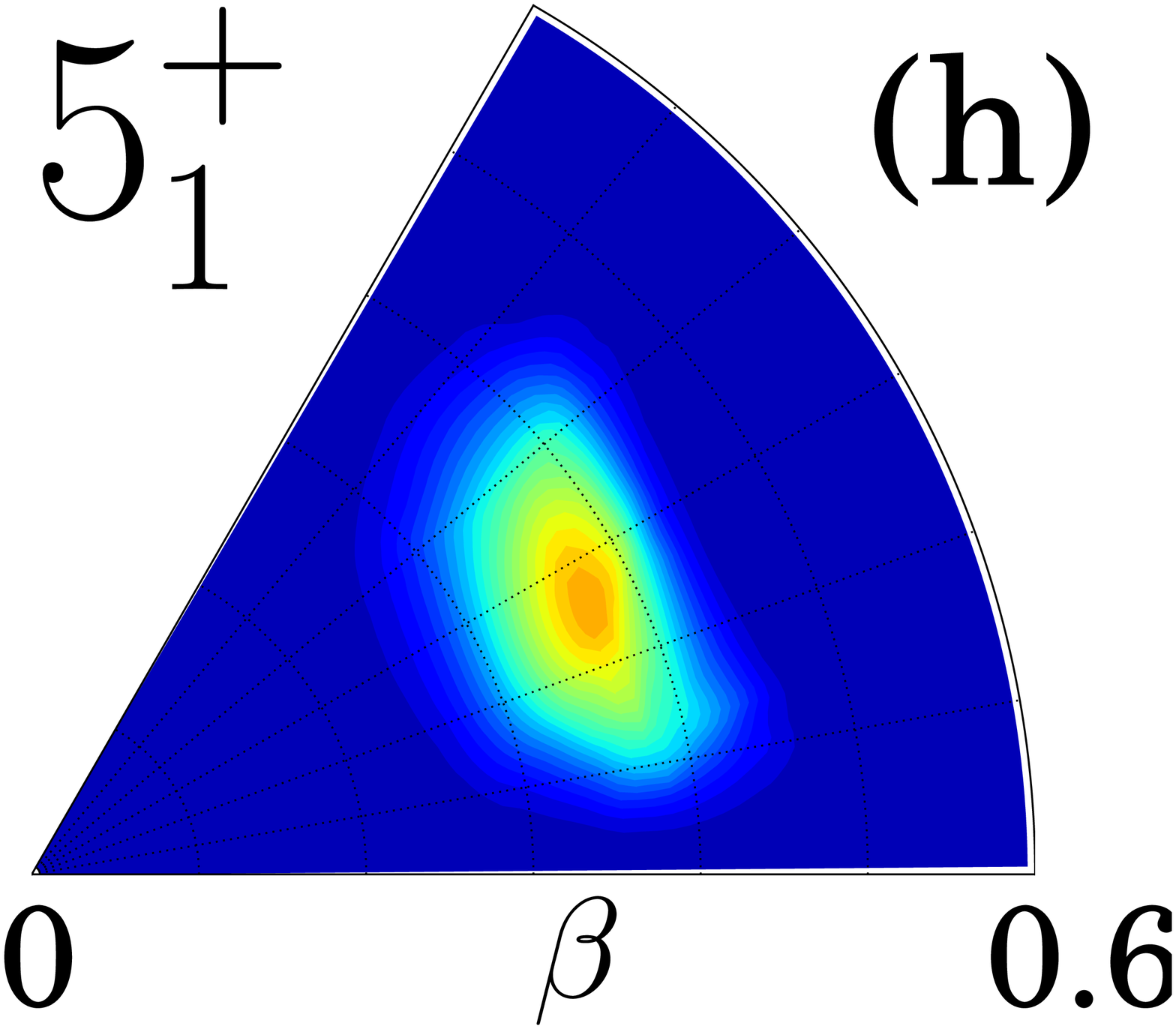} \\ &
\includegraphics[width=25mm]{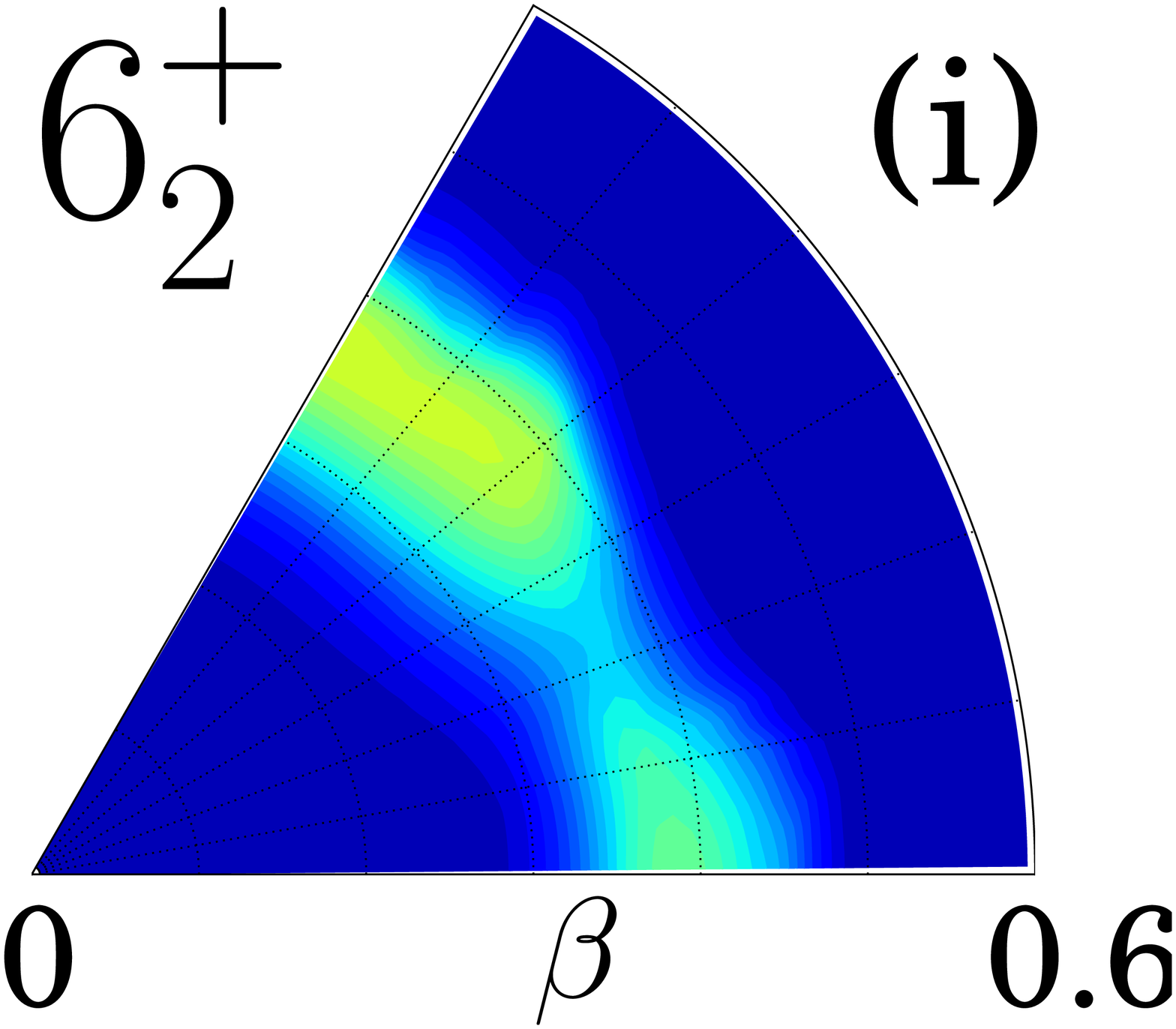}
\end{tabular}
\caption{\label{fig:26Mg-wave}
(Color online) Vibrational wave functions squared
$\beta^4|\Phi_{\alpha I}(\beta,\gamma)|^2$ calculated for $^{26}$Mg.
}
\end{figure}

The vibrational wave functions squared are shown in
Fig.~\ref{fig:26Mg-wave}.
A significant difference from $^{24}$Mg is seen in the
deformation property of the vibrational wave functions of the ground $0_1^+$ state.
In contrast to the well-developed prolate structure of the $^{24}$Mg yrast band,
the $0_1^+$ state of $^{26}$Mg spreads over the triaxial region from
oblate to prolate ones, although the shallow potential minimum is located at the oblate region.
This indicates the very $\gamma$-soft character of the ground state.
The members of the yrast band tend to localize around the prolate shape 
as the angular momentum increases.
As for the excitation energies of the yrast band,  
the $E_x(4^+_1)/E_x(2^+_1)$ ratio is 2.64 in theoretical calculation, 
which explains the experimental value 2.71 very well.
As seen in Table~\ref{table:26Mg-Q}, 
the spectroscopic quadrupole moments of the yrast band are consistent with the prolate
deformation, but the absolute values are relatively smaller than 
those of $^{24}$Mg.
Moreover, $Q(6_1^+)$ is almost twice of $Q(2_1^+)$, and this is much
larger than the value of $K=0$ rotational collective model,
indicating that the development of the prolate deformation with
increase of the angular momentum.
This feature is also seen in the ratio
$B(E2;4_1^+\rightarrow 2_1^+)/B(E2;2_1^+\rightarrow 0_1^+)$.
The theoretical value of the ratio is 1.7, which is larger than
the collective model value 1.43.
However, the experimental value of the ratio 1.05 again indicates
the effect of the single-particle alignment.

The quantum states in the side band distribute widely in the $\gamma$ direction.
This character is remarkably different from that of $^{24}$Mg,
where all the members of the $K=2$ side band are localized in 
the triaxial region close to the prolate local minimum (Fig.~\ref{fig:24Mg-wave}).
In particular, an oblate character develops in 
even angular momentum states of the side band, and the $4_2^+$ and $6_2^+$ states form the
two-peak structure in the oblate and prolate region.
This two-peak structure 
indicates the $\gamma$-soft character of the collective potential
as discussed in Ref.~\cite{PTP.123.129}.

\begin{table}[htbp]
\caption{\label{table:26Mg-E2} 
The values of $B(E2)$ for $^{26}$Mg
listed in units of $e^2$fm$^4$.  
See also caption in Table~\ref{table:24Mg-E2}.
Experimental values are taken from Refs.~\cite{ENSDF,ZPA324_187},
and the assignment of the levels are given in Ref.~\cite{ZPA324_187}.}
\begin{tabular}{ccccc} \hline\hline
                         & EXP & CHB+LQRPA &  neutron & proton \\ \hline

$2_1^+ \rightarrow 0_1^+$ & 61.3 &  52.870 &  11.131 &  13.953 \\
$4_1^+ \rightarrow 2_1^+$ & 64.1 &  90.302 &  18.379 &  24.070 \\
$6_1^+ \rightarrow 4_1^+$ & - & 112.596 &  23.016 &  29.975 \\ \hline
$3_1^+ \rightarrow 2_2^+$ & 41.2 &  74.890 &  13.707 &  20.568 \\
$4_2^+ \rightarrow 3_1^+$ & 37.0\footnote{Data taken from Ref.~\cite{ZPA324_187}.} &  20.437 &   1.366 &   6.887 \\
$5_1^+ \rightarrow 4_2^+$ & - &  32.073 &   4.387 &   9.470 \\
$6_2^+ \rightarrow 5_1^+$ & - &  19.052 &   1.654 &   6.156 \\
$4_2^+ \rightarrow 2_2^+$ & 23.8$^a$ &  58.789 &  16.303 &  14.180 \\
$5_1^+ \rightarrow 3_1^+$ & - &  70.150 &  14.724 &  18.530 \\
$6_2^+ \rightarrow 4_2^+$ & - &  88.693 &  23.078 &  21.876 \\ \hline
$4_1^+ \rightarrow 2_2^+$ & 16.0$^a$ &   2.073 &   0.131 &   0.704 \\
$2_2^+ \rightarrow 2_1^+$ & 28.4 &  62.940 &  19.792 &  14.486 \\
$2_2^+ \rightarrow 0_1^+$ & 1.60 &   0.765 &   2.040 &   0.011 \\
$3_1^+ \rightarrow 4_1^+$ & - &  18.259 &   5.971 &   4.138 \\
$3_1^+ \rightarrow 2_1^+$ & 0.23$^a$ &   0.456 &   3.234 &   0.022 \\
$4_2^+ \rightarrow 4_1^+$ & - &  28.948 &  12.298 &   5.847 \\
$6_1^+ \rightarrow 4_2^+$ & - &   0.635 &   0.022 &   0.232 \\
$4_2^+ \rightarrow 2_1^+$ & - &   1.536 &   1.755 &   0.148 \\
$6_1^+ \rightarrow 5_1^+$ & - &  13.227 &   4.769 &   2.879 \\
$5_1^+ \rightarrow 4_1^+$ & - &   0.773 &   2.999 &   0.000 \\
$6_2^+ \rightarrow 6_1^+$ & - &  18.167 &   9.774 &   3.238 \\
$6_2^+ \rightarrow 4_1^+$ & - &   1.113 &   1.006 &   0.136 \\ \hline\hline
\end{tabular}
\end{table}

\begin{table}
\caption{\label{table:26Mg-Q}
Spectroscopic quadrupole moment for $^{26}$Mg listed in units of $e$ fm$^2$.
See also caption in Table~\ref{table:24Mg-E2}.
Experimental data are taken from Ref.~\cite{ENSDF}.}
\begin{tabular}{ccccc} \hline\hline
 & EXP & CHB+LQRPA & neutron & proton \\ \hline
$Q(2_1^+)$ & $-$13.5 & $-$9.07 & $-$2.33 & $-$5.27 \\
$Q(4_1^+)$ & -       & $-$14.9 & $-$4.44 & $-$8.48 \\
$Q(6_1^+)$ & -       & $-$17.9 & $-$5.61 & $-$10.1 \\ \hline
$Q(2_2^+)$ & -       &    7.17 &    2.51 &    3.95 \\
$Q(3_1^+)$ & -       &    0    &    0    &    0    \\
$Q(4_2^+)$ & -       &    2.74 &    2.84 &    0.88 \\
$Q(5_1^+)$ & -       & $-$7.96 & $-$1.91 & $-$4.67 \\
$Q(6_2^+)$ & -       & $-$3.25 &    1.60 & $-$2.70 \\ \hline\hline
\end{tabular}
\end{table}

\subsubsection{$^{24}$Ne}

\begin{figure}
\includegraphics[width=80mm]{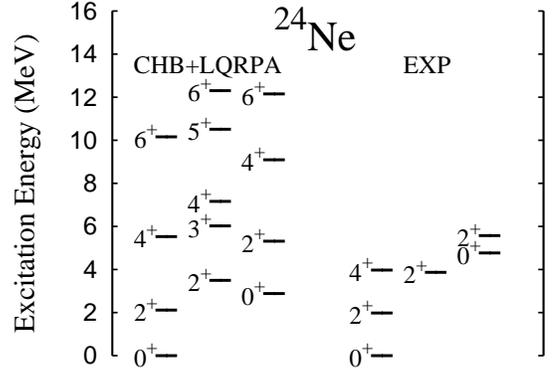}
\caption{\label{fig:24Ne-energy}
Excitation spectra calculated for $^{24}$Ne
by means of the CHB+LQRPA method and experimental data
~\cite{ENSDF}.}
\end{figure}

\begin{figure}[htbp]
\begin{tabular}{ccc}
\includegraphics[width=25mm]{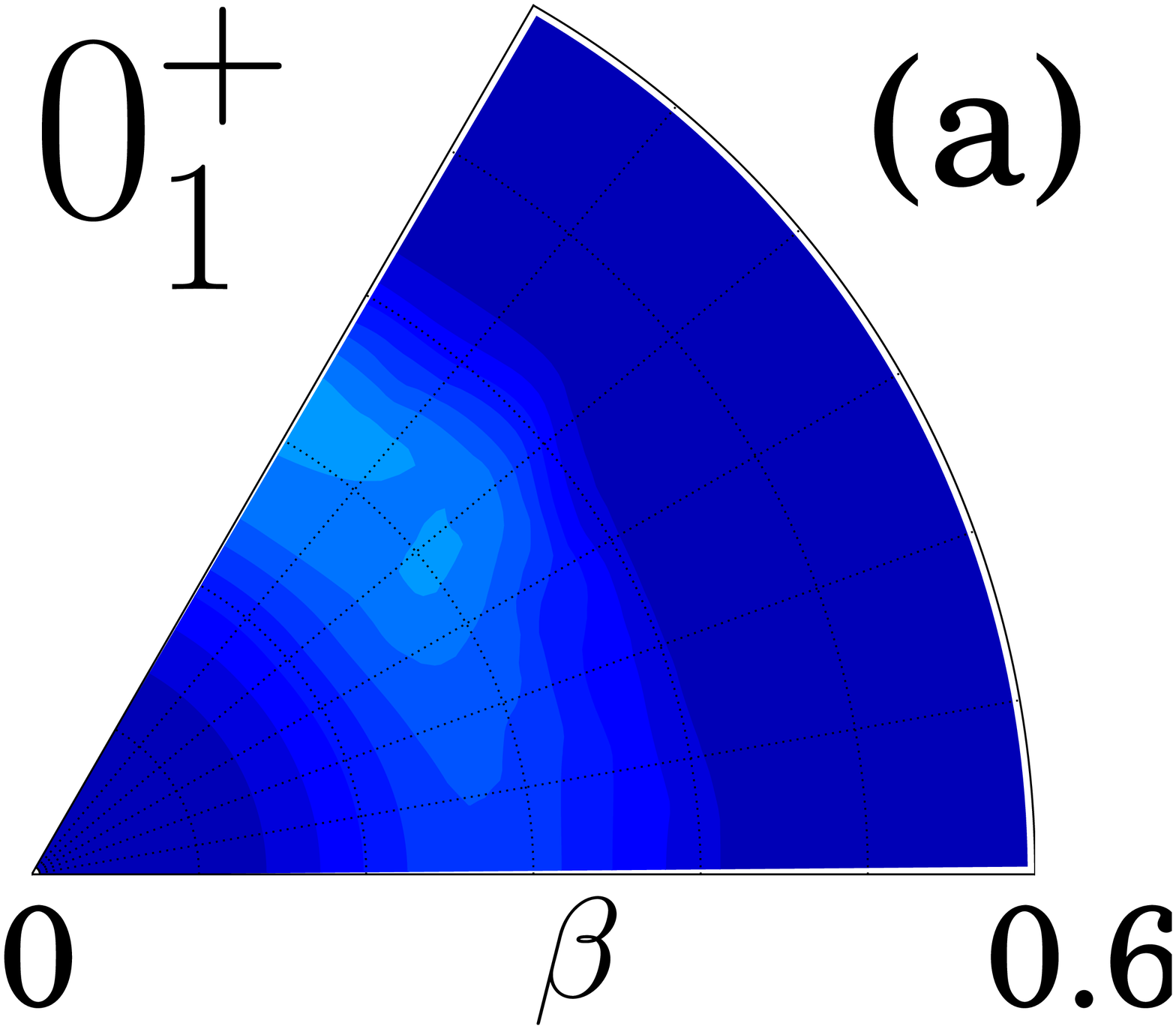} &
\includegraphics[width=25mm]{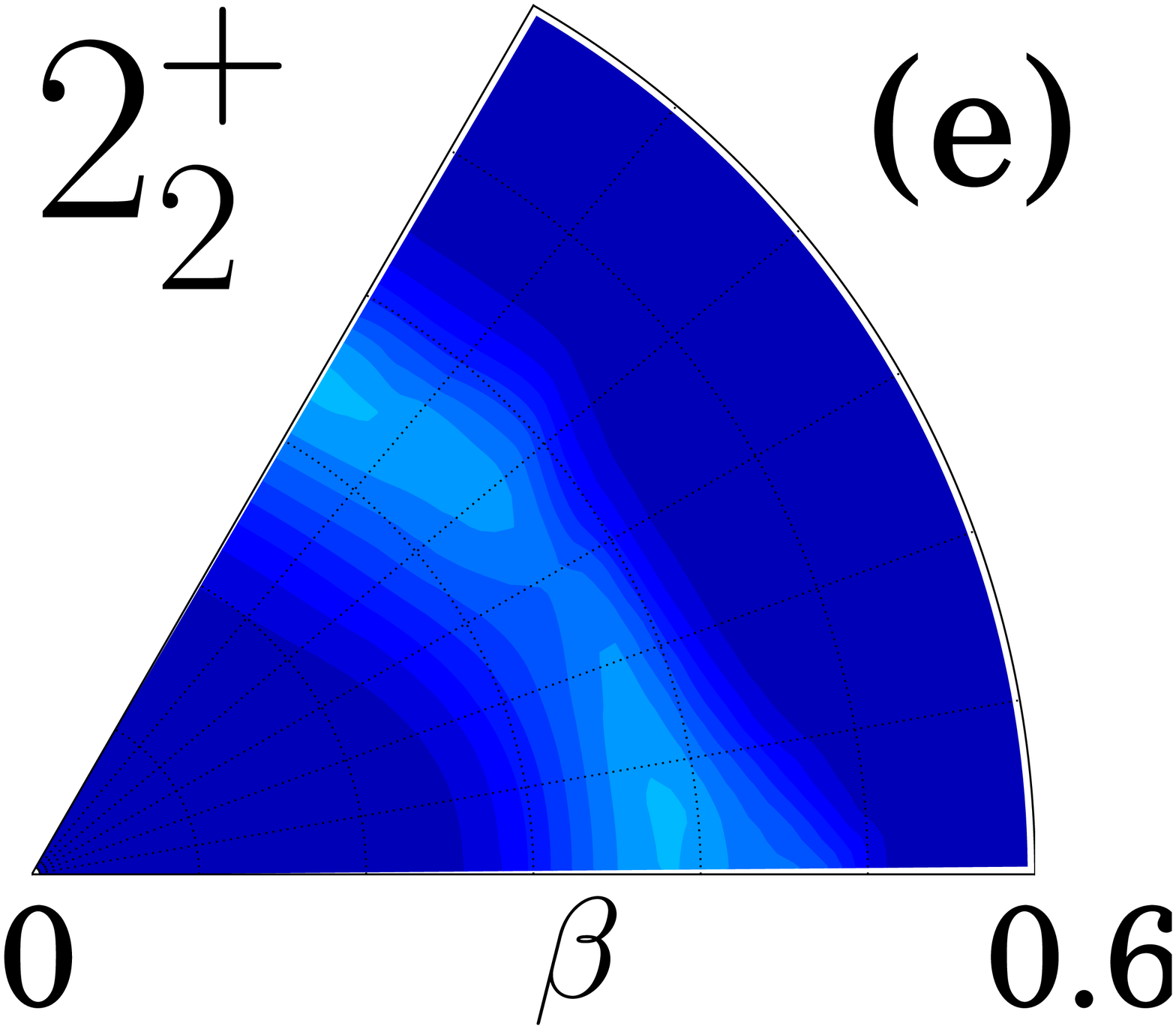} &
\includegraphics[width=25mm]{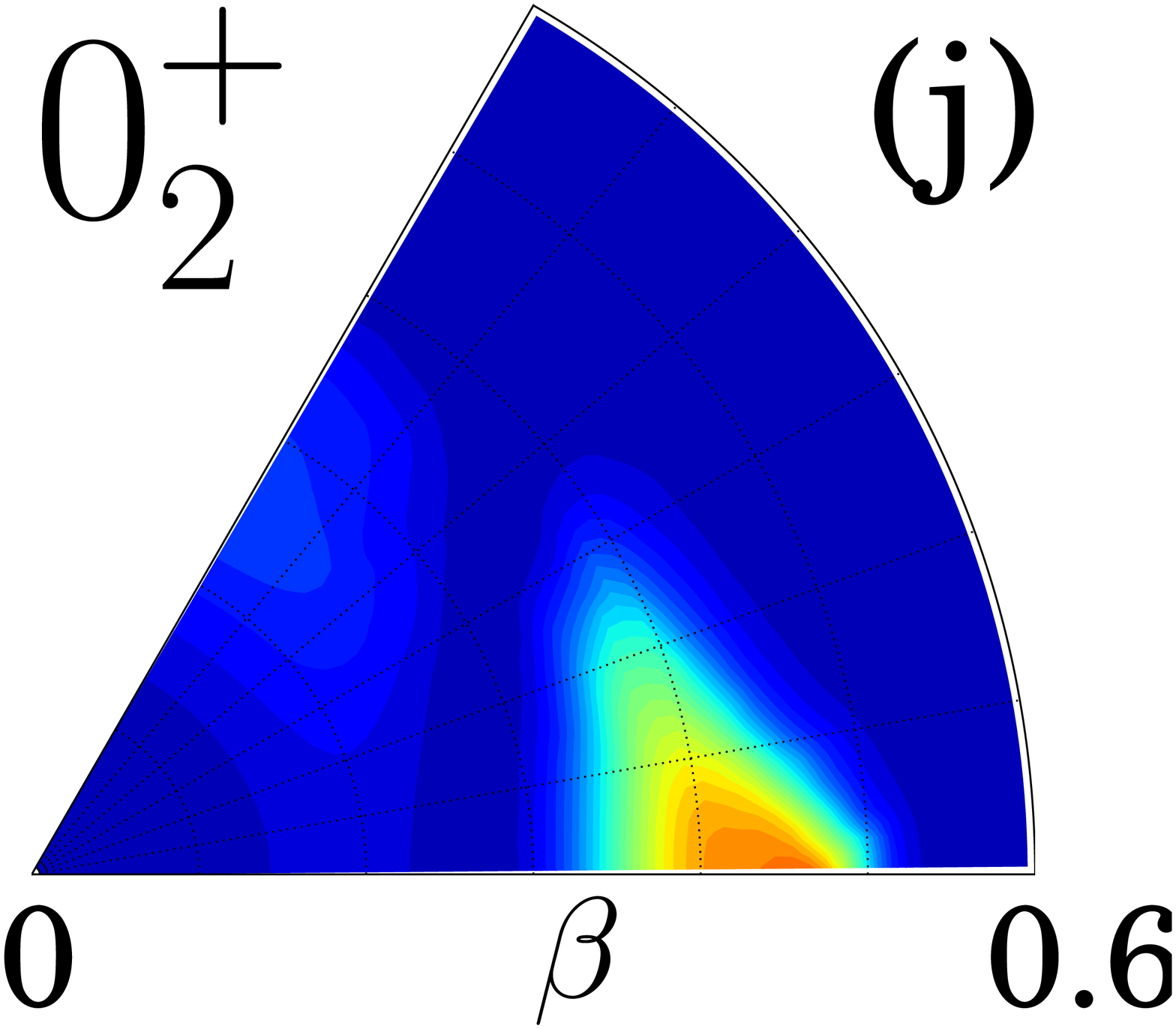} \\
\includegraphics[width=25mm]{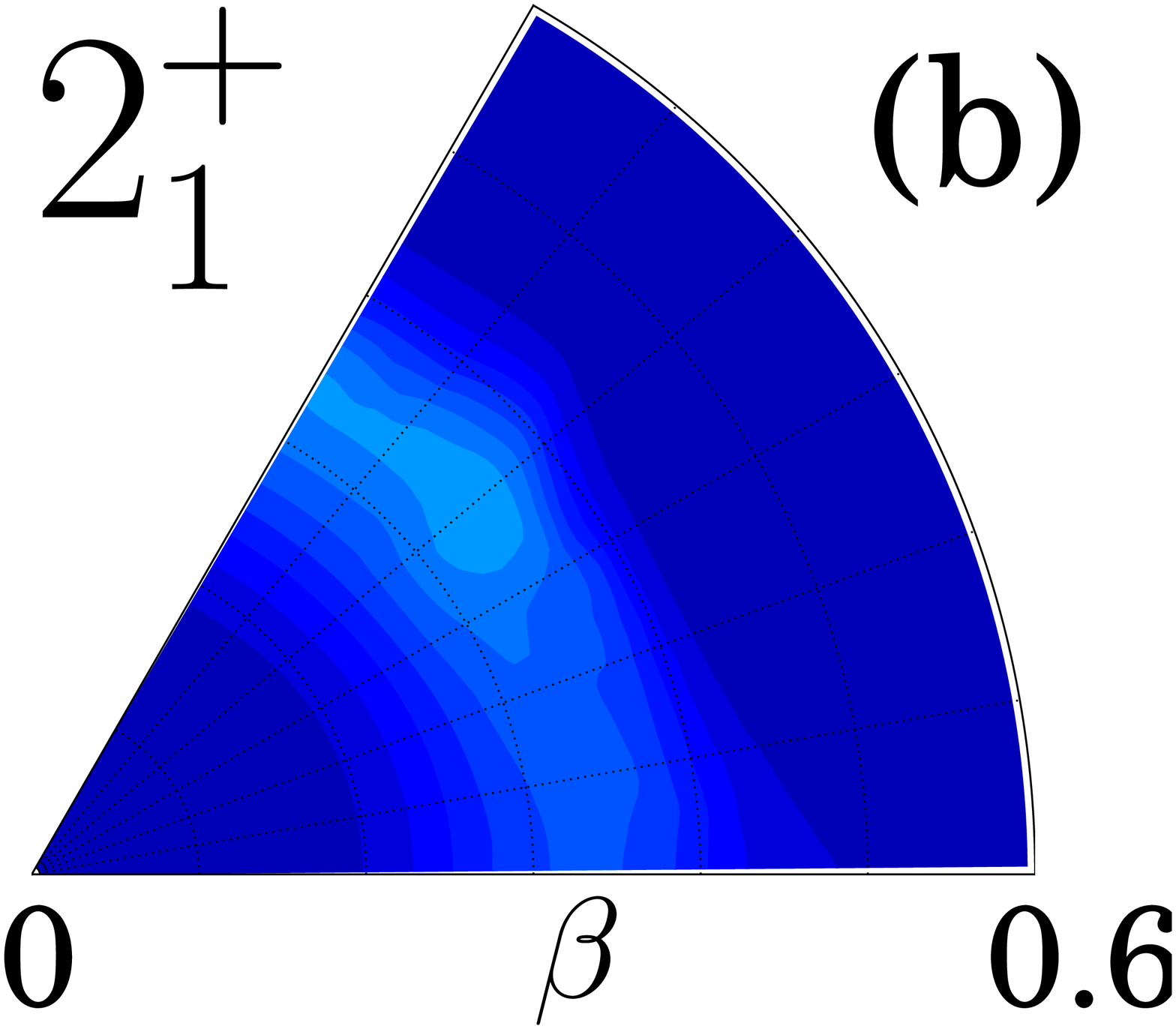} &
\includegraphics[width=25mm]{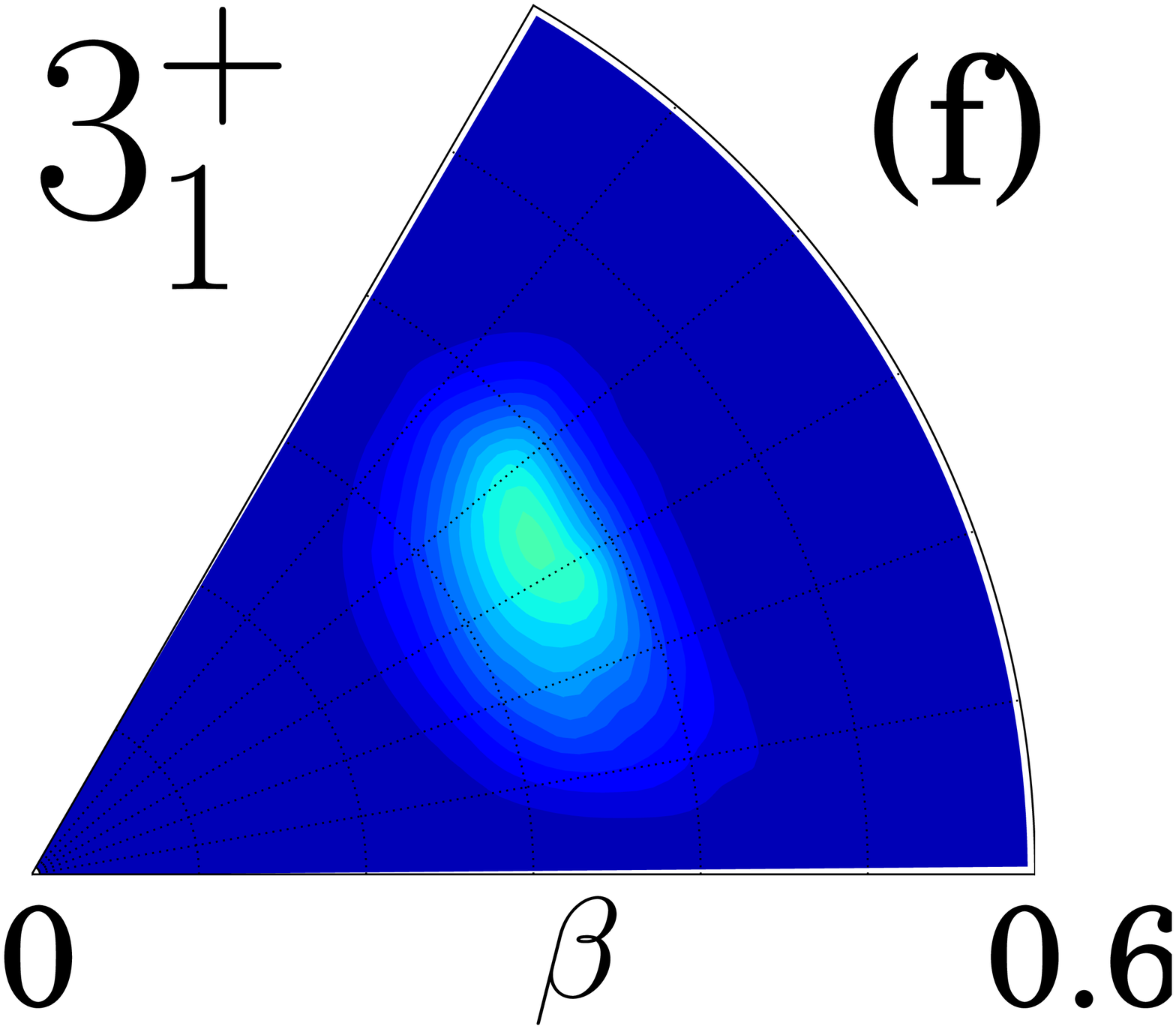} &
\includegraphics[width=25mm]{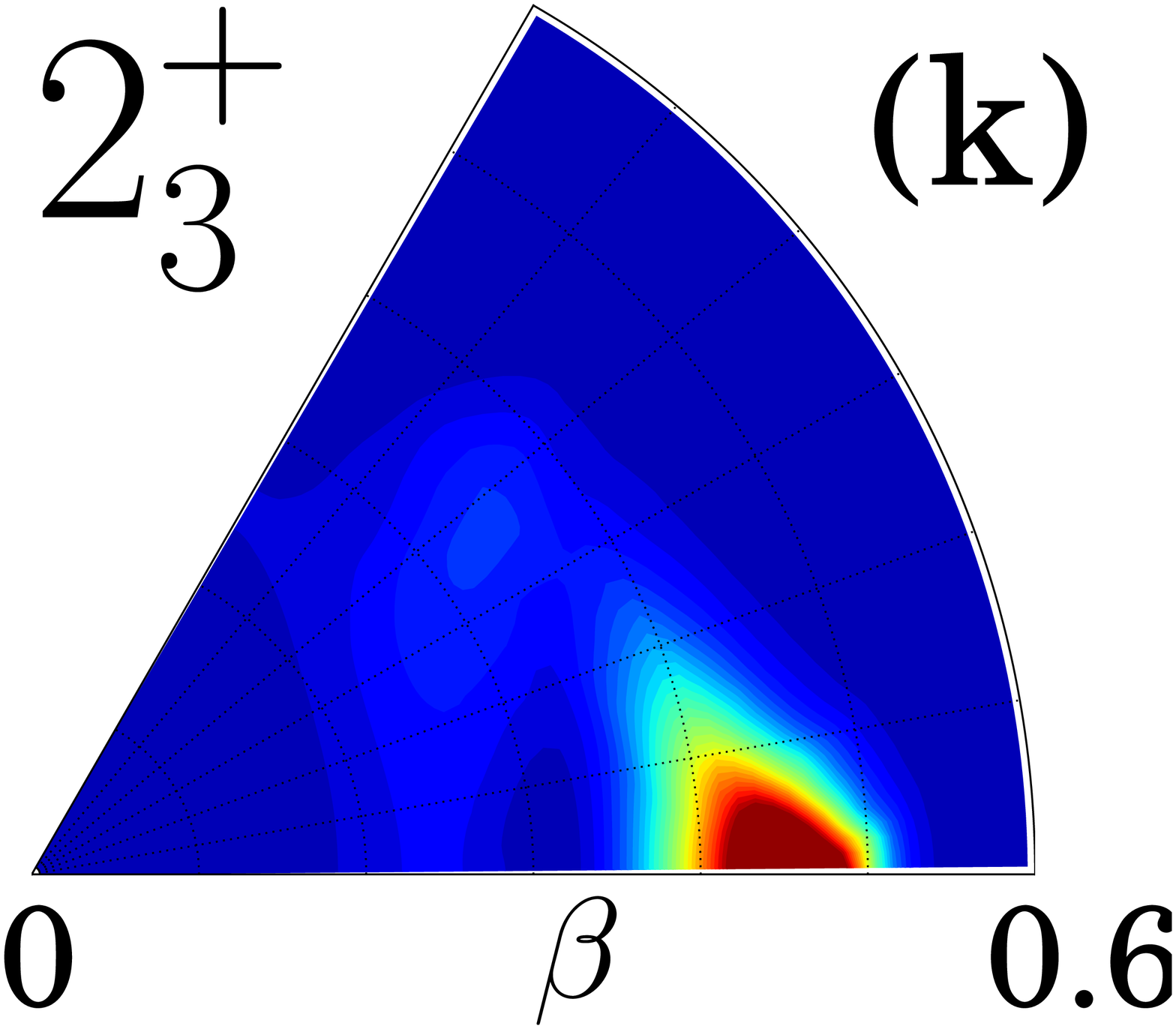} \\
\includegraphics[width=25mm]{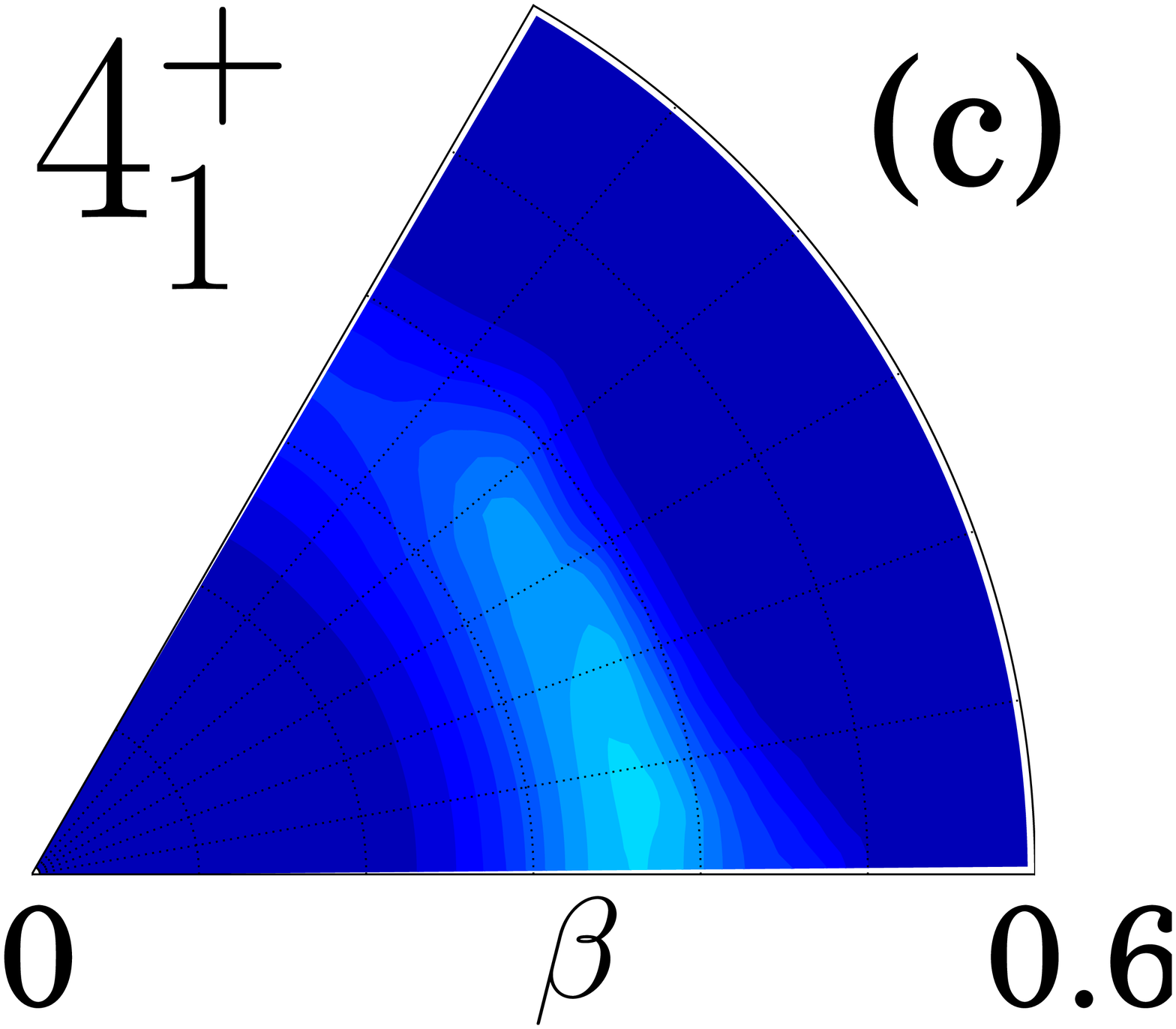} &
\includegraphics[width=25mm]{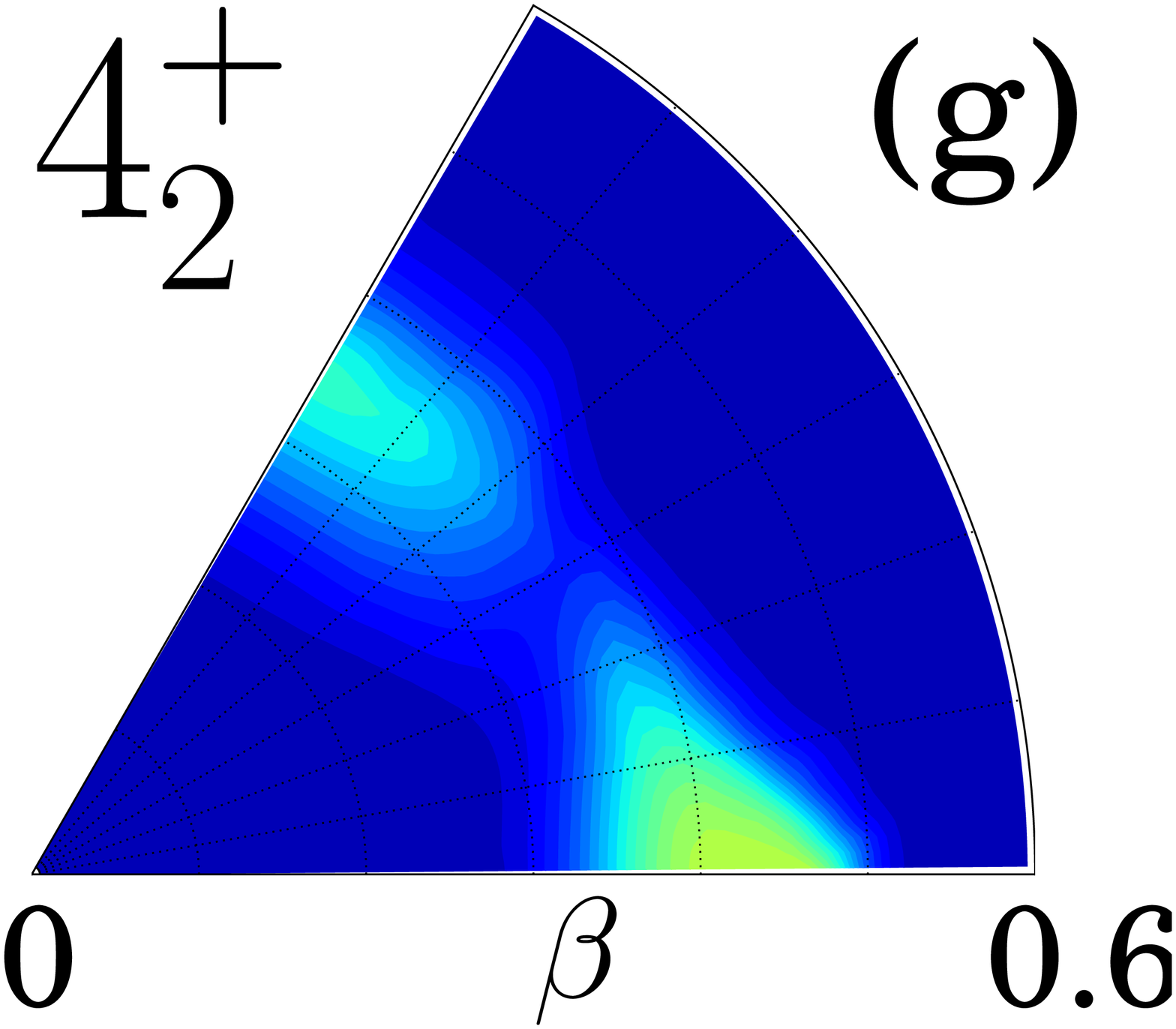} & \\
\includegraphics[width=25mm]{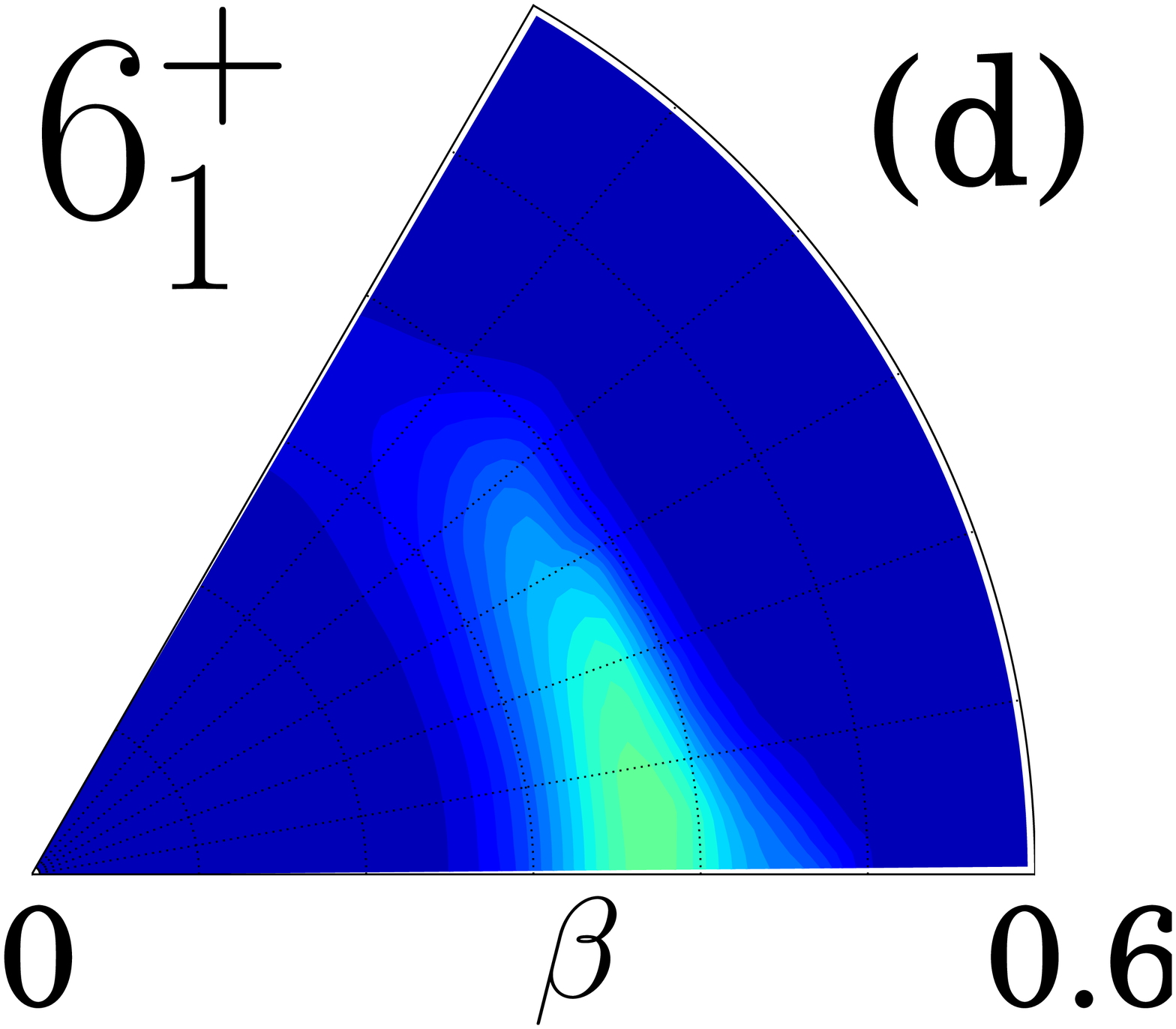} &
\includegraphics[width=25mm]{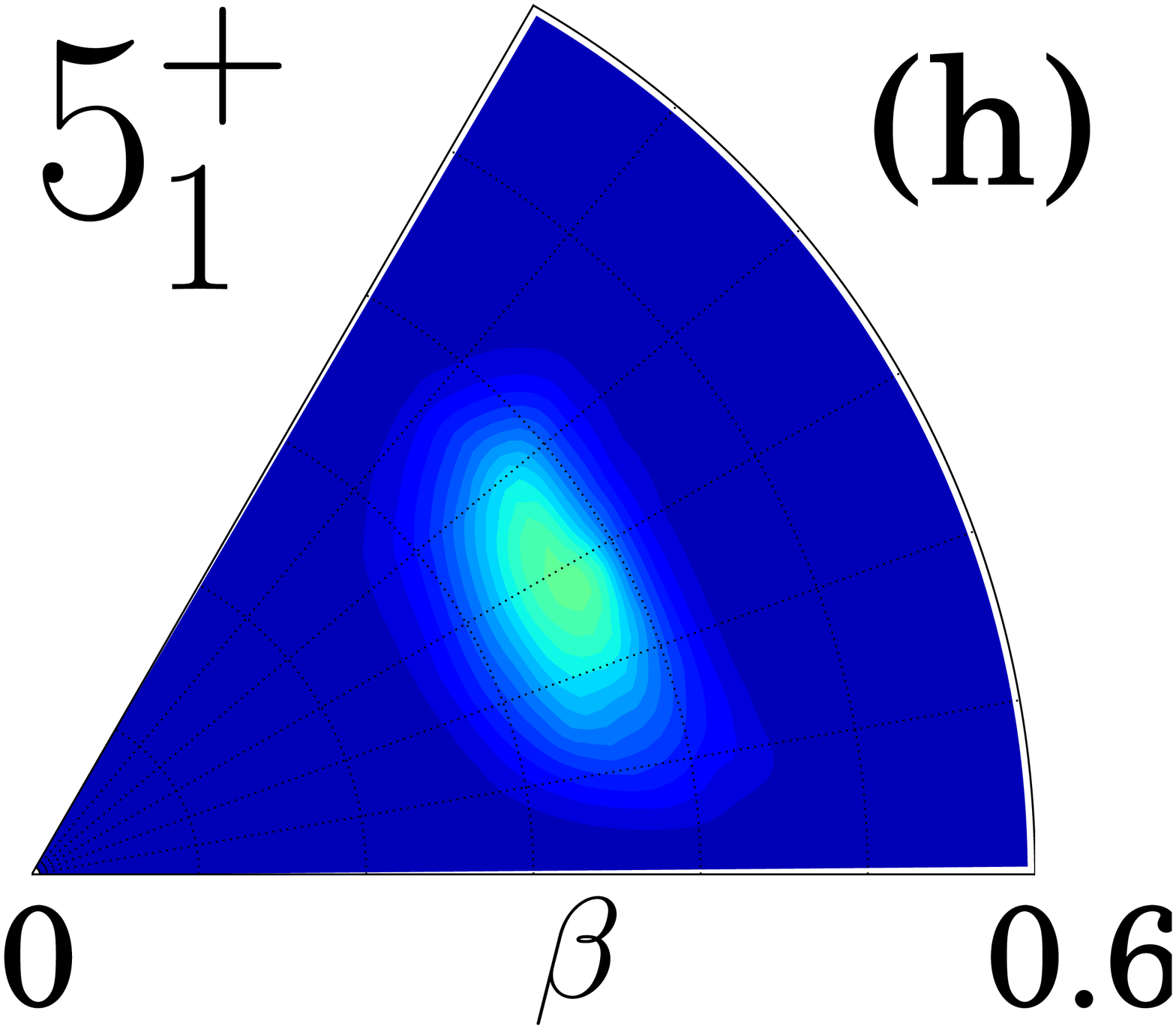} & \\
& 
\includegraphics[width=25mm]{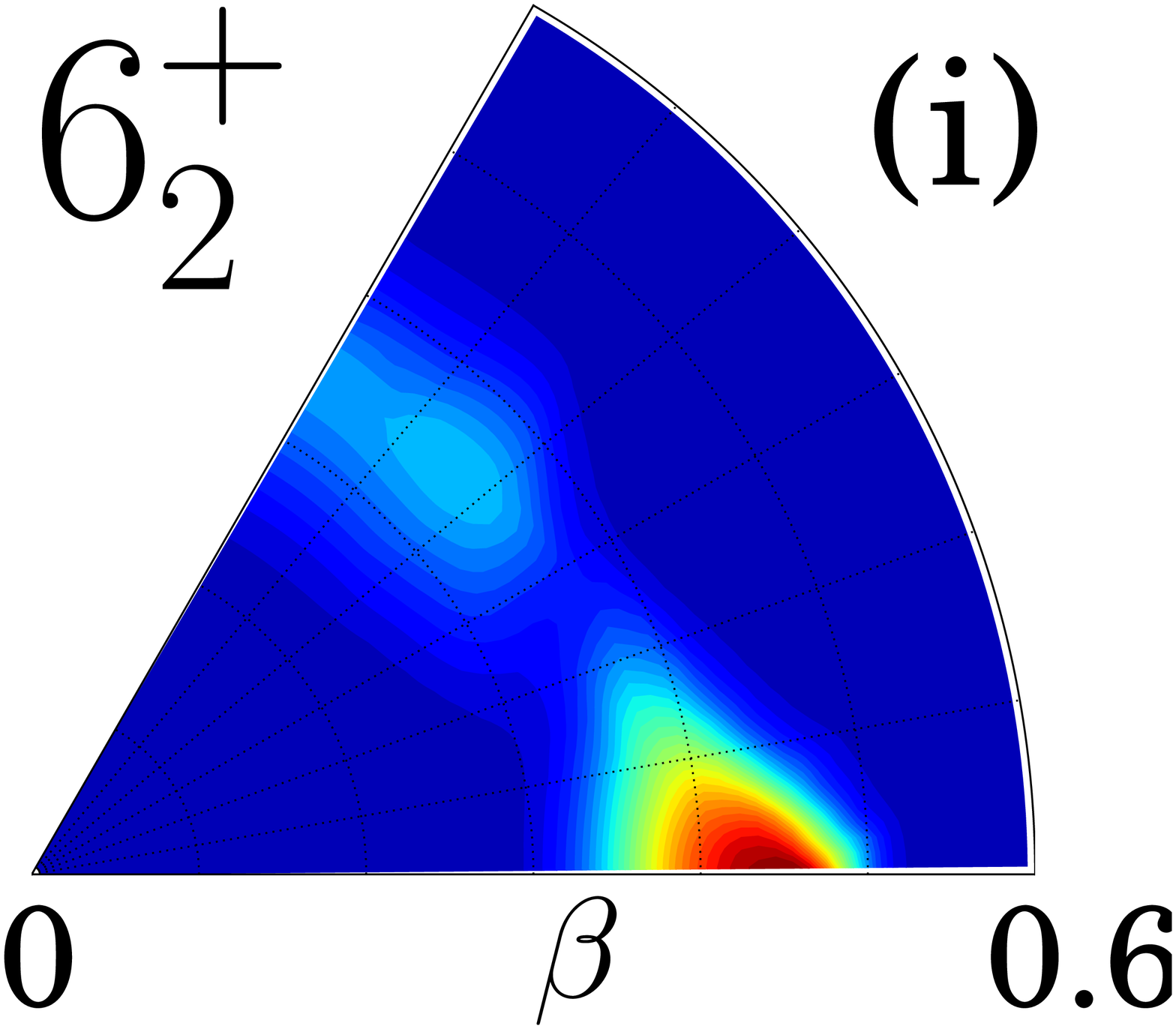} & 
\end{tabular}
\caption{\label{fig:24Ne-wave}
(Color online)
Vibrational wave functions squared
 $\beta^4|\Phi_{\alpha I}(\beta,\gamma)|^2$
calculated for $^{24}$Ne.}
\end{figure}

Figures \ref{fig:24Ne-energy} and \ref{fig:24Ne-wave}
display the energy spectra and vibrational wave functions 
squared for $^{24}$Ne, and the $B(E2)$ and the spectroscopic quadrupole
moments are listed in Tables~\ref{table:24Ne-E2} and \ref{table:24Ne-Q}.
The theoretical yrast band $0_1^+$, $2_1^+$, $4_1^+$, and $6_1^+$
has the $\gamma$-soft character similar to that of $^{26}$Mg;
the vibrational wave function of the $0_1^+$ state has a peak
at the oblate region, but spreads over the triaxial region.
Moreover, the gradual shape change in the yrast states 
with the increase of angular momentum
is also found in the calculation.
The oblate peak of the vibrational wave function in the $0_1^+$ state
shifts to the prolate region in the $4_1^+$ state.

The spectroscopic quadrupole moments of the yrast band are consistent
with the prolate deformation, but the absolute values are 
smaller than $^{26}$Mg.

The two-peak structure in the $2_2^+, 4_2^+$, and $6_2^+$ states 
in the side band is also seen in $^{24}$Ne
showing again the $\gamma$-soft character as well as $^{26}$Mg.

The excitation energy of the $2_1^+$ state and $B(E2;2_1^+\rightarrow
0_1^+)$
are reproduced by the theoretical calculation.
However, the experimental energy spectrum is much more vibrational than
the theoretical one.

\begin{table}[htbp]
\caption{\label{table:24Ne-E2} 
The values of $B(E2)$ for $^{24}$Ne listed in units of
 $e^2$fm$^4$.
See also caption in Table~\ref{table:24Mg-E2}.
Experimental data are taken from Ref.~\cite{ENSDF}}
\begin{tabular}{ccccc} \hline \hline
 & EXP  & CHB+LQRPA & neutron & proton \\ \hline
$2_1^+ \rightarrow 0_1^+$ & 28.0 &  33.576 &  14.124 &   6.813 \\
$4_1^+ \rightarrow 2_1^+$ & - &  56.446 &  21.854 &  11.906 \\
$6_1^+ \rightarrow 4_1^+$ & - &  74.510 &  27.395 &  16.080 \\ \hline
$3_1^+ \rightarrow 2_2^+$ & - &  45.159 &  17.889 &   9.426 \\
$4_2^+ \rightarrow 3_1^+$ & - &   6.841 &   0.171 &   2.579 \\
$5_1^+ \rightarrow 4_2^+$ & - &  15.843 &   4.637 &   3.747 \\
$6_2^+ \rightarrow 5_1^+$ & - &   7.551 &   0.550 &   2.512 \\
$4_2^+ \rightarrow 2_2^+$ & - &  39.297 &  15.416 &   8.239 \\
$5_1^+ \rightarrow 3_1^+$ & - &  44.036 &  16.045 &   9.540 \\
$6_2^+ \rightarrow 4_2^+$ & - &  61.481 &  23.999 &  12.919 \\ \hline
$2_3^+ \rightarrow 0_2^+$ & - &  29.908 &  12.458 &   6.098 \\
$4_1^+ \rightarrow 2_2^+$ & - &   3.298 &   0.419 &   0.990 \\
$2_2^+ \rightarrow 2_1^+$ & - &  53.242 &  22.961 &  10.675 \\
$2_2^+ \rightarrow 0_1^+$ & - &   0.361 &   0.861 &   0.504 \\
$3_1^+ \rightarrow 4_1^+$ & - &  20.227 &   9.280 &   3.932 \\
$4_1^+ \rightarrow 3_1^+$ & - &  15.732 &   7.218 &   3.058 \\
$3_1^+ \rightarrow 2_1^+$ & - &   1.487 &   1.542 &   1.505 \\
$4_2^+ \rightarrow 4_1^+$ & - &  23.575 &  14.274 &   3.911 \\
$6_1^+ \rightarrow 4_2^+$ & - &   2.375 &   0.325 &   0.701 \\
$4_2^+ \rightarrow 2_1^+$ & - &   0.116 &   1.450 &   0.030 \\
$5_1^+ \rightarrow 6_1^+$ & - &  16.505 &   8.537 &   3.008 \\
$5_1^+ \rightarrow 4_1^+$ & - &   0.642 &   2.095 &   1.034 \\
$6_2^+ \rightarrow 6_1^+$ & - &  13.261 &  11.282 &   1.711 \\
$6_2^+ \rightarrow 4_1^+$ & - &   0.115 &   0.785 &   0.005 \\ \hline
$2_2^+ \rightarrow 0_2^+$ & - &  16.025 &   4.635 &   3.807 \\
$0_2^+ \rightarrow 2_2^+$ & - &  80.124 &  23.175 &  19.034 \\
$0_2^+ \rightarrow 2_1^+$ & - &  23.593 &   5.891 &   5.901 \\
$2_3^+ \rightarrow 0_1^+$ & - &   0.464 &   0.044 &   0.274 \\
$2_3^+ \rightarrow 2_1^+$ & - &   1.843 &   0.567 &   0.428 \\
$2_3^+ \rightarrow 4_1^+$ & - &   6.076 &   2.253 &   1.306 \\ \hline\hline
\end{tabular}
\end{table}

\begin{table}
\caption{\label{table:24Ne-Q} 
Spectroscopic quadrupole moments for $^{24}$Ne 
listed in units of $e$ fm$^2$.
See also caption in Table~\ref{table:24Mg-E2}.}
\begin{tabular}{ccccc} \hline\hline
                   &  EXP    & CHB+LQRPA & neutron & proton \\ \hline
$Q(2_1^+)$ & -     & $-$2.71 &    1.28 & $-$2.23 \\
$Q(4_1^+)$ & -     & $-$8.10 & $-$0.22 & $-$5.33 \\
$Q(6_1^+)$ & -     & $-$11.6 & $-$1.44 & $-$7.24 \\ \hline
$Q(2_2^+)$ & -     & $-$0.21 & $-$1.68 &    0.42 \\
$Q(3_1^+)$ & -     &  0      &    0    &    0    \\
$Q(4_2^+)$ & -     & $-$4.68 & $-$1.75 & $-$2.54 \\
$Q(5_1^+)$ & -     & $-$5.16 & $-$0.47 & $-$3.28 \\
$Q(6_2^+)$ & -     & $-$9.38 & $-$2.80 & $-$5.32 \\ \hline
$Q(2_3^+)$ & -     & $-$8.16 & $-$4.94 & $-$3.79 \\ \hline\hline
\end{tabular}
\end{table}

\section{\label{sec:discussion}Discussion}

\subsection{Rotational hindrance of shape mixing in $^{26}$Mg and $^{24}$Ne}

Here, we discuss the character of the vibrational wave functions in more
detail.
What is commonly seen in the calculated results for $^{26}$Mg and $^{24}$Ne is
the localization of the vibrational wave functions squared
as the increase of the angular momentum.
While the ground $0_1^+$ state spreads over the $\gamma$ direction,
the yrast band tends to localize in the ($\bg$) plane around the prolate region,
even though the collective potential has the shallow oblate minimum.

\begin{figure}[htbp]
\begin{tabular}{cc}
\includegraphics[width=45mm]{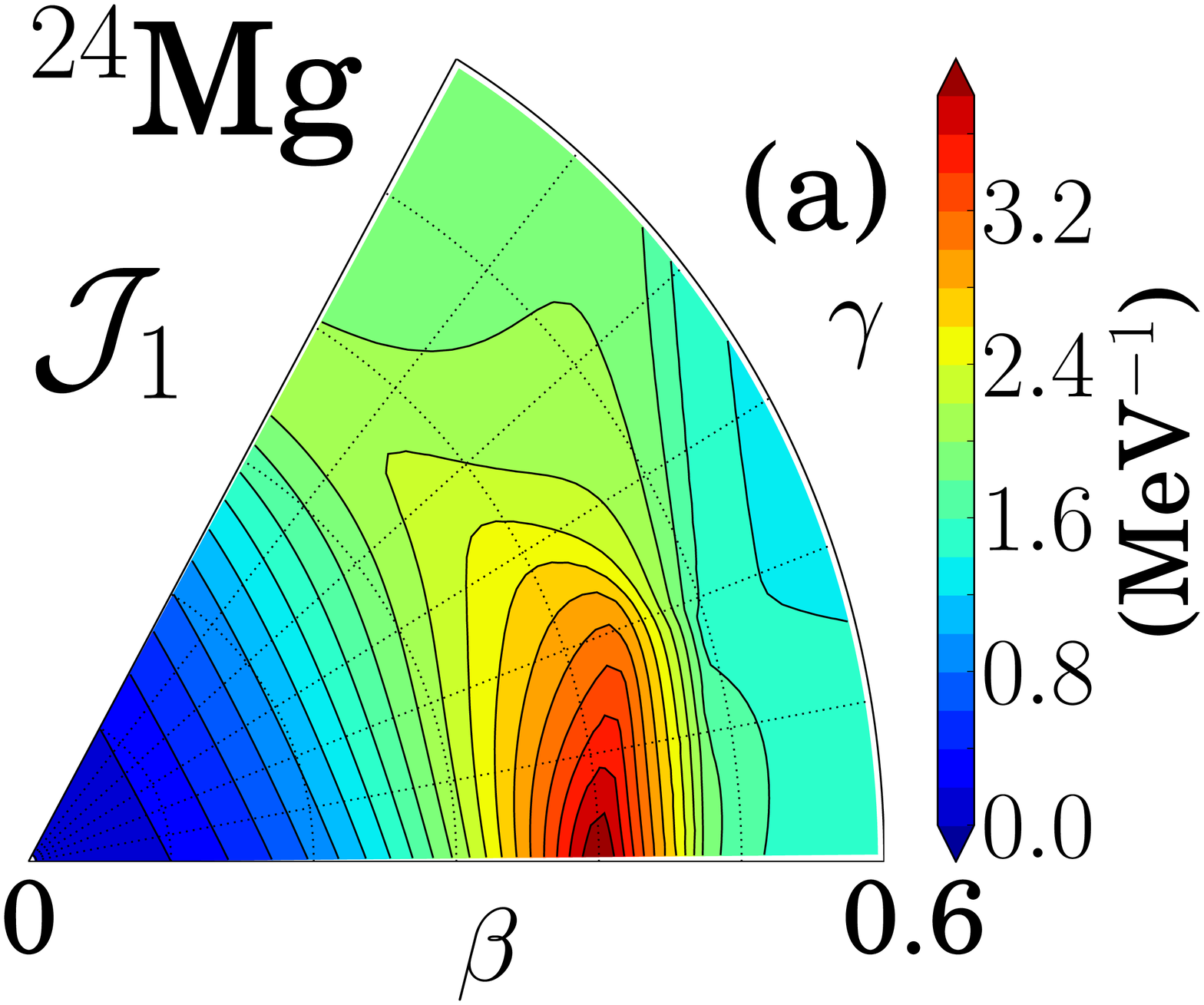} &
\includegraphics[width=45mm]{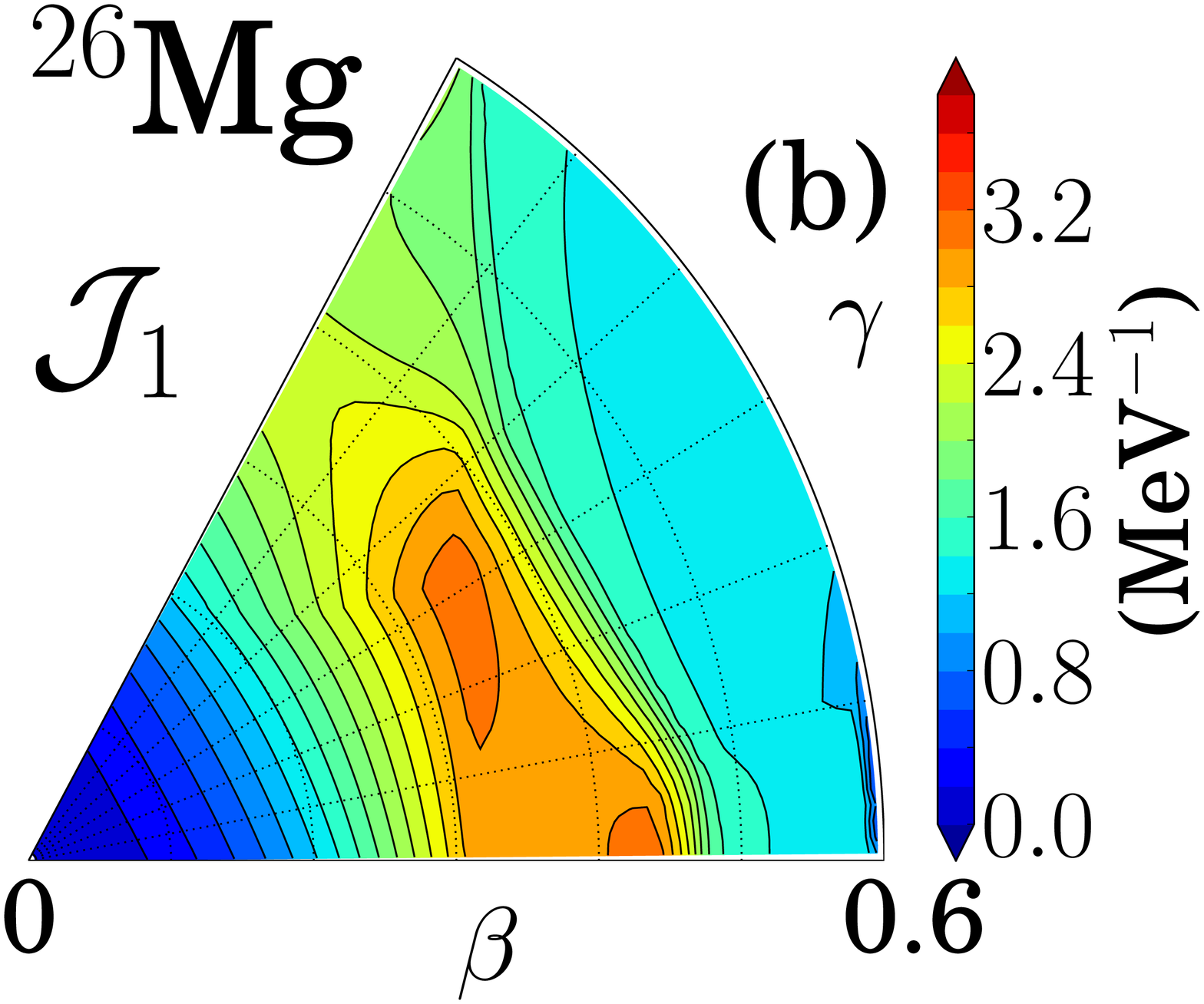} \\
\includegraphics[width=45mm]{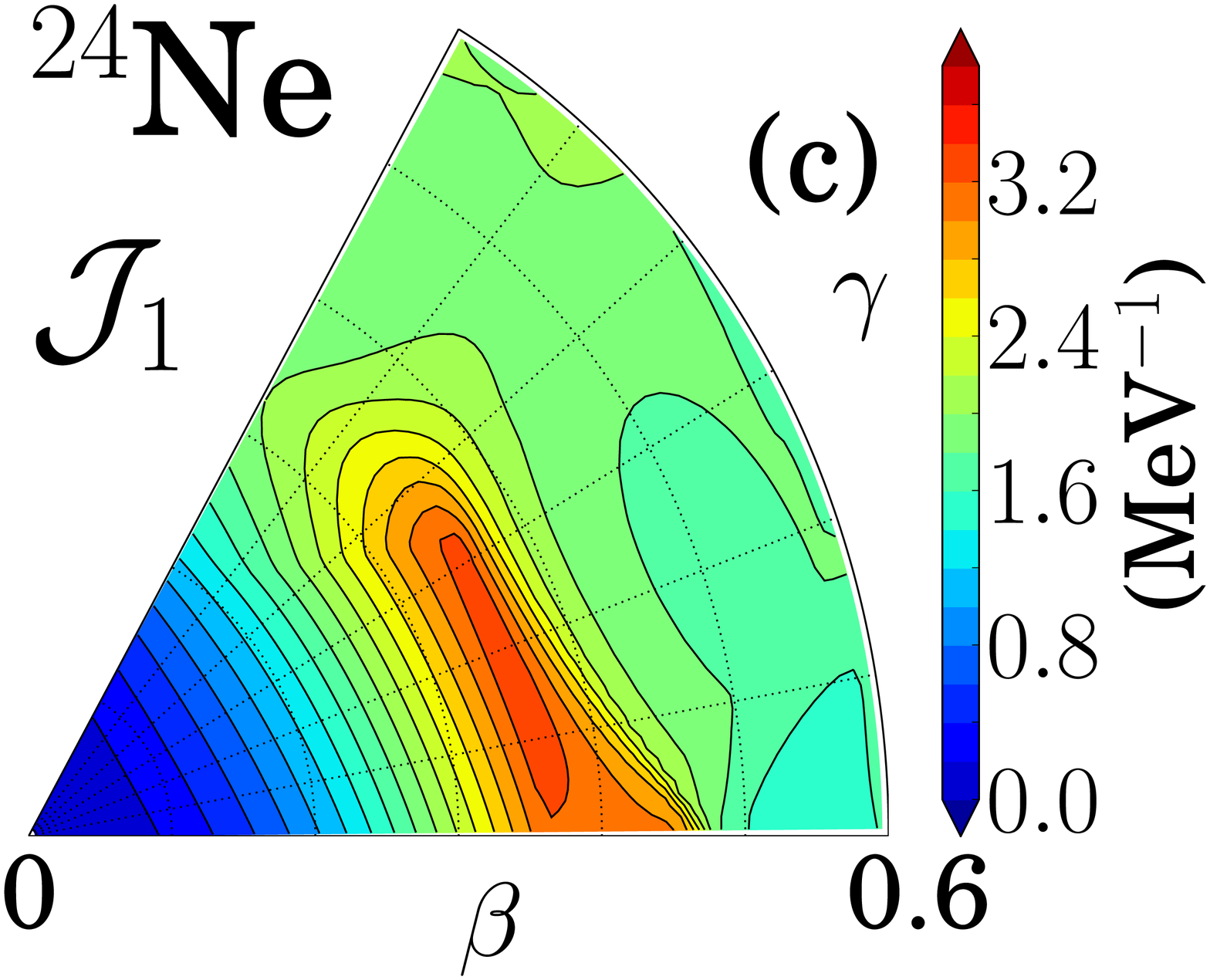} &
\includegraphics[width=45mm]{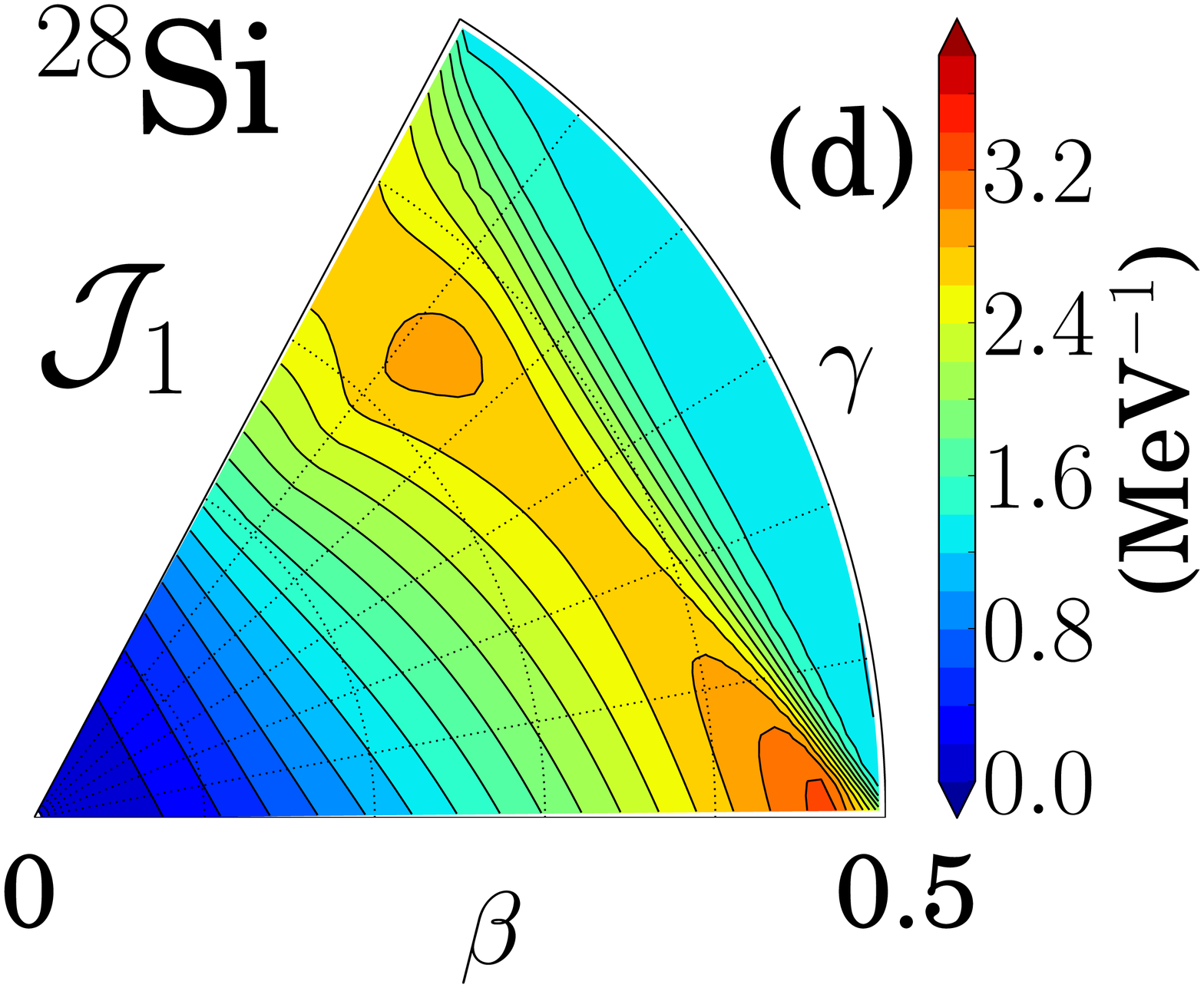} 
\end{tabular}
\caption{\label{fig:MOI}
(Color online) Rotational moments of inertia about the intermediate
 axis, $\Jc_1(\bg)$.}
\end{figure}

This rotational hindrance of shape mixing
is also seen in the cases of oblate-prolate shape coexistence
\cite{hinohara:014305,PTP.123.129}
and can be understood from the deformation dependence of the 
rotational moments of inertia.
Figure~\ref{fig:MOI} shows the rotational moments of inertia 
about the intermediate axis, $\Jc_1(\bg)$, for $^{24,26}$Mg, $^{24}$Ne, and $^{28}$Si.
An oblate-prolate asymmetry is seen in the rotational moments of inertia;
$\Jc_1(\bg)$ for $^{26}$Mg and $^{24}$Ne 
becomes larger in the prolate side than the oblate side for the constant $\beta$ value.
This is the reason why the prolate shape is favored especially for 
high angular momentum states.
Moreover, due to the strong shell effect
there exists a maximum point of the rotational moment of inertia
in the $(\bg)$ plane, which is inconsistent with
the ideal irrotational moments of inertia proportional to the $\beta^2$. 
As is well known, the pairing correlation decreases the moment of inertia \cite{Ring-Schuck},
and in the region where the pairing gap vanishes,
the moment of inertia becomes larger.
In the case of $^{26}$Mg,
the neutron and proton pairing gaps vanish in the prolate region
where the moment of inertia becomes large in this region (Fig.~\ref{fig:gap}).
Because of the behavior of the rotational moments of inertia,
this prolate region is favored in rotational kinetic energy, while 
it is unfavored in collective potential energy, which 
increases as the deformation increases.
As a result,
the vibrational wave function for higher angular momentum states
localizes, and the $\beta$- or $\gamma$-soft nature 
of the vibrational wave function is hindered.

The change of the yrast state structure discussed above is also seen in $^{28}$Si.
In the case of $^{28}$Si, one can see that the deformation of the vibrational wave function 
grows as angular momentum increases. The minimum of the collective
potential locates around $\beta=0.26$,
while the maximum of the moment of inertia locates around $\beta=0.36$ in the oblate side.
This is the reason why the deformation increases in the yrast band of $^{28}$Si.
In case of $^{24}$Mg, however, the minimum of the collective potential  
and the maximum of the moment of inertia coincide around $\beta=0.41$,
and such a change in the structure of the yrast band does not occur.

\subsection{Analysis with mirror nucleus method}

One of the interesting issues in $N\ne Z$ nuclei in the middle of $sd$-shell
region is the properties of the neutron and proton
quadrupole transition matrix elements. In the mirror
nucleus method \cite{PhysRevLett.42.425}, the proton matrix element $M_p$ is
determined from the $E2$ transitions as 
$B(E2;I\rightarrow I') = M^2_p/(2I + 1)$, 
while the neutron one $M_n$ is determined
from the same $E2$ transitions of the mirror nucleus as
$B_{\rm mirror}(E2;I\rightarrow I') = M^2_n/(2I + 1)$. 
Although the $M_n/M_p$ ratio should be equal to
the ratio $N/Z$ in the simple collective model,
 it has been experimentally suggested that the $M_n/M_p$ ratio
for the ground band transitions deviates from $N/Z$ in
some $N\ne Z$ nuclei \cite{Bernstein1981255, PTPS.146.575}
indicating possible difference
between proton and neutron shapes or shape dynamics.
Moreover, the $M_n/M_p$ ratio for the $2^+_2\rightarrow 0^+_1$ transition
in $^{26}$Mg is known to be extremely larger than the expected value
$N/Z$ in mirror nucleus method and also in the analysis of $(p,p')$ scattering reactions
\cite{Alons198141,PhysRevC.31.736}.
The value of $(M_n/M_p)/(N/Z)$ is $1.83 \pm 0.34$ \cite{ENSDF}.
This clearly indicates the
dominance of the neutron matrix element in this transition.
In this subsection, we discuss the origin of the neutron dominance
in the $2_2^+ \rightarrow 0_1^+$ transition in $^{26}$Mg
in terms of the large-amplitude triaxial shape dynamics.
We also discuss the yrast transition $2_1^+\rightarrow 0_1^+$ in 
$^{26}$Mg and $^{24}$Ne.

\subsubsection{$2_2^+\rightarrow 0_1^+$ transition in $^{26}$Mg}

In Table~\ref{table:22to01}, the experimental and theoretical values for 
$B(E2;2_2^+\rightarrow 0_1^+)$ transition are summarized.
The theoretical value $(M_n/M_p)/(N/Z)=2.15$ reproduces the neutron dominance very well.
Actually, the bare proton contribution to this transition in $^{26}$Mg
is more than hundred times smaller than the neutron one.
As seen from Table~\ref{table:24Mg-E2} and \ref{table:28Si-E2}, such difference 
cannot be found in $N=Z$ nuclei.
The shell model value 2.10 \cite{PhysRevC.26.2247} given with the effective charges $(e_n,e_p)=(0.35,1.35)$ also explains the neutron dominance of this transition.

\begin{figure*}[htbp]
\begin{tabular}{ccc}
\includegraphics[width=50mm]{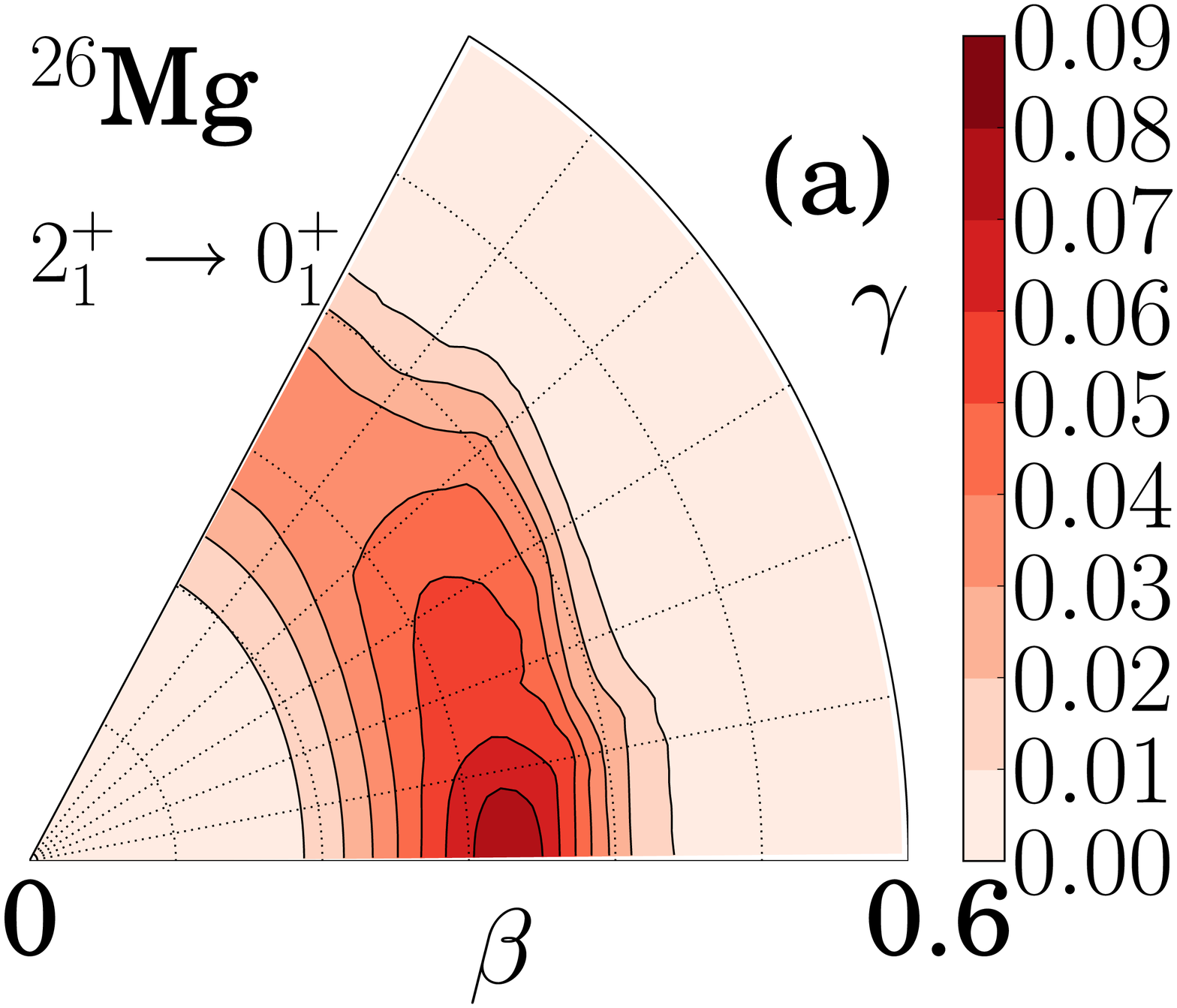}   &
\includegraphics[width=50mm]{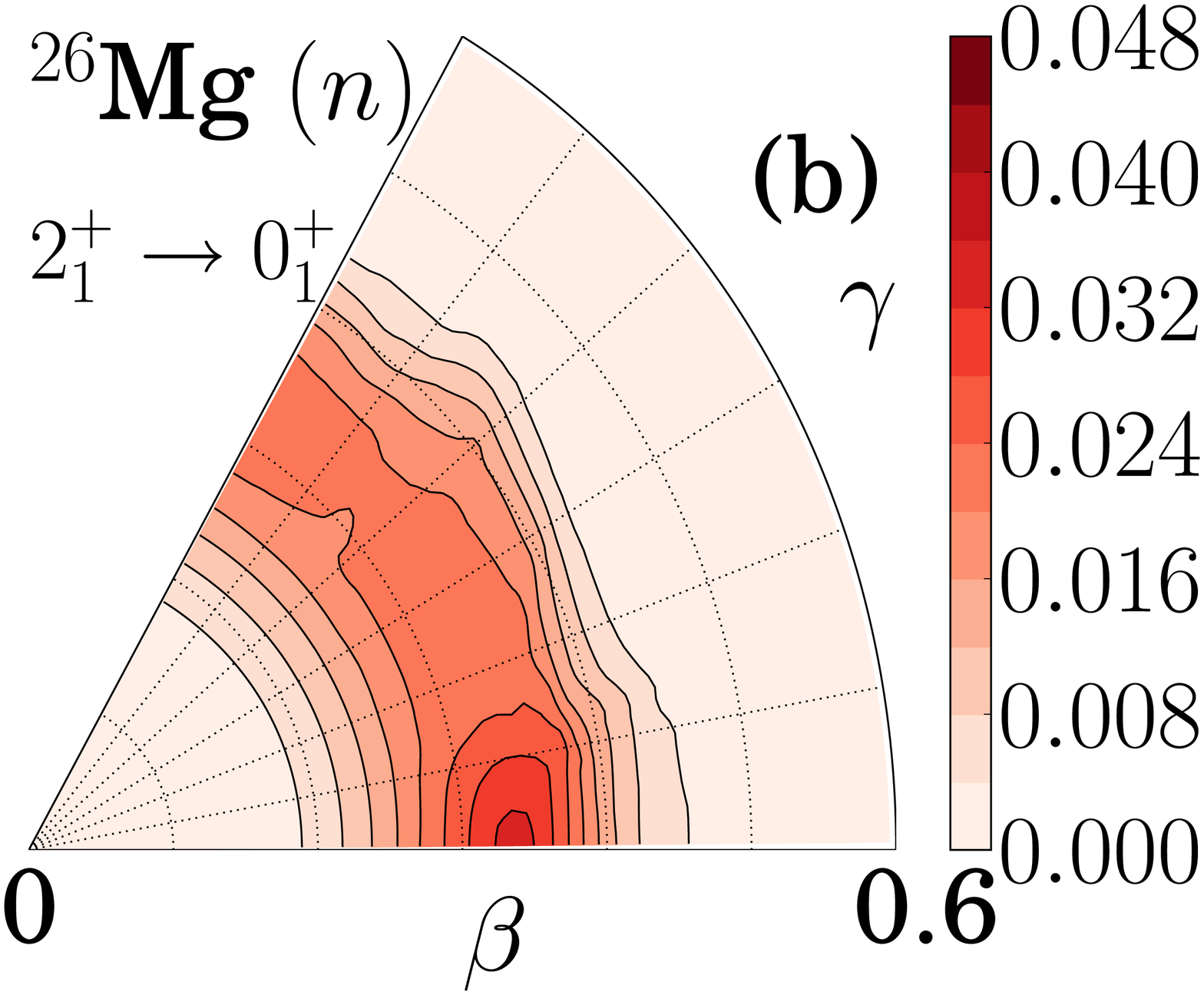} &
\includegraphics[width=50mm]{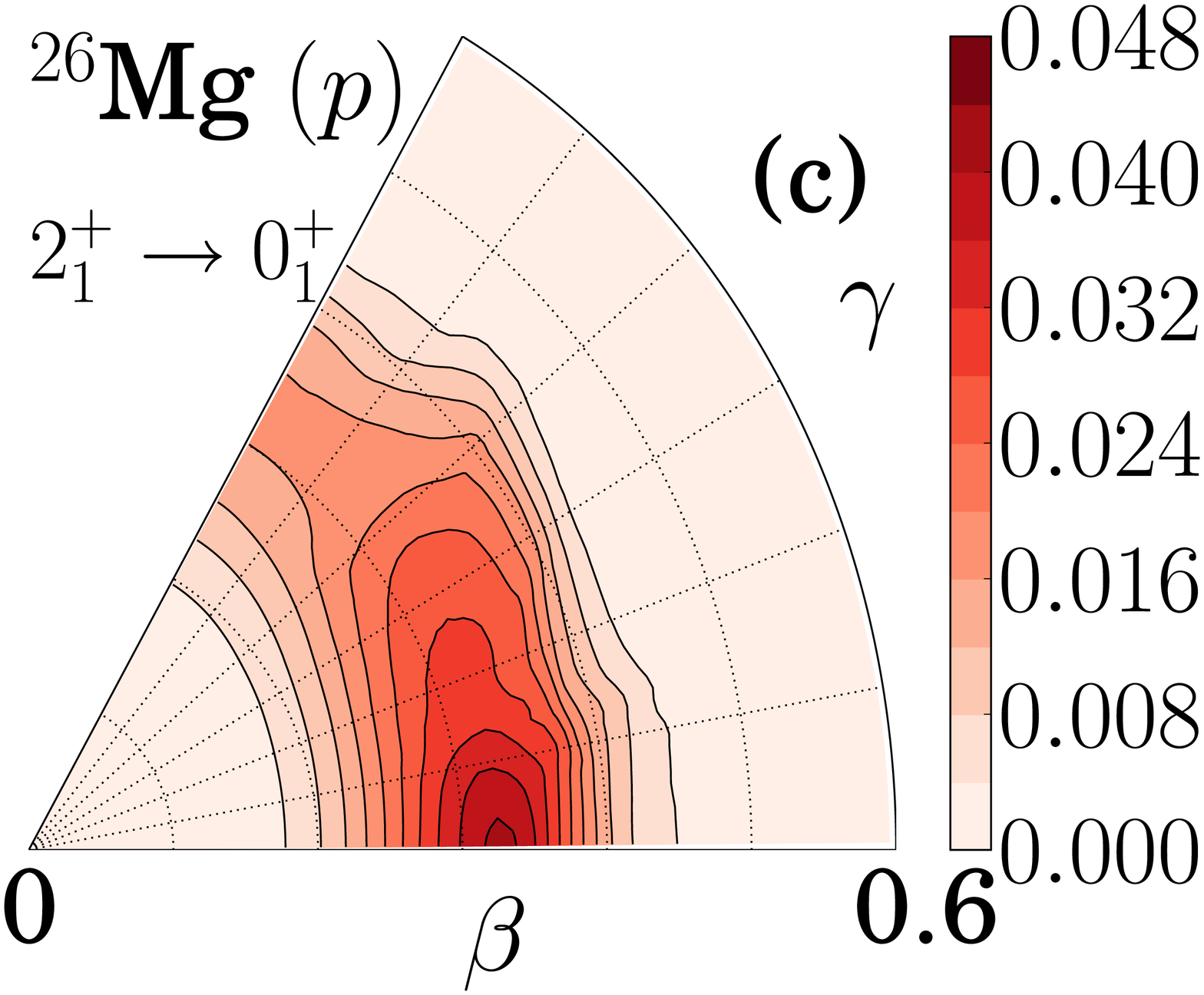} \\
\includegraphics[width=50mm]{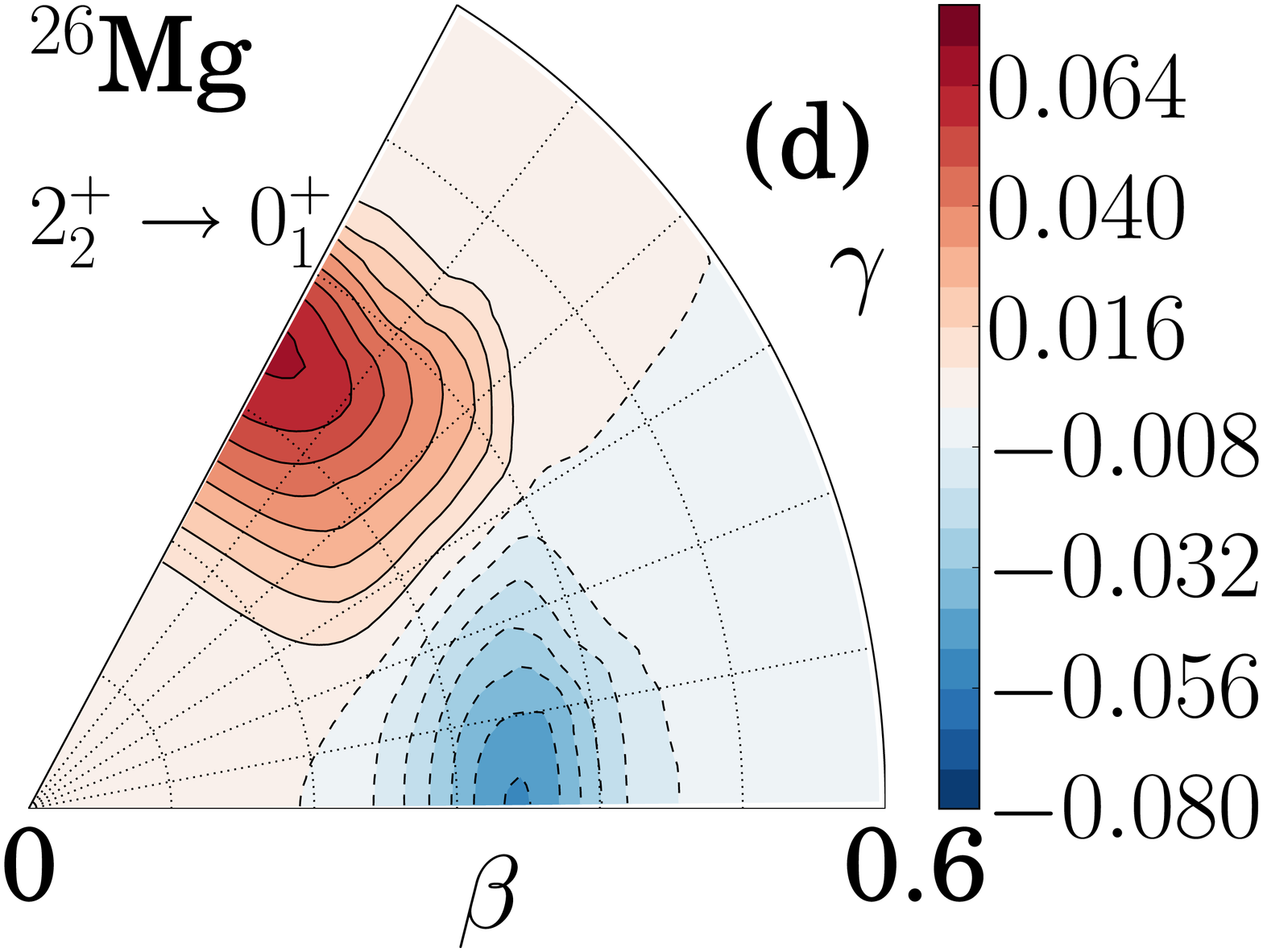}   &
\includegraphics[width=50mm]{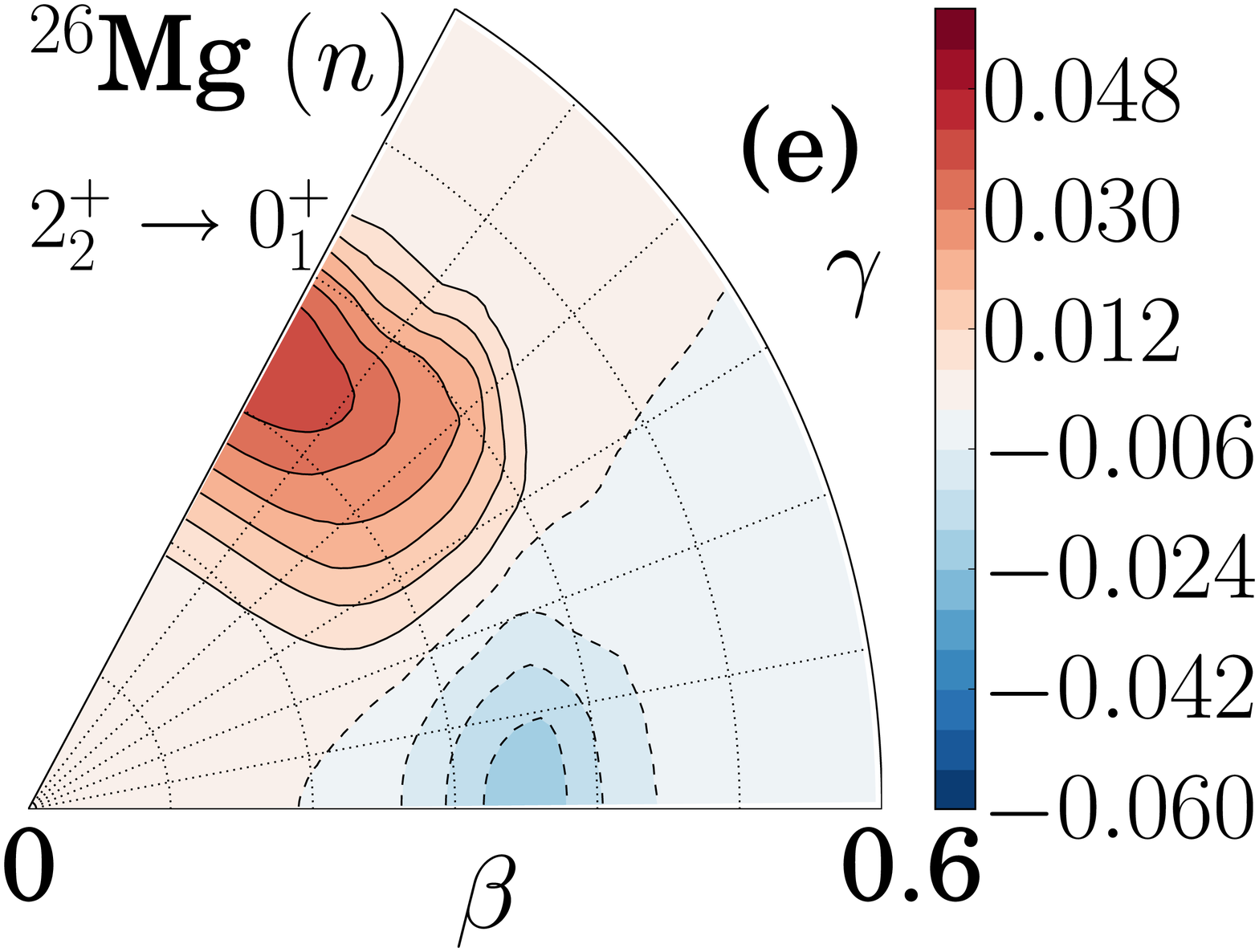} &
\includegraphics[width=50mm]{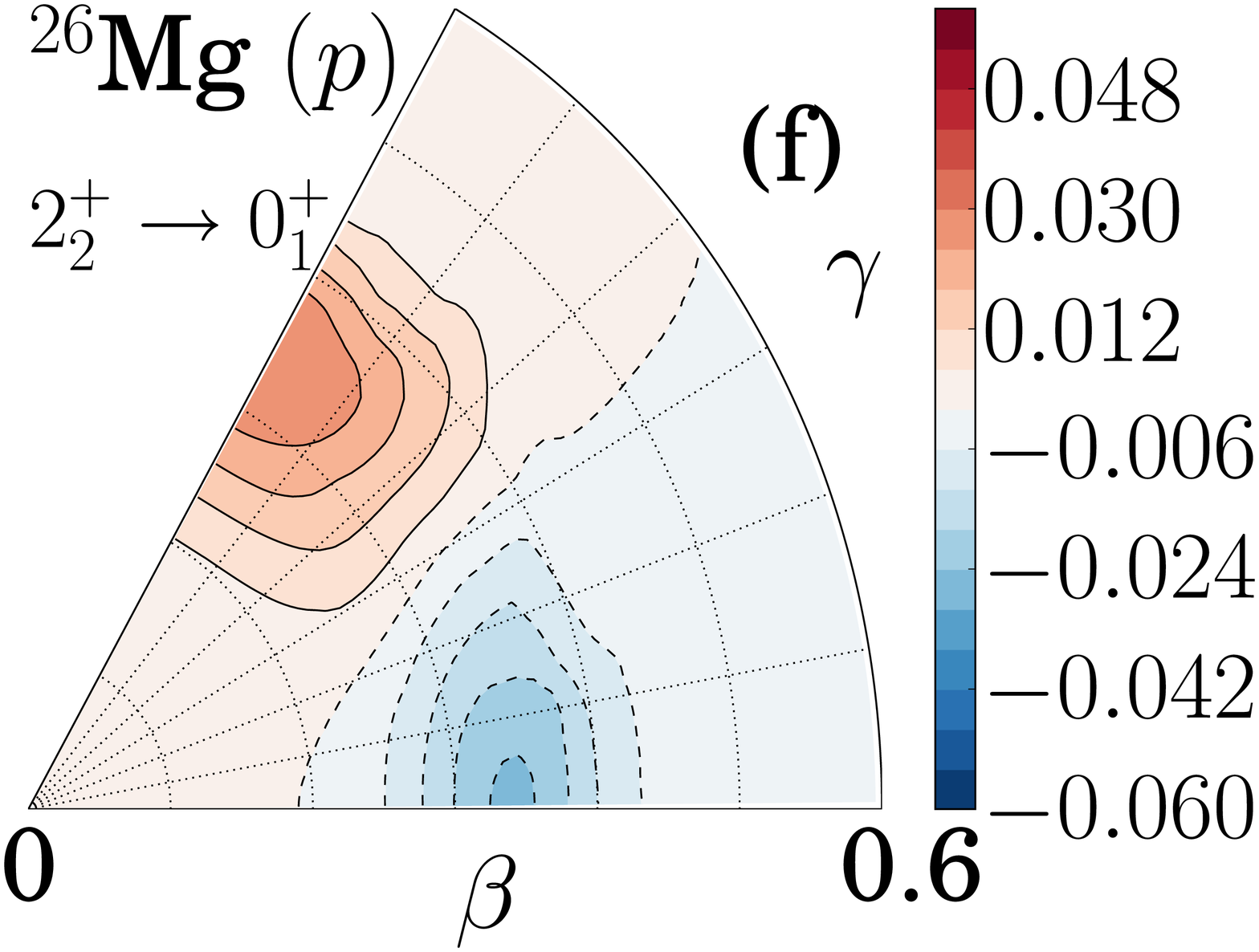}
\end{tabular}
\caption{\label{fig:26Mg-E2}
(Color online)  $E2$ transition density 
$\beta^4 \rho^{(E2)}_{\alpha I \alpha' I'}(\bg)$ for 
$2_1^+\rightarrow 0_1^+$ and $2_2^+ \rightarrow 0_1^+$ transitions in $^{26}$Mg.
Three sets of effective charges $(e_{\rm eff}^{(n)}, e_{\rm eff}^{(p)})$
=(0.5,1.5), (1,0) and (0,1) are used for left, middle and right panels, respectively.}
\end{figure*}

To analyze the mechanism of the neutron dominance in this transition,
we present the $E2$ transition density $\rho_{\alpha I \alpha' I'}^{(E2)}(\bg)$ defined in Eq.~(\ref{eq:E2density})
for $^{26}$Mg in Fig.~\ref{fig:26Mg-E2}.
Let us first compare $E2$ transition densities for $2_1^+\rightarrow 0_1^+$ and
$2_2^+\rightarrow 0_1^+$.
Because of the structure of the yrast vibrational wave functions,
the sign of the transition density for the former in-band transition 
is positive
all over the $(\bg)$ deformation,
while the transition density for the latter transition
changes its sign in the $(\bg)$ plane,
since the vibrational wave function of an excited state
has nodes in the $(\bg)$ plane.
In the case of the $2_2^+\rightarrow 0_1^+$ transition as seen in Fig.~\ref{fig:26Mg-E2},
the sign of the transition density is opposite in the prolate region 
and oblate region, and this results in the cancellation after the $(\bg)$ integration.
In the case of $^{26}$Mg, the proton matrix element
is almost completely canceled after the integration of the transition density.
Concerning the neutrons, the contribution to this transition density is
relatively larger in the oblate region than in the prolate region, since the neutron favors the oblate
deformation for $N=14$ system. This situation produces the large $M_n/M_p$ ratio.

\begin{figure}[htbp]
\includegraphics[width=70mm]{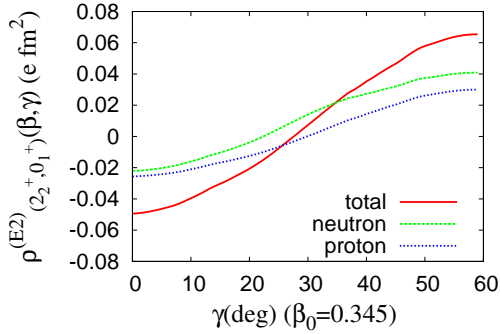}
\caption{\label{fig:26Mg-E2-gamma}
(Color online) $E2$ transition density 
$\beta^4 \rho^{E2}_{\alpha I \alpha' I'}(\bg)$
for
$2_2^+\rightarrow 0_1^+$ transition in $^{26}$Mg at $ \beta = 0.345$
is plotted as function of $\gamma$.
The red solid, green dashed, and blue dotted lines show the calculations with three sets
of effective charges,
$(e^{(n)}_{\rm eff}, e^{(p)}_{\rm eff})$= (0.5,1.5), (1,0), and
 (0,1), respectively.}
\end{figure}

This cancellation taking place in the $(\bg)$ plane is the result of the large-amplitude 
collective dynamics in the $(\bg)$ plane,
 especially in the $\gamma$ direction.
The importance of triaxial degree of freedom is clearly seen from
Fig.~\ref{fig:26Mg-E2-gamma}, where
the $E2$ transition density is plotted as a function of $\gamma$ for a constant
value of $\beta$.
It is seen that the proton transition density is almost anti-symmetric with
respect to $\gamma=30^\circ$, while the asymmetry 
is present for neutron transition density.

\begin{table}[htbp]
\caption{\label{table:22to01}
The values of $B(E2;2_2^+\rightarrow 0_1^+)$ for $^{26}$Mg and $^{26}$Si
listed in units of $e^2$fm$^4$.
Theoretical values for $^{26}$Mg are calculated with the effective charges
$(e^{(n)}_{\rm eff}, e^{(p)}_{\rm eff})=(0.5,1.5)$, while 
those for $^{26}$Si are calculated assuming the mirror symmetry.
The bare neutron and proton contributions calculated with 
$(e^{(n)}_{\rm eff}, e^{(p)}_{\rm eff})=(1,0)$ and (0,1) are also listed.
Experimental data and shell model calculation are taken from Refs.~\cite{ENSDF}, and \cite{PhysRevC.26.2247}, respectively.}
\begin{tabular}{cccccc} \hline\hline
          &    EXP        & CHB+LQRPA & neutron & proton & SM \\ \hline
$^{26}$Mg & 1.60$\pm$0.32 &   0.765   & 2.040   &  0.011 & 3.28 \\ 
$^{26}$Si & 7.32$\pm$2.28 &   4.82    & 0.011   &  2.040 & 19.7 \\ \hline\hline
\end{tabular}
\end{table}

\subsubsection{$2_1^+\rightarrow 0_1^+$ transition in $^{26}$Mg and $^{24}$Ne}

In contrast to the large $(M_n/M_p)/(N/Z)$ for the $2_2^+\rightarrow 0_1^+$ transition discussed above, 
the experimental value of the ratio $(M_n/M_p)/(N/Z)$ for the $2_1^+\rightarrow 0_1^+$ transition
in $^{26}$Mg is $0.92\pm0.05$, which is close to unity. 
This indicates that a simple collective model picture
with the usual assumption that the radius and the deformation for neutrons are consistent with those for protons
is expected to hold.
In the case of $^{24}$Ne, the experimental value of this ratio is $0.59\pm0.11$.
The possible suppression of the ratio from unity may suggest that the simple picture does not hold 
for neutrons and protons in this system,
and the smaller neutron deformation
than the proton deformation is expected in the ground state of $^{24}$Ne \cite{PhysRevC.71.014303}.

We evaluate the $M_n/M_p$ ratio by using the mirror nucleus method.
In Table~\ref{table:mirror}, $B(E2;2_1^+\rightarrow 0_1^+)$ values for 
$^{26}$Mg, $^{26}$Si, $^{24}$Ne, and $^{24}$Si are summarized.
The relative magnitudes of $E2$ transition probability for mirror pairs
cannot be satisfactory reproduced by the theoretical calculation both for 
$A=24$ and 26 systems.

The theoretical value of the ratio $(M_n/M_p)/(N/Z)$ for $^{26}$Mg is 0.81.
The calculated value is smaller than unity,
and qualitatively reproduces the tendency of the experimental value.
The shell model value \cite{PhysRevC.26.2247} gives smaller ratio 0.69
than the present calculation.
For $^{24}$Ne, the ratio is calculated to be 0.86.
The results for $^{24}$Ne fail to quantitatively describe the suppression of the ratio extracted from the
central values of the experimental $B(E2)$.
For more detailed discussions, precise measurements of the $E2$ transition strengths for $^{24}$Ne and $^{24}$Si 
are required.

\begin{table}[htbp]
\caption{\label{table:mirror}
The values of $B(E2;2_1^+\rightarrow 0_1^+)$ for 
$^{26}$Mg, $^{26}$Si, $^{24}$Ne, and $^{24}$Si are summarized in
 units of $e^2$fm$^4$.
Theoretical values for $^{26}$Si and $^{24}$Si are calculated assuming
 the mirror symmetry. See also caption in Table~\ref{table:22to01}.
Experimental values and shell model values are taken from Refs.~\cite{PTPS.146.575,PhysRevC.64.057304,ENSDF}, and Ref.~\cite{PhysRevC.26.2247}, respectively.}
\begin{tabular}{cccccc} \hline\hline
      & EXP & CHB+LQRPA & neutron & proton  & SM \\ \hline
$^{26}$Mg & 61.4$\pm$ 1.8 & 52.870 & 11.131 & 13.953 & 57.1 \\
$^{26}$Si & 70.5$\pm$ 6.9 & 47.226 & 13.953 & 11.131 & 36.5 \\ \hline
$^{24}$Ne & 28.0$\pm$ 6.6 & 33.576 & 14.124 &  6.813 & - \\
$^{24}$Si & 19.1$\pm$ 5.9 & 48.197 &  6.813 & 14.124 & - \\ \hline\hline
\end{tabular}
\end{table}

\section{\label{sec:summary}Summary}

Large-amplitude triaxial quadrupole deformation dynamics in the low-lying states 
of $sd$-shell nuclei, $^{24}$Mg, $^{28}$Si, $^{26}$Mg, and $^{24}$Ne
are analyzed on the basis of the quadrupole collective Hamiltonian derived microscopically from the
CHFB + LQRPA method.

As for the $N=Z$ systems, the calculation reproduces the prolate rotational band and the $\gamma$ vibrational band in $^{24}$Mg, and the oblate rotational band and the $\beta$ vibrational band in $^{28}$Si.
As for $N\neq Z$ systems, $^{26}$Mg and $^{24}$Ne, the collective potentials are shown to be soft against the $\beta$ and $\gamma$ deformations,
and the large shape-fluctuation in the $(\bg)$ plane is found in the
vibrational wave functions of the ground states.

The yrast bands show rotational hindrance of the shape mixing,
and the states localize around the prolate region as the angular momentum increases.
The neutron and proton quadrupole matrix elements are analyzed for $N\neq Z$ systems.
The neutron dominance in the $2_2^+\rightarrow 0_1^+$ transition in
$^{26}$Mg is explained in terms of the large-amplitude collective
dynamics in the $\gamma$-direction.
The neutron and proton matrix elements for $2_1^+\rightarrow 0_1^+$ yrast
transition are analyzed with use of the mirror nucleus method for $^{26}$Mg and $^{24}$Ne,
Also in other $N\ne Z$ nuclei,
differences in the behavior of neutrons and protons in
large-amplitude shape dynamics are expected to be interesting.

\begin{acknowledgments}
The authors are grateful to K. Sato, T. Nakatsukasa, M. Matsuo and 
K. Matsuyanagi for helpful discussions.
One of the authors (N.~H.) is supported 
by the Special Postdoctoral Researcher Program of RIKEN.
The numerical calculations were carried out on 
Altix3700 BX2 at Yukawa Institute for Theoretical Physics
in Kyoto University, and RIKEN
Cluster of Clusters (RICC) facility.
This work is supported by Grants-in-Aid for Scientific
Research (No.~22540275) from the Japan Society for the
Promotion of Science (JSPS) and the JSPS Core-to-Core Program
``International Research Network for Exotic Femto
Systems (EFES),'' and also by the Grant-in-
Aid for the Global COE Program ¡ÈThe Next Generation
of Physics, Spun from Universality and Emergence¡É
from the Ministry of Education, Culture, Sports, Science
and Technology (MEXT) of Japan.
\end{acknowledgments}

%\appendix

%\section{Appendix}

\bibliographystyle{apsrev} 
\bibliography{../../../Bibtex/paper}% Produces the bibliography via BibTeX.

\end{document}